%% file: S4_EAH.tex
\pdfoutput=1

\documentclass[aps,prd,showpacs,amssymb,amsmath,amsfonts,superscriptaddress,
twocolumn,floatfix,preprintnumbers,altaffilletter]{revtex4}
\usepackage{graphicx}
\usepackage{subfigure}
\usepackage{hyperref}
\usepackage{amssymb}
\usepackage{amsmath}
\usepackage{longtable}
\usepackage{rotating}
\usepackage{color}
\usepackage{epsfig}
\usepackage{epsf}
\usepackage{fancyhdr}

\newcommand{\Tobs}{T_{\mathrm{obs}}}
\newcommand{\Tspan}{T_{\mathrm{span}}}
\newcommand{\Tspanj}{T_{\mathrm{span},j}}
\newcommand{\Tsft}{T_\mathrm{SFT}}
\newcommand{\hours}{\mathrm{h}}
\newcommand{\Hz}{\mathrm{Hz}}
\newcommand{\second}{\mathrm{s}}
\newcommand{\muHz}{\mu\mathrm{Hz}}
\newcommand{\D}{\mathcal{D}}
\newcommand{\seg}{\mathrm{seg}}

\newcommand{\ee}[1]{\!\times\!10^{#1}}

\newcommand{\F}{\mathcal{F}}

\def\lineup{\def\0{\hbox{\phantom{\footnotesize\rm 0}}}%
    \def\m{\hbox{$\phantom{-}$}}%
    \def\-{\llap{$-$}}}

\input defs.tex

\DeclareGraphicsExtensions{.png}

\begin{document}
\pagestyle{fancy}

\rhead[]{}
\lhead[]{}

\title{The Einstein@Home search for periodic gravitational waves in LIGO S4 data}

\input{authorlist}

\date{\today}

\begin{abstract}
\noindent
A search for periodic gravitational waves, from sources such as
isolated rapidly-spinning neutron stars, was carried out using 510
hours of data from the fourth LIGO science run (S4).  The search was
for quasi-monochromatic waves in the frequency range from $50\,\Hz$ to
$1500\,\Hz$, with a linear frequency drift $\dot{f}$ (measured at the solar
system barycenter) in the range $-f/\tau < \dot f < 0.1\,f/\tau$, where the
minimum spin-down age $\tau$ was 1000~years for signals below
$300\,\Hz$ and 10\,000~years above $300\,\Hz$.  The main
computational work of the search was distributed over approximately
$100\,000$ computers volunteered by the general public.  This large
computing power allowed the use of a relatively long coherent
integration time of $30\,\mathrm{hours}$, despite the large
parameter space searched.  No statistically significant signals were
found.  The sensitivity of the search is estimated, along with the
fraction of parameter space that was vetoed because of contamination
by instrumental artifacts.  In the $100\,\Hz$ to $200\,\Hz$ band, more than
90\% of sources with dimensionless gravitational wave strain amplitude
greater than $10^{-23}$ would have been detected.
\end{abstract}
\preprint{LIGO-P080021-01-Z}
\maketitle
%

\section{Introduction\label{sec:introduction}}

Gravitational waves are a fundamental prediction of Einstein's General
Theory of Relativity \cite{Einstein1, Einstein2}.  But these waves are
very weak, so although there is compelling indirect evidence for their
existence \cite{Taylor}, direct detection has so far not been
possible.

In the past decade, advances in lasers, optics and control systems
have enabled construction of a new generation of gravitational-wave
detectors \cite{HoughRowan} that offer the first realistic promise of
a direct detection.  The Laser Interferometer Gravitational-wave
Observatory (LIGO) \cite{ligo1,ligo2} is currently the most sensitive
of these instruments.  LIGO consists of three kilometer-scale instruments.
Two are located in a common vacuum envelope in Hanford, Washington, USA and
the other is located in Livingston, Louisiana, USA.

This paper reports on the results of the Einstein@Home search for
``continuous wave'' sources in the data from the fourth LIGO science
run (S4).  The configuration of the LIGO detectors during the S4 run
is described in a separate instrumental paper
\cite{S4InstrPaper}.

\subsection{Continuous wave (CW) sources and detection methods}

``Continuous waves'' (CW) are quasi-monochromatic gravitational-wave
signals whose duration is longer than the observation time. They have
a well-defined frequency on short time-scales, which can vary slowly
over longer times.  These types of waves are expected, for example,
from spinning neutron stars with non-axisymmetric deformations. If the
system is isolated, then it loses angular momentum to the radiation.
The spinning slows down, and the gravitational-wave frequency
decreases.  Gravitational acceleration towards a large nearby mass
distribution can also produce such a frequency drift (of either sign).
Many possible emission mechanisms could lead the to the emission of
such waves by spinning neutron stars~\cite{Bildsten:1998ey,
  Ushomirsky:2000ax,
  Cutler:2002nw,Melatos:2005ez,Owen:2005fn,Owen:1998xg,Andersson:1998qs,
  Jones:2001yg,VanDenBroeck:2004wj}.

If there were no acceleration between the LIGO detectors and the GW
sources, then it would be possible to search for CW signals from unknown
sources using only ``standard'' computing resources, such as a high-end
workstation or a small computing cluster.  In this case the analysis
technique would be simple: compute the Fast Fourier Transform (FFT)
\cite{fft1,fft2} of the original time-series data, and search
along the frequency axis for peaks in the power spectrum.  Time-domain
resampling or similar techniques could be used to compensate for the
effects of a linear-in-time frequency drift.

However, this simple analysis is not possible because of the
terrestrial location of the LIGO detectors: signals that are purely
sinusoidal at the source are Doppler-modulated by the Earth's motion and thus
are no longer sinusoidal at the detector.  The Earth's rotation about
its axis modulates the signal frequency at the detector by
approximately one part in $10^6$, with a period of one sidereal day.
In addition, the Earth's orbit about the Sun modulates the signal
frequency at the detector by approximately one part in $10^4$, with a
period of one year.  These two modulations, whose exact form depends
upon the precise sky location of the source, greatly
complicate the data analysis when searching for unknown
sources.  The search becomes even more complicated if the CW-emitter
is part of a binary star system, since the orbital motion of the
binary system introduces additional modulations into the waveform.

The ``brute force'' approach to the data analysis problem
would employ matched filtering, convolving all available data with a
family of template waveforms corresponding to all possible putative
sources.  The resulting search statistic is called the
$\F$-statistic and was first described in a seminal paper of
Jaranowski, Kr\'olak, and Schutz~\cite{jks}.  But even for isolated
neutron stars (i.e.\ which are not in binary systems) the parameter
space of possible sources is four-dimensional, with two parameters
required to describe the source sky position using 
standard astronomical equatorial coordinates $\alpha$ (right ascension) 
and $\delta$ (declination), and additional coordinates
$(f, \dot{f})$ denoting the intrinsic frequency and frequency drift.
To achieve the maximum
possible sensitivity, the template waveforms must match the source
waveforms to within a fraction of a cycle over the entire observation
time (with current detectors this is months or years).  So one must
choose a very closely spaced grid of templates in this
four-dimensional parameter space, and the computational cost exceeds
all available computing resources on the planet~\cite{jks3}.  Thus the
direct approach is not possible in practice.

More efficient and sensitive methods for this type of search have been
studied for more than a decade and are under development
\cite{metric1, metric2, metric3, BCCS, BC00}.
In this paper, the frequency-domain method described 
in \cite{S1PulsarPaper,S2FstatPaper} is
used to calculate the $\F$-statistic.  In order to maximize the
possible integration time, and hence achieve a more sensitive coherent
search, the computation was distributed among
approximately $10^5$ computers belonging to $\sim 5 \times 10^4$
volunteers in $\sim 200$ countries.  This distributed
computation project, called Einstein@Home~\cite{EaHURL}, follows the model of a
number of other well-known volunteer distributed computing projects
such as SETI@home~\cite{SETI} and Folding@home~\cite{FOLDING}.

Other methods have also been employed for the CW
search of the S4 data \cite{S3S4TDPaper, S4IncoherentPaper} and
searches for other signal types (burst, inspiral, stochastic background)
have also been carried out \cite{S4second, S4third,
  S4fourth, S4fifth, S4sixth, S4seventh} with this data set.  The
results of these searches are all upper bounds, with no detections
reported.

\subsection{Outline of this paper}

The outline of this paper is as follows. Sections~\ref{sec:DataPrep}
and~\ref{sec:JobConst} describe the overall construction of the
search, including the data set preparation, regions of parameter space
searched, and the choices of thresholds and sensitivities.
Section~\ref{sec:PostProcess} describes the post-processing
pipeline. The level of sensitivity of the search is estimated in
Section~\ref{sec:ExpectSen}.  Section~\ref{sec:VetoMethod}
describes the vetoing of instrumental line artifacts and the fraction
of parameter space that was therefore excluded.
Section~\ref{sec:HardwareInjections} describes the end-to-end
validation of the search and the post-processing pipeline, which was
done by injecting simulated CW signals into the detector hardware.
Section~\ref{sec:Results} describes the final results of the search,
followed by a short conclusion.


\section{Data selection and preparation
\label{sec:DataPrep}}

\begin{figure}
	\includegraphics[scale=0.465,angle=0]{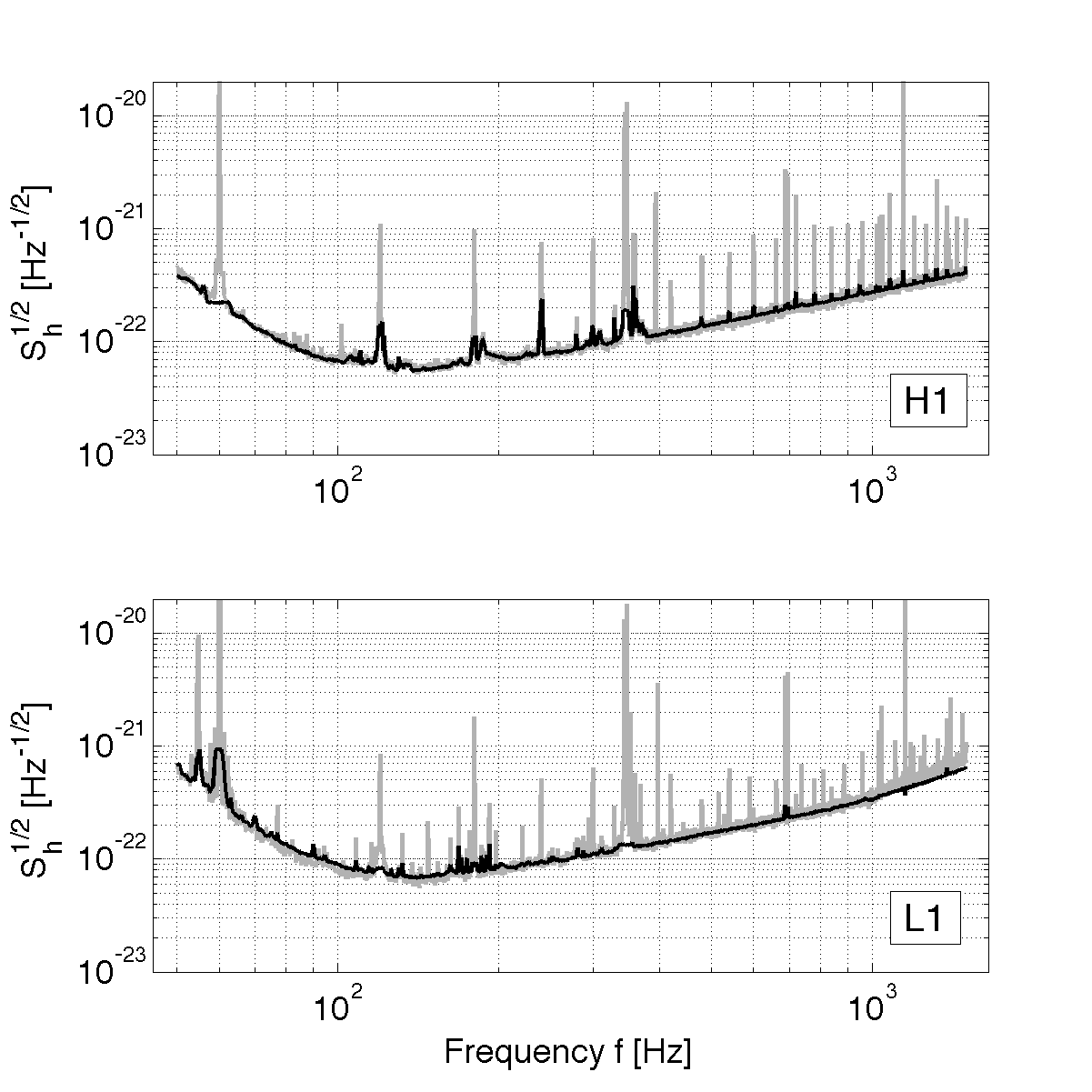}
	\caption{Strain amplitude spectral densities $\sqrt{S_{\rm
              h}(f)}$ of the S4 data from the LIGO detectors H1 (top)
          and L1 (bottom). The gray curves are medians of the {\it
            entire uncleaned} LIGO S4 science-mode data set with a
          frequency resolution of $0.125\,{\rm Hz}$.  The black curves
          show the {\it cleaned} S4 data used in this analysis with a
          frequency resolution of $0.5\,{\rm Hz}$.  The top (bottom)
          plot is the mean of the 10 H1 (7 L1) 30-hour data segments
          used in this Einstein@Home analysis.}
	\label{f:noisefloors}
\end{figure}

The data for the S4 run was collected between February 22, 2005 and
March 23, 2005.  The data analyzed consisted of 300~hours of data from
the LIGO Hanford 4-km (H1) detector and 210~hours of data from the
LIGO Livingston 4-km (L1) detector.

The search method used here (explained in detail in
Section~\ref{sec:JobConst}) consists of computing a coherent
$\F$-statistic over data segments of 30 hours each, and combining
these results via an incoherent coincidence scheme.  However, the
30-hour segments have time-gaps, and the number of templates needed for the
coherent $\F$-statistic step grows rapidly as the gaps get longer.
For this reason, the start and end times of the data segments were
selected based on the criteria that the gaps totaled no more than ten
hours: each data segment contains 30~hours of science-mode data and
lies within a total time span of less than 40~hours.  
Here and in the following the term
``segment'' is always used to refer to one of these time stretches of
data, each of which contains exactly $\Tobs = 30\,\hours$ of data.
The total time spanned by the data in segment $j$ is written
$\Tspanj$; $30\,\hours < \Tspanj < 40\,\hours$.

The data segments consist of \emph{uninterrupted} blocks
of $1800$\,s of \emph{contiguous} science-mode data. This is for technical reasons:
the $\F$-statistic code uses
Short Fourier Transforms (SFTs) over $\Tsft = 1800$\,s as input data,
(this data format is described in~\cite{SFTv1}).
To produce these SFTs, the data is first calibrated in the
time domain using the method described in \cite{hoftpaper, S4CalibrationNote}.
Then the data is windowed in $1800$\,s intervals using a Tukey
window with a characteristic turn-on/turn-off time of $500$\,ms,
followed by an FFT.

Applying the above constraints to the S4 data set yielded a total of
$N_\seg = 17$ data segments (10 from H1, 7 from L1), labeled by
$j = 1, \cdots, 17$.
The GPS start time of segment $j$ is denoted $t_j$, and these values
are listed in Table~\ref{t:datasegs}.
\begin{table}
\begin{center}
\begin{tabular}{cccccc}
\hline
$j$ & Detector  & $t_j$ [s] & $t_j^{\mathrm{end}}$[s] & $\Tspanj$ [s] \\
\hline\hline
1& H1& 794461805& 794583354& 121549\\
2& H1& 794718394& 794840728& 122334\\
3& H1& 795149016& 795285470& 136454\\
4& H1& 793555944& 793685616& 129672\\
5& H1& 795493128& 795626713& 133585\\
6& H1& 793936862& 794078332& 141470\\
7& H1& 794885406& 795015166& 129760\\
8& H1& 794244737& 794378322& 133585\\
9& H1& 794585154& 794714794& 129640\\
10& H1& 793766877& 793910817& 143940\\
11& L1& 795553209& 795678679& 125470\\
12& L1& 795115986& 795246307& 130321\\
13& L1& 795408715& 795545555& 136840\\
14& L1& 794625269& 794762685& 137416\\
15& L1& 794053883& 794197272& 143389\\
16& L1& 794206397& 794328337& 121940\\
17& L1& 794875424& 795016053& 140629\\
\hline
\end{tabular}
\caption{ \label{t:datasegs} Segments of S4 data
  used in this search, in order of decreasing sensitivity at
  $141.3\,\Hz$ for H1 and at $135.3\,\Hz$ for L1.  The
  columns are the data segment index $j$, the GPS start time $t_j$,
  the GPS end time $t^{\mathrm{end}}_j$, and the time spanned
  $\Tspanj = t^{\mathrm{end}}_j - t_j$.
  }
\end{center}
\end{table}

The maximum Doppler modulation (from the Earth's motion about the Sun)
is about one part in $10^4$.  Over the length of S4, and in the
parameter range considered, the frequency changes due to intrinsic
spin-down are smaller still. This means that the CW signals searched
for here always stays within a narrow frequency band, drifting no more than about $\pm0.15\,\Hz$ from some fiducial frequency.  For this reason
the input data, spanning the frequency range of $50\,\Hz$ to
$1500\,\Hz$, is partitioned in the frequency domain into 5800
``slices'' of $0.5$\,Hz plus wings of $0.175$\,Hz on either side.  The
size of one such input data slice is $7\,368\,000$~bytes for H1
(containing 600 SFTs from 10 segments) and $5\,157\,600$~bytes for L1 (containing 420
SFTs from 7 segments).

\input{linetable.tex}

The detector data contains dozens of narrow-band spectral lines whose
origin is instrumental, for example the harmonics of the $60\,{\rm
  Hz}$ mains frequency, and violin modes of the mirror suspensions in
the range from $342\,{\rm Hz}$~-~$350\,{\rm Hz}$ (H1) and $335\,{\rm
  Hz}$~-~$355\,{\rm Hz}$ (L1).  To simplify later analysis, line
features that are known to be instrumental artifacts are removed
(``cleaned'') from the data by replacing the frequency-domain data
bins with computer-generated random Gaussian values.  The frequencies of
these lines are shown in Table~\ref{t:lines}.  The cleaning algorithm
uses a moving-in-frequency median of the power in individual frequency
bins to determine the instrumental noise floor.
To prevent bias at the boundaries of the cleaned regions, the mean of
the random values to replace the line features interpolates linearly
between the noise floor at either side of the line feature.
The median noise strain amplitude
spectra of the final cleaned H1 and L1 data sets are shown in
Figure~\ref{f:noisefloors}.

\section{ Data Processing \label{sec:JobConst}}

\begin{figure}
	\includegraphics[scale=0.54,angle=0]{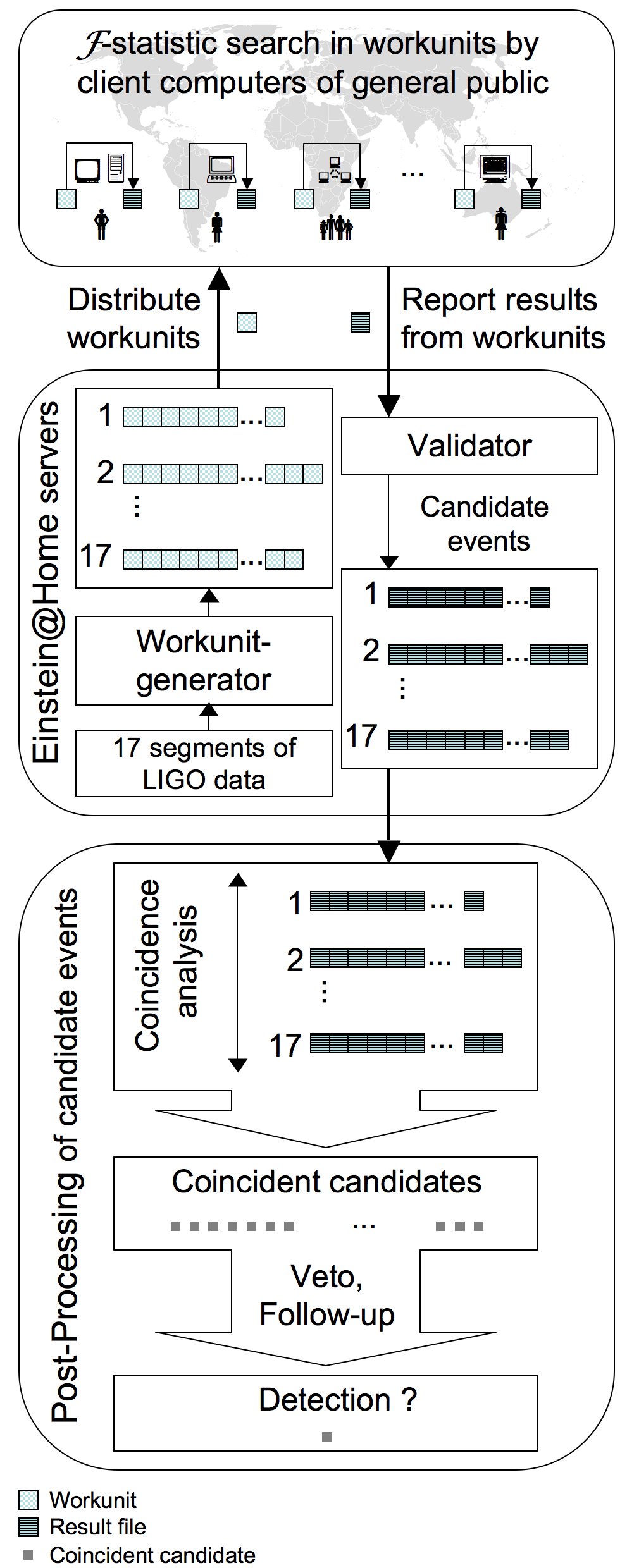}
	\caption{Schematic overview of the Einstein@Home 
	data-processing and subsequent post-processing.}
	\label{f:flowdiagram}
\end{figure}

Figure~\ref{f:flowdiagram} is a schematic flow-diagram of the
Einstein@Home data processing which is described in this section and
in the following section on post-processing.  It shows what parts of
the analysis were done by project participants, what parts were
done on project servers, and the relationships between these.

\subsection{BOINC workunit distribution and validation}
\label{sec:boinc-work-distr}

The computational work of the search is partitioned into
$6\,731\,410$~workunits (separate computing tasks) and processed using
the Berkeley Open Infrastructure for Network Computing
(BOINC)~\cite{BOINC,BOINC1,BOINC2}.  Because the work is done on
computers that are not owned or controlled by our scientific
collaboration or institutions, any individual result could be wrong.
Error sources include defective hardware (such as over-clocked
memory), defective software (erroneous system libraries), and
malicious users (faking correct results).  To identify and eliminate
such errors, BOINC was configured so that each workunit is done
independently by computers owned by at least three different
volunteers.

The most common
types of errors (lack of disk space, corrupted or missing input files,
inconsistent internal state, etc.) are detected during program
execution. If an error is detected during run-time, the program
reports to the Einstein@Home server that the workunit was {\it unsuccessful},
and BOINC generates another
instance of the workunit, to be sent to another volunteer's computer.
This behavior is repeated as necessary until three successful results
have been obtained.

The three successful results obtained for each workunit are then
compared by an automatic validator, which rejects results that do not
agree closely. The validation process is more complicated than simple
byte-by-byte comparison of output files, because Einstein@Home
supports multiple computing platforms (Windows, GNU/Linux, Mac OS X on
Intel and PPC, FreeBSD, and Solaris) and differences in CPU hardware,
compiler instruction ordering, and floating-point libraries mean that
correct and valid result files may exhibit small numerical
differences.  The automatic validation takes place in two steps.

The output files have a fixed five-column format and contain $13\,000$
candidate events, with one line per candidate event, as described in
Section~\ref{ss:WUoutput}.  The first validation step checks that the
file syntax is correct and that each value is within the allowed range
for that column.  This detects most file corruption.

Then the validator does comparison of all possible pairs of result
files.  For a given pair of result files, the validator checks that
corresponding candidate events lie on the same template grid-point and
have $\F$-statistic values that agree to within 1\%. Since each file
contains the $13\,000$ events with the largest values of the
$\F$-statistic, numerical fluctuations in determining the value of
$\F$ can lead to slightly different lists being returned on
different platforms. Hence the validator tolerates unpaired candidate
events if they lie within 1\% of the smallest value on the list.

A workunit is validated once it has three results that agree with one
another to within these tolerances.  If the three results do not pass
this validation step, the Einstein@Home server generates more
instances of this workunit until three \emph{valid} results have been
obtained.  For the search described in this paper, a ``post-mortem''
analysis of the computation shows that the probability of a successful
but invalid result is small (0.36\%), and the errors which make a
successful result invalid are typically unique and irreproducible.
Hence we estimate that it is highly improbable that even a single
incorrect result has been marked as ``valid'' by the automatic
validator.

\subsection{Workunit design}
\label{sec:templ-grids-work}

The different workunits cover (search) different parts of parameter
space.  A key design goal of these workunits is that they should have
roughly equal computational run times, of the order of $\sim 8$~hours,
and that the computational effort to carry out the entire search
should last about 0.5--1 years.  Another key design goal is that
each workunit uses only a small re-usable subset of the total data
set. These allow Einstein@Home volunteers to do useful computations on
the one-day timescale, and minimizes the download burden on their
internet connections and on the Einstein@Home data servers.

Each workunit uses only one segment of data over a narrow frequency-range,
but covers the whole sky and the full range of frequency-derivatives
$\dot{f}$ at that frequency.
Therefore, the entire search is divided into computational units over
different data segments and frequency-bands.
In the following it will be useful to label the workunits by three
indices $(j,k,\ell)$, where $j=1, \cdots, 17$ denotes the data segment,
$k=1,\cdots, 2900$ labels the $0.5\,\Hz$ band covered by the
input data file, and $\ell=1, \cdots,M(j,k)$ enumerates individual
workunits associated with data segment $j$ and frequency band $k$.
Note that each workunit uses a frequency band that is \emph{smaller}
than the $0.5\,\Hz$ covered by the input data files, i.e.\ $M(j,k) \ge 1$.

\subsubsection{Search grid in parameter space}
\label{sss:search-grid}

The parameter space is gridded in such a way that no point has a
``squared-distance'' from its nearest grid point that exceeds a certain
``maximal mismatch''.  The distance is defined by a metric on parameter
space, first introduced in \cite{metric1,metric2}.
The squared distance is 
the fractional loss of squared signal-to-noise ratio (${\rm SNR}^2$) due to
waveform mismatch between the putative signal and the template.
The search grid was constructed based on the projected metric on
the subspace orthogonal to the frequency direction $\partial_f$ with $\dot f=0$..
For any given workunit, the parameter-space grid is a
Cartesian product of uniformly-spaced steps $d f$ in frequency,
uniformly-spaced steps $d \dot{f}$ in frequency derivative,
and a two-dimensional sky-grid, which has non-uniform spacings
determined by the projected metric.
For frequencies in the range
$[50,290)\,\Hz$, the maximal mismatch was chosen as $m = 0.2$
(corresponding to a maximal loss in ${\rm SNR}^2$ of $20\%$), while in the range
$[300,1500)\,\Hz$, the maximal mismatch was $m = 0.5$.
Due to a bug in the script generating the sky-grids,
the range $[290,300)\,\Hz$, was covered by frequency and spin-down steps
corresponding to $m = 0.2$, whereas the sky-grids were
constructed for $m=0.5$. The distribution of actual mismatches in this
frequency range will therefore be somewhat in between those of the
low-frequency and high-frequency workunits.

It can be shown~\cite{EHDoc} that these relatively large mismatches
give near-optimal sensitivity for a coherent search at fixed CPU
power.  Choosing finer grid spacings (i.e.\ a smaller mismatch) would
require searching more grid-points, thus reducing the maximal
possible coherent integration time. A coarser search grid would
allow longer integrations but at a larger average loss in SNR.
Because of these two competing tendencies, the sensitivity as a
function of mismatch $m$ has a maximum in the range
\mbox{$m \sim 0.25$--0.7}, depending on the choice of false-dismissal
rate from the grid mismatch.  Full details of the parameter-space grid and
workunit construction are given in~\cite{EHDoc}; a short summary follows.

\subsubsection{Search grid in frequency and frequency-derivative}

The step-size in frequency was determined using the metric-based expression
\begin{equation}
  \label{eq:1}
  df_j = \frac{2\sqrt{3m}}{\pi \,\Tspanj}\,,
\end{equation}
so the frequency-spacing depends on $\Tspanj$ of the data segment $j$.
For the low-frequency workunits (\mbox{$f < 300\,\Hz$}), this results
in frequency steps in the range \mbox{$d f_j \in [ 3.43,\,4.06]\,\muHz$},
while for high-frequency workunits \mbox{$d f_j \in [5.42,\,6.41]\,\muHz$}.

The range of frequency-derivatives $\dot f$ searched is defined in
terms of the ``spin-down age'' \mbox{$\tau \equiv - f/\dot f$}, namely
\mbox{$\tau \ge 1000\,$years} for low-frequency and \mbox{$\tau \ge 10\,000$\,years}
for high-frequency workunits.
Younger neutron stars than the limited range of the search
probably would have left a highly visible (Sedov phase) supernova remnant
or a pulsar wind nebula, and thus our search for unknown neutron stars
targeted older objects, which also resulted in less computational cost.
The search also covers a small ``spin-up'' range, so the actual ranges
searched are \mbox{$\dot{f} \in [-f/\tau,\,0.1 f/\tau]$}.
In $\dot f$ the grid points are spaced according to
\begin{equation}
  \label{eq:2}
  d\dot{f}_j = \frac{12 \sqrt{5m}}{\pi\,\Tspanj^2}\,,
\end{equation}
resulting in resolutions $d\dot{f}_j \in [1.84,\,2.59]\times 10^{-10}\,\Hz/\second$
for low-frequency workunits, and
$d\dot{f}_j \in [2.92,\, 4.09]\times 10^{-10}\,\Hz/\second$ for high-frequency
workunits, depending on the duration $\Tspanj$ of different segments $j$.

\subsubsection{Search grid in the sky parameters}

\begin{figure}
	\subfigure{\includegraphics[scale=0.21,angle=0]{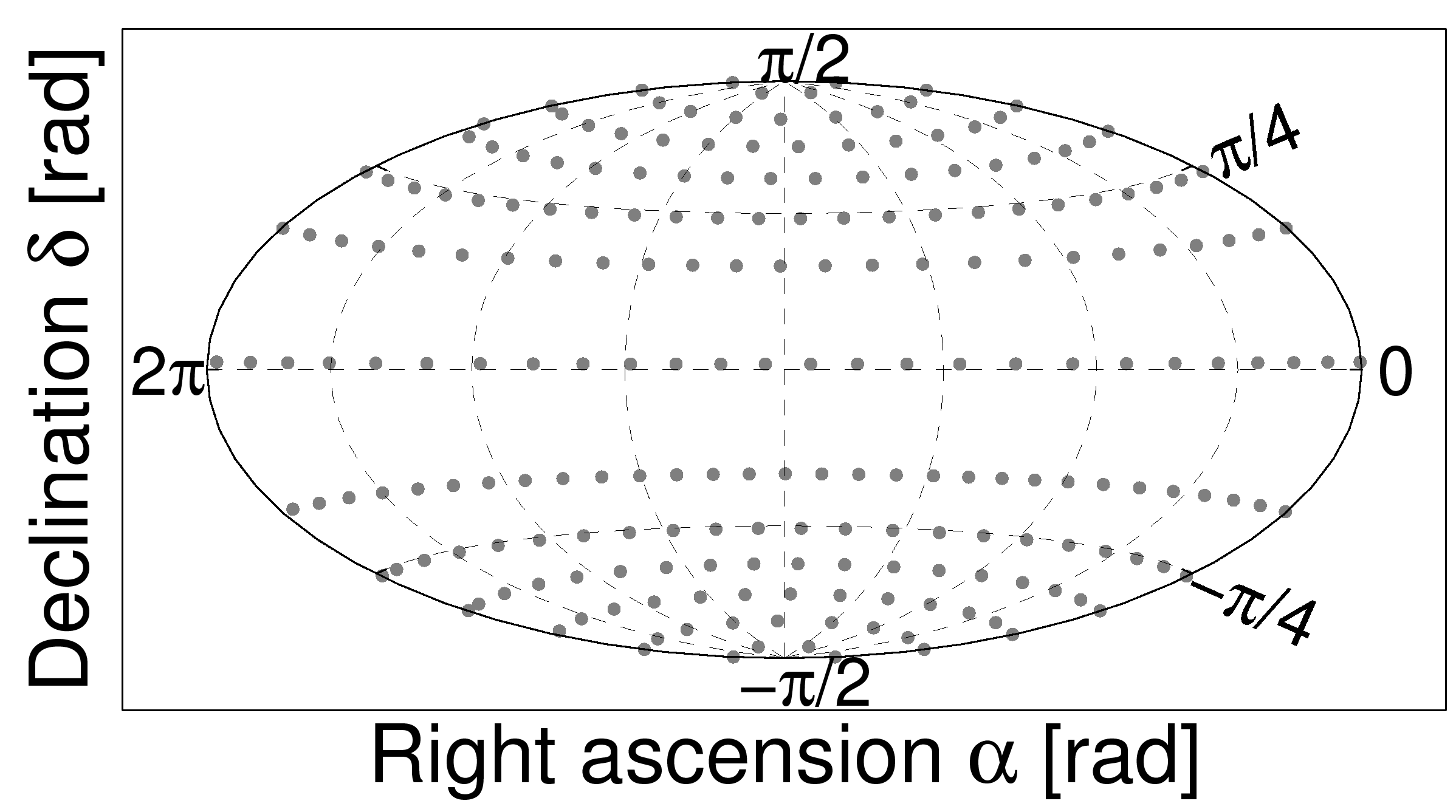}}
	\subfigure{\includegraphics[scale=0.21,angle=0]{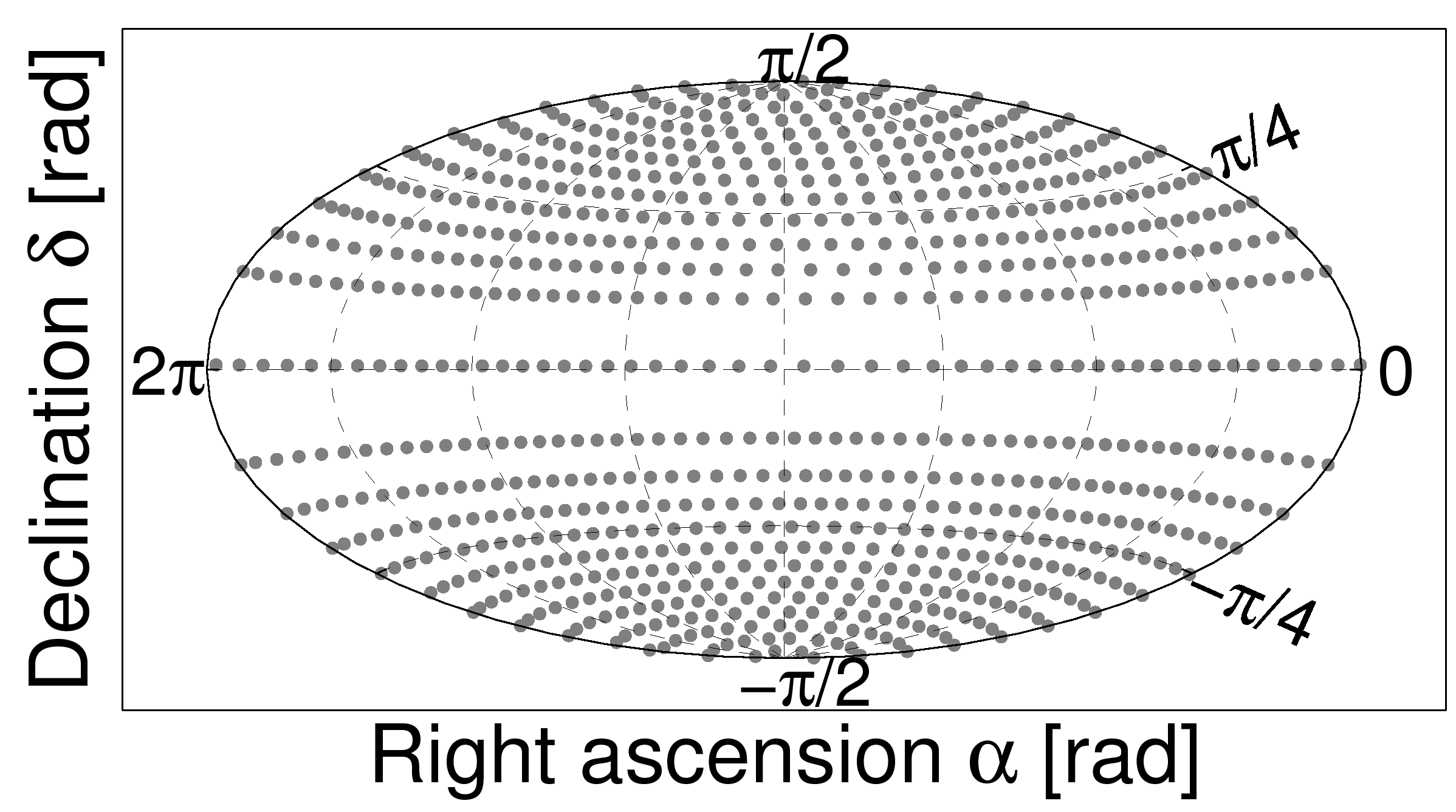}}\\
	\subfigure{\includegraphics[scale=0.21,angle=0]{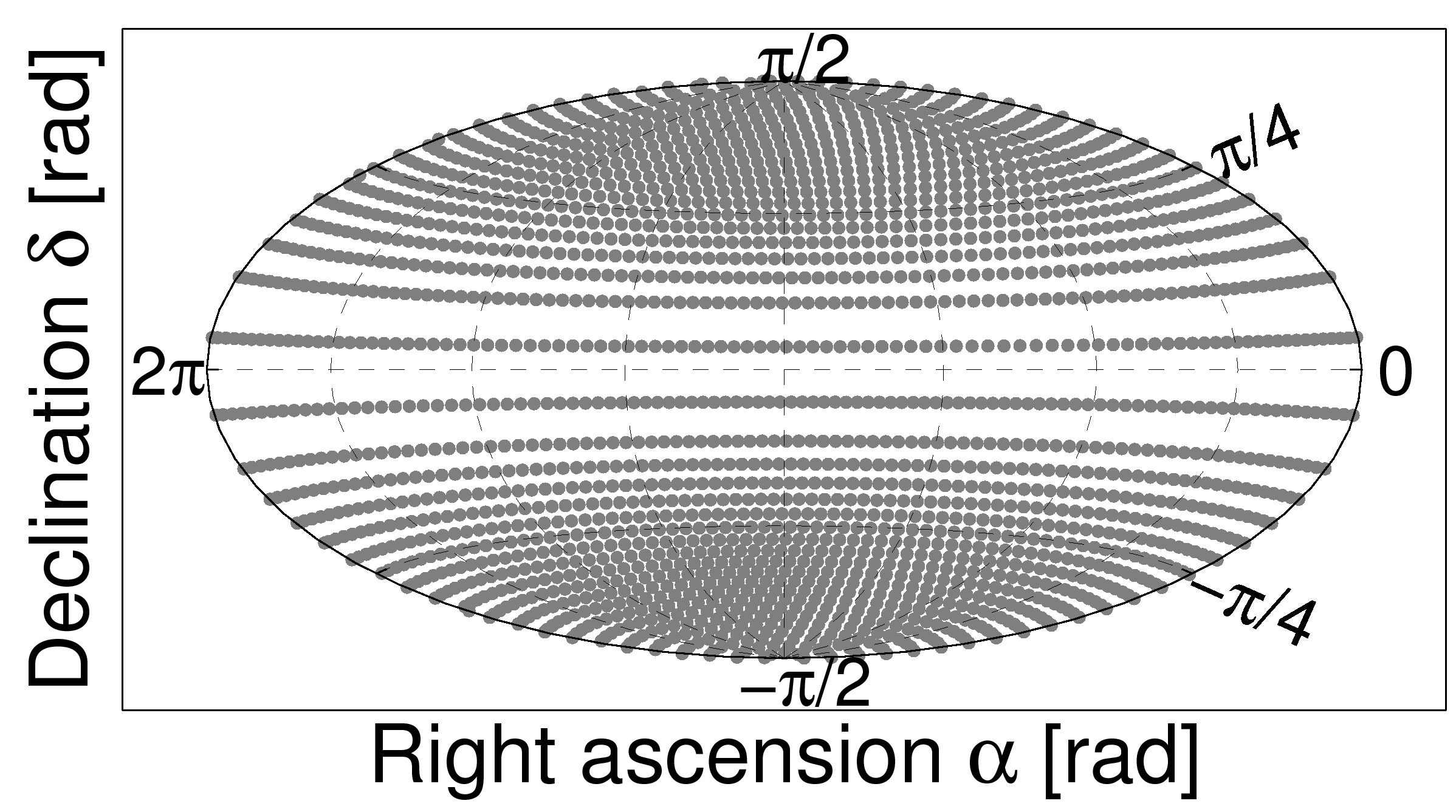}}
	\subfigure{\includegraphics[scale=0.21,angle=0]{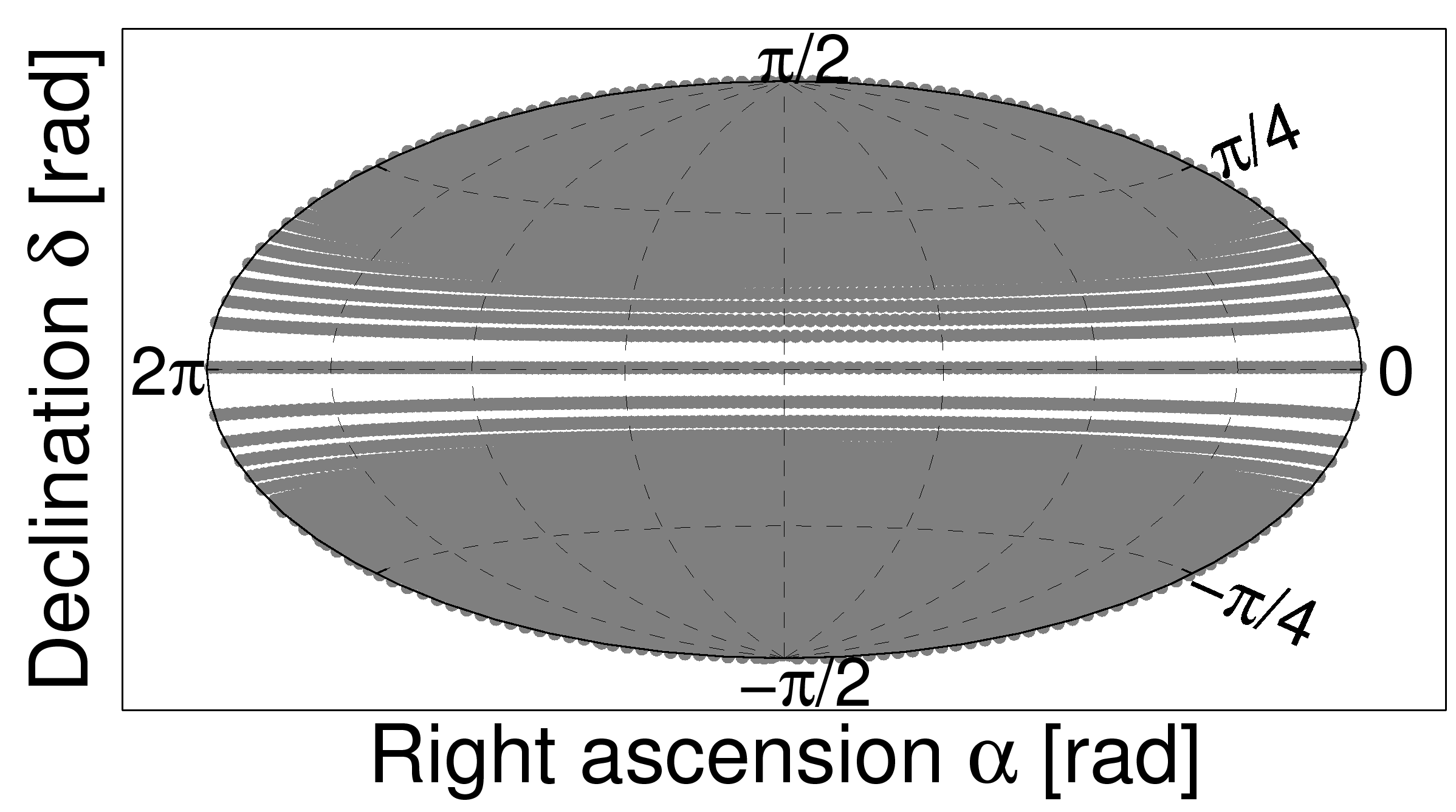}}\\
	\caption{Four different sky-grids in Hammer-Aitoff
          \cite{HammerAitoff} projection.
          The top row is for frequency $f = 60\,\Hz$ and the bottom
          row is for $f =310\,\Hz$. The left column shows data segment
          $j=1$ (from H1) with a spanned time of $T_{\mathrm{span},1} = 33.8\,\hours$,
          while the right column shows
          data segment $j=15$ (from L1) with a spanned time of
          $T_{\mathrm{span},15}=39.8\,\hours$.  The grid points are
          spaced more closely for a longer spanned time, and for a
          higher frequency.}
    	\label{f:skygrids}
\end{figure}

The resolution of the search grid in the sky-directions
depends both on the start-time $t_j$ and
duration $\Tspanj$ of the segment, as well as on the frequency. The
number of grid points on the sky scales as $\propto f^2$,
and approximately as $\propto \Tspanj^{2.4}$
for the range of $\Tspanj \sim 30--40$~h used in this search.
Contrary to the simple uniform spacings in $f$ and $\dot{f}$, the
sky-grids are computed beforehand and shipped together with the
workunits. In order to simplify the construction of workunits and
limit the amount of different input-files to be sent, the sky-grids are fixed
over a frequency range of $10\,\Hz$, but differ for each data
segment $j$. The sky-grids are computed at the higher end of each
$10\,\Hz$ band, so they are slightly ``over-covering'' the sky instead
of being too coarse. The search covers a frequency range of
$1450\,\Hz$, and so there are 145 different sky-grids for each segment.
To illustrate this, four of these
sky-grids are shown in Figure~\ref{f:skygrids} corresponding to two
different data segments at two distinct frequency bands.

To ensure that each workunit takes a similar amount of CPU time, the
total number of template grid points of each workunit is chosen to be
\emph{approximately} constant for all workunits.  However, in practice,
this number can vary by up to a factor of 2 due to discretization effects.
The number of points in the sky-grids grows with frequency as $f^2$
and the number of points in the spin-down grid grows linearly with $f$.
Thus, to keep the number of templates (and therefore the CPU time)
approximately constant, the frequency range covered by each workunit
decreases as $f^{-3}$.
Hence for fixed $j$, $M(j,k)$ is roughly proportional to~$k^3$.

\subsection{The output of a workunit \label{ss:WUoutput}}

The result from completing one workunit on an Einstein@Home host
computer is a ZIP-compressed ASCII text file containing the
$13\,000$~candidate events with the largest values of the
$\F$-statistic found over the parameter-space grid points analyzed by
that workunit.  Each line of the output file contains five columns:
frequency (Hz), right ascension angle (radians), declination angle
(radians), spin-down-rate (Hz/s) and $2\F$ (dimensionless).  The
frequency is the frequency at the Solar System Barycenter (SSB) at the
instant of the first data point in the corresponding data segment.

The number $13\,000$ was decided in advance,
when the workunits were first launched on the
Einstein@Home project, which was about one year before the
post-processing pipeline was developed. 
The network bandwidth required to retain more than $13\,000$ candidates per workunit,
and the storage space required to preserve them, would have exceeded the capacity
of the Einstein@Home server and its internet connection.
For frequency-band and data
segment combinations with small numbers of workunits, for example the
$j=1$ data set from 301.0~Hz to 301.5~Hz, almost all of the $13\,000$
candidate events are later used in the post-processing pipeline.
However (as can be seen later in Figure~\ref{f:ncseg}) for most
frequency-bands the post-processing pipeline only needed and used a
fraction of the events that were returned.

Returning the ``loudest'' $13\,000$~candidate events effectively
corresponds to a floating threshold on the value of the
$\F$-statistic.  This avoids large lists of candidate events being
produced in regions of parameter space containing non-Gaussian noise,
such as instrumental artifacts that were not removed a priori
from the input data.

\section{Post-Processing\label{sec:PostProcess}}

As shown previously in Figure~\ref{f:flowdiagram}, after result files
are returned to the Einstein@Home servers by project participants,
further post-processing is carried out on those servers and on
dedicated computing clusters.  The goal of this post-processing
pipeline is to identify consistent candidate events that appear in
many of the 17~different data segments.

In this paper, a consistent (coincident) set of ``candidate events''
is called a ``candidate''.  Candidate events from different data
segments are considered coincident if they cluster closely together in
the four-dimensional parameter space.  A clustering method using a
grid of ``coincidence cells'' will reliably detect strong CW signals,
which would produce candidate events with closely-matched parameters.

The post-processing
pipeline operates in $0.5\,{\rm Hz}$-wide frequency-bands, and
can be summarized in three steps.
In {\it step one}, the coincident false alarm probability is fixed.
In {\it step two}, the frequency values of candidate events are shifted to
the same fiducial time. 
In {\it step three}, a grid of cells is constructed in the four-dimensional
parameter space, and each candidate event is assigned to a 
particular cell. In the following the details involved in each step
are described.

\subsection{Preparation and selection of candidate events}
\label{ssec:select-cands}

In the {\it first step} the individual result files are prepared for the later analysis
by uncompressing them and keeping only a subset of the candidate
events: from the $(j,k,\ell)$'th workunit only the $\mathcal{E}(j,k,\ell)$
candidate events with the largest values of $2\F$ are retained.

The number of these candidate events is chosen a priori to obtain a
pre-determined fixed false alarm probability.  The false
alarms should be approximately uniformly distributed among the workunits, 
since each workunit examines a similar number of
independent grid points in parameter space.  The number of candidate
events is chosen so that in a $0.5\,{\rm Hz}$-wide frequency-band the
probability that one or more coincidence cells after doing the clustering
(in step three) has $\mathcal{C}_{\rm max} =
7$~or~more coincidences is $P_{\rm F} = 0.001$.  Thus, in the
analysis of 2900~such frequency bands, in random noise one would
expect to find only about three candidates with seven or more
coincidences.  (As explained later in Section~\ref{sss:coincalg}, this overall probability for the
entire search is somewhat increased because the coincidence cell grids
are also shifted by half their width in 16 possible combinations).

\begin{figure}
	\includegraphics[scale=0.46,angle=0]{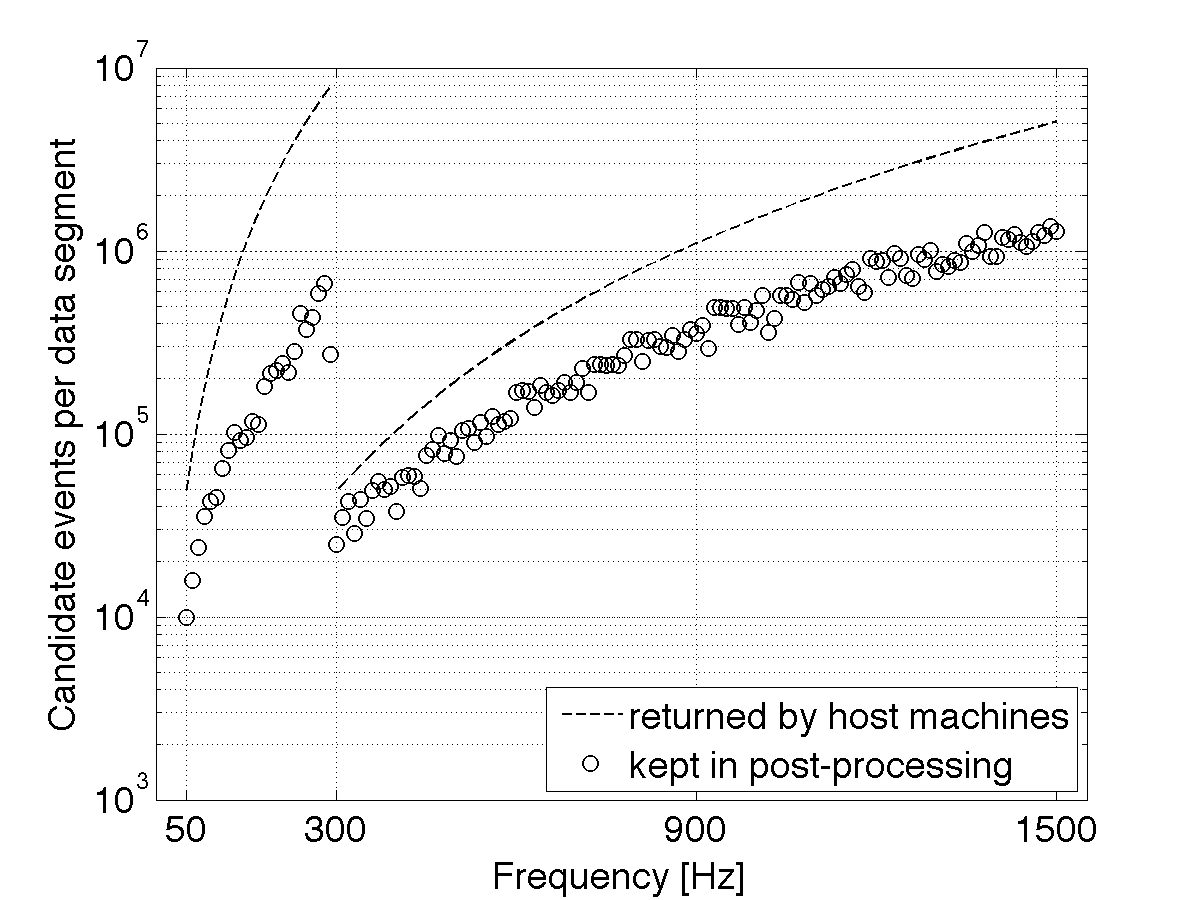}
  	\caption{\label{f:ncseg} The circles show the number of
          candidate events  $\mathcal{E}_{\rm seg}(k)$ retained per data segment and
          per $0.5\,{\rm Hz}$ frequency band in the post-processing in
          each $10\,{\rm Hz}$ band. The dashed curve represents the
          number of candidate events which are returned from
          volunteering hosts participating in Einstein@Home. The
          strange location of the point at $290\,{\rm Hz}$ is
          explained in Sec.~\ref{sss:search-grid}.}
\end{figure}

In terms of the notation introduced in the previous section, the
number of candidate events kept from the $(j,k,\ell)$'th workunit takes
the form
\begin{equation}
	 \mathcal{E}(j,k,\ell) = {\mathcal{E}_{\rm seg}(k) \over M(j,k)}\;,
\end{equation}
where $\mathcal{E}_{\rm seg}(k)$ is shown in Figure~\ref{f:ncseg}. Because the
individual workunits are constructed to use approximately the same 
amount of CPU time, each workunit examines approximately 
the same number of templates in parameter space, so the same number 
of candidate events are retained from all workunits which have the same 
input data file and data segment.  This implies that the number of candidate 
events that are kept per data segment $j$ and per frequency band is 
independent of the data segment~$j$:
\begin{equation}
 	\sum_{\ell = 1}^{M(j,k)} \mathcal{E}(j,k,\ell) = \mathcal{E}_{\rm seg}(k)\;.
\end{equation}
Since the sky-grids are fixed in $10\,{\rm Hz}$ intervals, $\mathcal{E}_{\rm seg}(k)$
takes the same value for all values of $k$ in the range of $20p+1,\cdots,20(p+1)$ 
where $p$ labels the sky-grids by an integer in the range $p \in 0,\cdots,144$.

It is illustrative to look at a specific case.  For example consider
the 0.5~Hz band covering $[301.0,301.5)\,\Hz$, this band is labeled by
$k=503$.  As is shown in Figure~\ref{f:ncseg}, in this band the
post-processing pipeline retains $\mathcal{E}_{\rm seg}(k=503) =
24\,960$ candidate events from each of the 17 different $30$-hour~data
segments.  The $30$-hour~data segment from H1 with the {\it shortest} 
time span ($j=1$) has approximately $4.3\times 10^8$ templates divided 
among just $M(j=1,k=503)=2$ workunits, so $12\,480$ candidate events 
are retained from each of these workunits.  
The $30$-hour~data segment from L1 with the {\it longest}
time span ($j=15$) has approximately $1.7 \times 10^9$ templates 
divided among $M(j=15,k=503)=7$ workunits, so $3\,565$ candidate events
are retained from each of these workunits.  In the later stage of the
post-processing, this ensures that each of the different data segments
contributes equally to the probability of generating false alarms in
the coincidence step.

\subsection{Number of cells in the post-processing coincidence grid}

It is important to calculate the number of coincidence cells in the
coincidence grid. Together with the desired false alarm probability,
this determines the number of candidate events to retain in the
post-processing pipeline.

The number of coincidence cells $N_{\rm cell}(k)$ contained in each
$0.5\,\Hz$ frequency band $k$ (while doing the clustering in step
three) is determined by the sizes of the cells. This is given by
\begin{equation}
  N_{\rm cell}(k) = \biggl({0.5\,\Hz \over \Delta f} \biggr) \biggl( { 1.1 \,f \over \tau \Delta \dot f} \biggr) 
  \int_{-\pi/2}^{\pi/2} { d\delta \over \Delta \delta(\delta)} \int_0^{2\pi} {d\alpha  \over \Delta \alpha(\delta) }\;,
  \label{e:Ncell}
\end{equation}
where 
$\Delta f$ denotes the coincidence cell width in frequency, 
$\Delta \dot f$ denotes the width in spin-down, and 
$\Delta \alpha(\delta)$ and $\Delta \delta(\delta)$
denote the coincidence cell widths in right ascension and declination (both of which vary
with declination $\delta$). The choice of the coincidence
cell sizes will be explained in detail later when step three will be described.

\subsection{False alarm rate and the number of candidate events retained}
\label{ssec:cands-retained}

The number of candidate events that must be retained is determined by
the number of cells in the coincidence grid $N_{\rm cell}(k)$ and by
the desired probability of false alarm $P_{\rm F}$ for false
coincidence of candidate events from $\mathcal{C}_{\rm max}$ or more data
segments in each $0.5\,\Hz$ band.  To relate these quantities, consider
the case of random instrumental noise, in which the candidate events
are distributed uniformly about the coincidence grid.  Concentrate on
a single $0.5\,\Hz$ band $k$, and consider the first of the $N_{\rm seg}=17$ data
segments.  A total of $\mathcal{E}_{\rm seg}(k)$ candidate events must be
distributed uniformly among $N_{\rm cell} (k)$ coincidence cells.  Each
candidate event falls in a random coincidence cell, independent of the
locations of the previous candidate events. The probability that the
first candidate event falls in the first coincidence cell is $1/N_{\rm cell}(k)$, 
and hence the probability that the first coincidence cell
remains empty is $1-1/N_{\rm cell}(k)$.  If the remaining $\mathcal{E}_{\rm seg}(k) -1$
candidate events fall independently and at random into the coincidence
cells, then this generates a binomial distribution, and the
probability that the first coincidence cell contains no candidate
events is 
\begin{equation}
  p_k(0) = \left(1- \frac{1}{N_{\rm cell}(k)}\right)^{ \mathcal{E}_{\rm seg}(k)} \;.
\end{equation}  
Since the first
coincidence cell is equivalent to any other, the probability that the
candidate events from the first data segment populate any given
coincidence cell with one or more candidate events is thus given by
\begin{equation}
  \epsilon(k) = 1-p_k(0) = 1 - \biggl(1-{1 \over N_{\rm cell}(k)} \biggr)^{\mathcal{E}_{\rm seg}(k)}\;.
\end{equation}
In random noise, the candidate events produced by each different data
segment are independent, so that the coincidence cells that are
``marked'' by one or more candidate events are also described by a
(different) binomial distribution.  Without loss of generality, again
consider the first coincidence cell.  The probability that it contains
candidate events from $n$ distinct data segments is then given by
\begin{equation}
{N_{\rm seg} \choose n} \bigl[\epsilon(k)\bigr]^n \bigl[1-\epsilon(k)\bigr]^{N_{\rm seg} - n}\;,
\end{equation}
where ${a \choose b} = {\frac{a!}{b! (a-b)!}}$ is the binomial
coefficient. Thus the probability per coincidence cell of finding
$\mathcal{C}_{\rm max}$ or more coincidences is given by
\begin{equation} 
 \frac{P_{\rm F} }{ N_{\rm cell}(k) } 
 = \sum_{n=\mathcal{C}_{\rm max}}^{N_{\rm seg}} {N_{\rm seg} \choose n} 
 [\epsilon(k)]^n [1-\epsilon(k)]^{N_{\rm seg} - n}\;. 
\end{equation}
For the desired $P_{\rm F} = 0.1\% = 10^{-3}$ per $0.5\,\Hz$ band $k$, 
this equation is solved numerically to find $\mathcal{E}_{\rm seg}(k)$.
The results for $\mathcal{E}_{\rm seg}(k)$ are shown in Figure~\ref{f:ncseg}.

\subsection{Choice of false alarm probability and detection threshold}

The goal of this work is to make a {\it confident detection}, not to
set upper limits with the broadest possible coverage band.  This is
reflected in the choice of the {\it expected false alarm probability}
and the choice of a {\it detection threshold}.

The detection threshold of 12 events was chosen because, as described
in Section~\ref{sec:HardwareInjections}, the hardware injections are
only ``turned on'' for 12 of the 17 data segments.  The detection
threshold ensures that these simulated signals are properly detected
by the post-processing pipeline.

The choice of false alarm probability ($P_{\rm F} = 0.1\% = 10^{-3}$
per $0.5\,\Hz$ band to have coincidences in $\mathcal{C}_{\rm max} =
7$ or more data segments) is a pragmatic choice, which leads to an
extremely small false alarm rate at the detection threshold.  For
actual data, the probability of finding 7 or more coincidences in a
given $0.5\,\Hz$ band can be somewhat larger than the target value of
$0.1\%$ because the candidate events are not uniformly distributed
over the grid of coincidence cells and because (as described in
Section~\ref{sss:coincalg}) 16 sets of coincidence cells are used for
each 0.5~Hz band.

In random noise, the probability of reaching the detection threshold
of 12 coincidences depends on the number of cells in the coincidence
grid, which is a function of frequency.  Some representative numbers
are given in Table~\ref{t:FalseProbs}; they vary from about $10^{-15}$
to $10^{-13}$ depending upon the 0.5~Hz band.  The false alarm
probabilities decrease very rapidly with increasing coincidence
number.  For example the probability of finding 14 or more
coincidences in random noise varies from about $10^{-18}$ to
$10^{-21}$.

\begin{table}
\begin{center}
\begin{tabular}{rrrccc}
\hline
 $f_{k}$[Hz] & $k$ &$N_{\rm cell}(k)$    &  $P_{\rm F}(\mathcal{C} \ge7)$ & $P_{\rm F}(\mathcal{C} \ge 12)$
 &  $P_{\rm F}(\mathcal{C} \ge 14)$ \\
\hline\hline
$50.0$ &1&  $734\,500$     & $10^{-3}$  & $1.5 \times 10^{-13}$     & $3.0 \times 10^{-18}$  \\
$290.0 $ &481&  $35\,198\,800$  & $10^{-3}$  & $8.7 \times 10^{-15}$     & $5.7 \times 10^{-20}$ \\
$301.0 $ &503&  $2\,161\,284$   & $10^{-3}$  & $6.7 \times 10^{-14}$     & $9.9 \times 10^{-19}$  \\
$1499.5$&2900&  $233\,345\,981$  & $10^{-3}$  & $2.2 \times 10^{-15}$     & $8.4 \times 10^{-21}$  \\
\hline
\end{tabular}
\caption{ \label{t:FalseProbs} False alarm probabilities $P_{\rm F}$ in four
  different $0.5\,\Hz$ frequency bands labeled by the integer $k$.  
  The frequency at the lower boundary of the $0.5\,\Hz$ band $k$ 
  is denoted by $f_{k}$.
  The number of coincidence cells in the $k$'th half-Hz frequency band is
  denoted by $N_{\rm cell}(k)$.
  The probability of finding 7 or more coincidences $(\mathcal{C} \ge7)$ in
  randomly-distributed noise is fixed to be 0.1\%.  The probability of
  finding 12 or more coincidences (the detection threshold, $\mathcal{C} \ge12$) 
  in random noise varies over two orders of magnitude, from about $10^{-15}$ to
  $10^{-13}$.  The probability of finding 14 or more coincidences 
  $(\mathcal{C} \ge 14)$ in random noise varies from about 
  $10^{-18}$ to $10^{-21}$.}
\end{center}
\end{table}

Once might ask why we chose to specify a uniform false alarm
probability, across all frequencies, of $0.1\%$ for $\mathcal{C}_{\rm
  max}=7$, rather than directly specify a much lower false alarm
probability at the detection threshold $\mathcal{C}=12$.  This was
because we wanted the most significant coincident events due to noise
alone to have $\mathcal{C}$ values a few less than our detection
threshold, and we wanted \emph{these} candidates to be uniformly
distributed over frequency bands.  Any detected signals could then be
compared against fairly uniform fields of noise candidates in adjacent
frequency bands.  If a uniform false alarm probability had been
specified at the $\mathcal{C}=12$ level, then the expected noise
candidates with $\mathcal{C}\sim 7$ would \emph{not} have been
uniformly distributed over frequency, due to the differing numbers of
coincidence cells in each frequency band.

The choice of detection threshold and false alarm probability
sacrifice a small amount of sensitivity compared with a higher values,
but ensures that high numbers of coincidences are extremely improbable
in random noise.  A strong signal (say a factor of 1.5 above the upper
curve in Figure~\ref{f:estULs}) would be expected to produce 15 or
more coincidences in this detection pipeline.  With the thresholds
that we have adopted, this would stand out very strongly: the
probability of having even one such an event appear in coincidence in
random noise is about $10^{-22}$ per 0.5~Hz band.

\subsection{Shifting candidate event frequencies to a fixed fiducial time}

In the {\it second step} of the post-processing, the frequency value of each
retained candidate event is shifted to the same fiducial time: the GPS
start time of the earliest ($j=4$) data segment,
$t_{\rm{fiducial}} = t_4 = 793\,555\,944\,{\rm s}$.  This shifting is
needed because a CW source with non-zero spin-down would produce
candidate events with different apparent frequency values in each data
segment.  This step would shift these candidate events back to the
same frequency value:
\begin{equation}
f(t_{\rm{fiducial}}) =  f(t_{j}) + 
 [t_{\rm{fiducial}} - t_{j}] \, \dot{f}  \label{fixed fiducial time} \; ,
\end{equation}
where $\dot{f}$ and $f(t_j)$ are the spin-down-rate and frequency of a
candidate event reported by the search code in the result file, and
$t_j$ is the timestamp of the first datum in the data segment.  At the
end of the second step, all candidate events for the $0.5\,{\rm Hz}$
band are merged into one file.

\begin{figure}
\includegraphics[scale=0.46,angle=0]{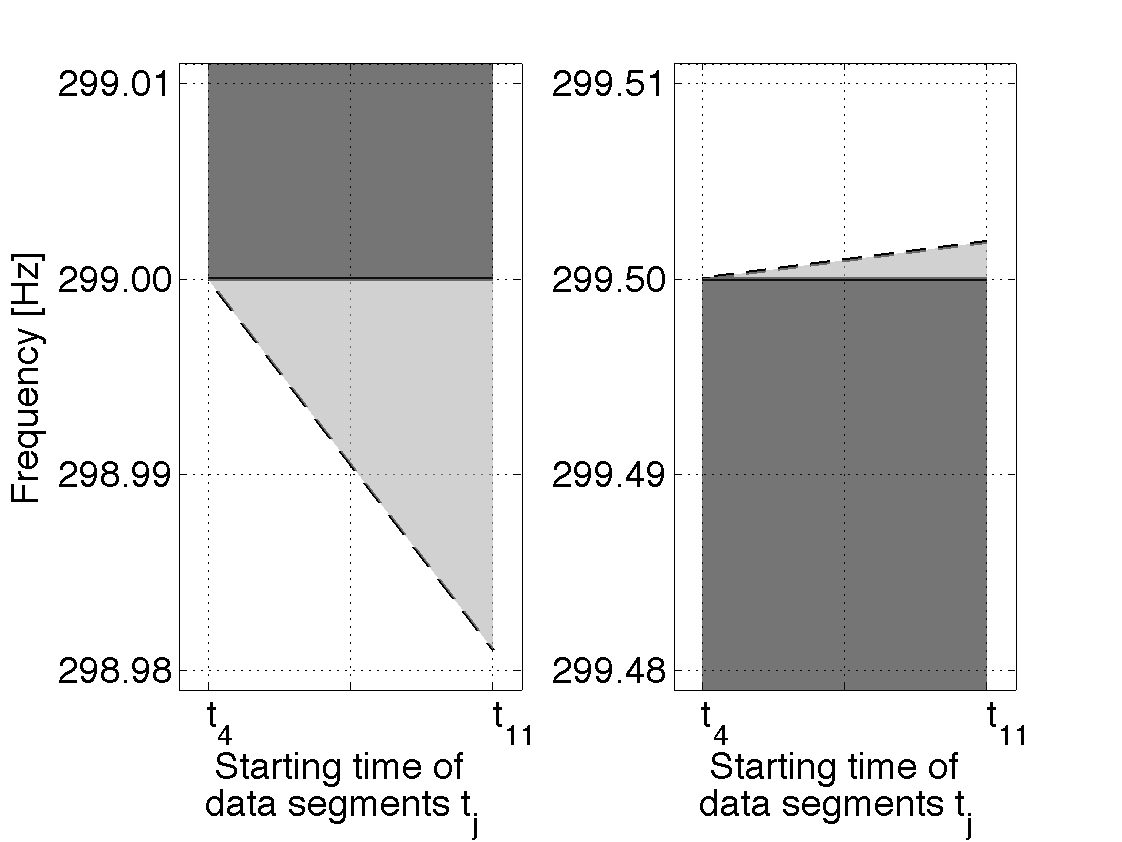}
\caption{\label{f:fbands} Additional ``wings'' at the boundaries of each
  $0.5\,{\rm Hz}$ frequency band must be included in the coincidence
  analysis stage of the post-processing.  This is because spin-down
  can carry a source below this half-Hz band, and spin-up can carry it
  above the band.  To illustrate this, the frequency band $k=498$
  (covering $[299,299.5)\,\Hz$) is (partly) shown by the dark-gray
  shaded area.  The dashed sloped lines show the boundaries of the
  small additional regions (light gray) in frequency space whose candidate events
  must also be considered in the post-processing.  Because the allowed
  spin-up range is ten times smaller than the allowed spin-down range,
  the upper boundary has a slope ten times smaller than the lower
  boundary.}
\end{figure}

These candidate events are collected from a frequency interval that
is slightly wider than 0.5~Hz.  To see why this is necessary, consider
a potential source whose frequency in the first data segment ($j=4$)
is at the lower (or upper) boundary of the 0.5~Hz interval.  If the
source has the minimum (or maximum) allowed value of $\dot{f}$, then
in the {\it later} data segments it moves into, or is recorded in, the
previous (or next) 0.5~Hz band.  This effect is most apparent for the
last $j=11$ data segment, as illustrated in Figure~\ref{f:fbands}.  So
in collecting the candidate events for analysis of a given 0.5~Hz
band, the frequency range is enlarged slightly for events coming from
later and later data segments, as shown in Figure~\ref{f:fbands}.

\subsection{Search for coincident candidate events}

The {\it third step} and final stage of the post-processing is to search for
parameter-space coincidence among the candidate events.  If a CW
source is present that is strong enough to be confidently detected,
then it would produce large $\F$-statistic values (i.e.\ candidate
events) in many or all of the 17~data segments. In addition, the
values of the frequency at the fiducial time $f(t_{\rm{fiducial}})$,
sky position (given by right ascension~$\alpha$ and declination~$\delta$),
and spin-down~$\dot{f}$ for these candidate events would
agree among all data segments (within some coincidence ``window'' or
``cell'').

\subsubsection{Coincidence search algorithm}
\label{sss:coincalg}

To find coincidences, a grid of cells is constructed in
four-dimensional parameter space, as described previously.  This analysis uses rectangular
cells in the coordinates $(f, \dot f, \alpha \cos \delta, \delta)$.
The dimensions of the cells are adapted to the parameter space search.
Each candidate event is assigned to a
particular cell.  In cases where two or more candidate events from the
same data segment $j$ fall into the same cell, only the candidate
event having the largest value of $2\F$ is retained in the cell.  Then
the number of candidate events per cell coming from distinct data
segments is counted, to identify cells with more coincidences than
would be expected by random chance.

The search for coincident cells containing large numbers of candidate
events is done with an efficient code that uses linked-list data
structures, $N \log N$ sort algorithms, and $\log N$ bisection search
algorithms.  To ensure that candidate events located on opposite sides
of a cell border are not missed, the entire cell coincidence grid is
shifted by half a cell width in all possible $2^4 = 16$~combinations
of the four parameter-space dimensions.  Hence 16~different
coincidence cell grids are used in the analysis.

The cells in the coincidence grid are constructed to be as small as
possible to reduce the probability of coincidences due to false
alarms.  However, since each of the 17~different data segments uses a
different parameter space grid, the coincidence cells must be chosen
to be large enough that the candidate events from a CW source (which
would appear at slightly different points in parameter space in each
of the 17~data segments) would still lie in the same coincidence cell.

\subsubsection{Frequency and spin-down coincidence windows}

In frequency, the spacing of the parameter-space grid is largest for
the data segment with the smallest value of $T_{{\rm span},j}$, which
is the first data segment $j=1$.  At first, this would appear to be
the correct size $\Delta f$ for the coincidence cell in the frequency
direction.  However since the frequency of a candidate event must be
shifted to a fixed fiducial time according to its spin-down value, and
since that spin-down value can not be more accurate than the $\dot f$
spacing, the size of the coincidence cell must be somewhat larger to
accommodate the effects of this discretization error in $\dot f$.  The
coincidence window in the frequency direction is thus determined by
\begin{equation}
  \Delta f = \; \max_{j} \, \left( df_j +  \Delta t \; d\dot f_j \right) \; ,
\end{equation}
where the maximization over $j$ selects the data segment with the
smallest $T_{{\rm span},j}$ (which is $j=1$) and
\begin{equation}
	\Delta t = |\max_{j} \, t_{j}  - \min_{j} \, t_{j} | = t_{11} - t_{4}  \;  = 1\,997\,256 \, {\rm s}
	\label{e:Deltat}
\end{equation}
is the total time span between the latest and earliest data segments.
For safety, e.g.\ against noise fluctuations that could shift a candidate peak, $\Delta f$ has been increased 
by a further 40\% below $300\,{\rm Hz}$, so that the width of the coincidence 
cell in frequency is $\Delta f = 0.77 \times 10^{-3}\,{\rm Hz}$, 
and by 30\% above $300\,{\rm Hz}$, so that
$\Delta f = 1.08 \times 10^{-3}\,{\rm Hz}$.

For the spin-down parameter, the size of the coincidence cell
is given by the largest $d\dot f_j$ spacing in the parameter space
grid, which is also determined by the smallest value of $T_{{\rm
    span},j}$.  For safety this is also increased by 40\% below
$300\,\rm Hz$ giving $\Delta\dot f = 3.7 \times 10^{-10} \,\textrm{ Hz
  s}^{-1} $, and by 30\% above $300\,{\rm Hz}$ giving
$\Delta\dot f = 5.18 \times 10^{-10}\, \textrm{ Hz s}^{-1}$.

\subsubsection{Coincidence windows in apparent sky position}

\begin{figure}
  \subfigure{\includegraphics[scale=0.4,angle=0]{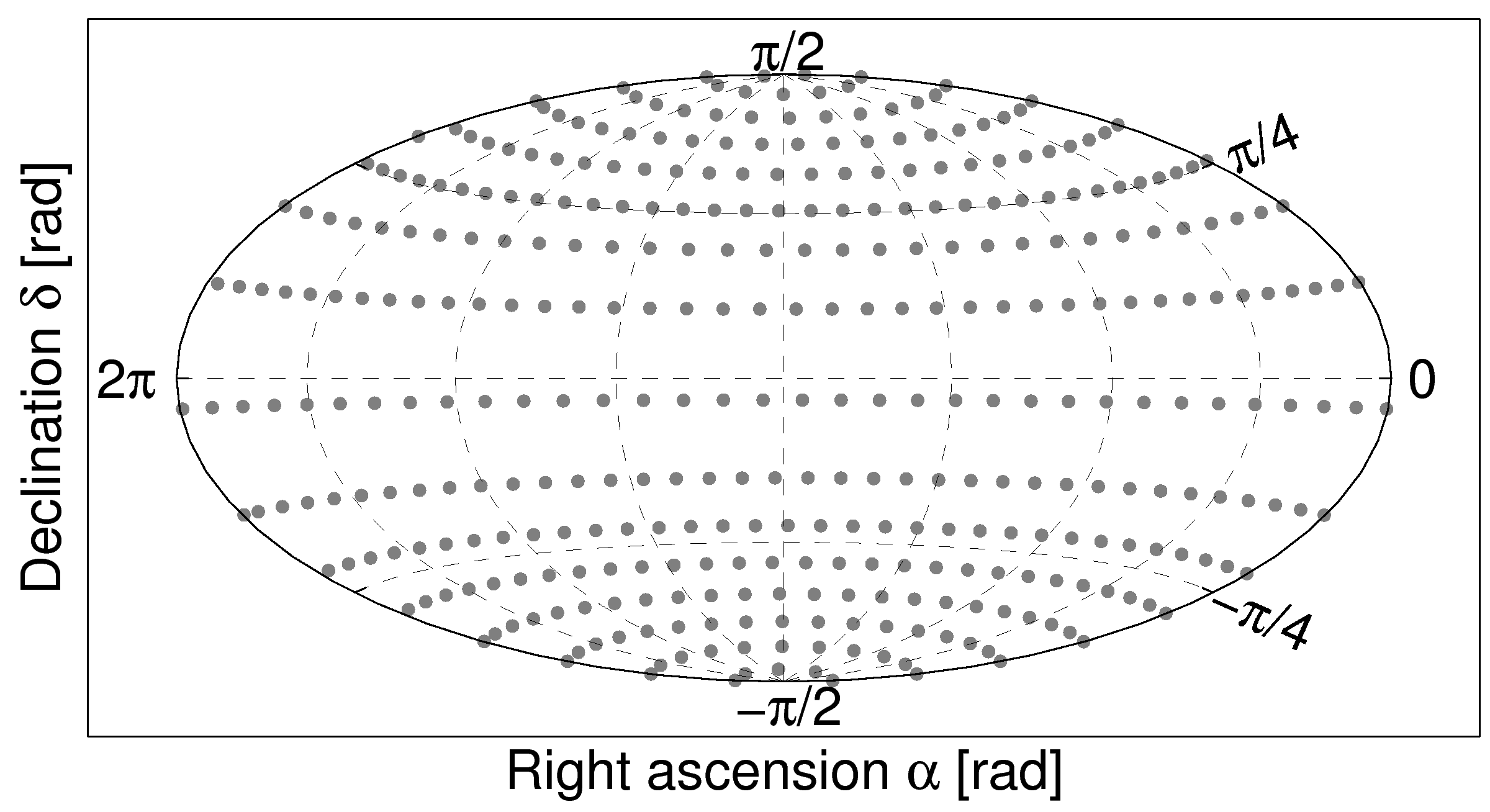}}\\
  \subfigure{\includegraphics[scale=0.46,angle=0]{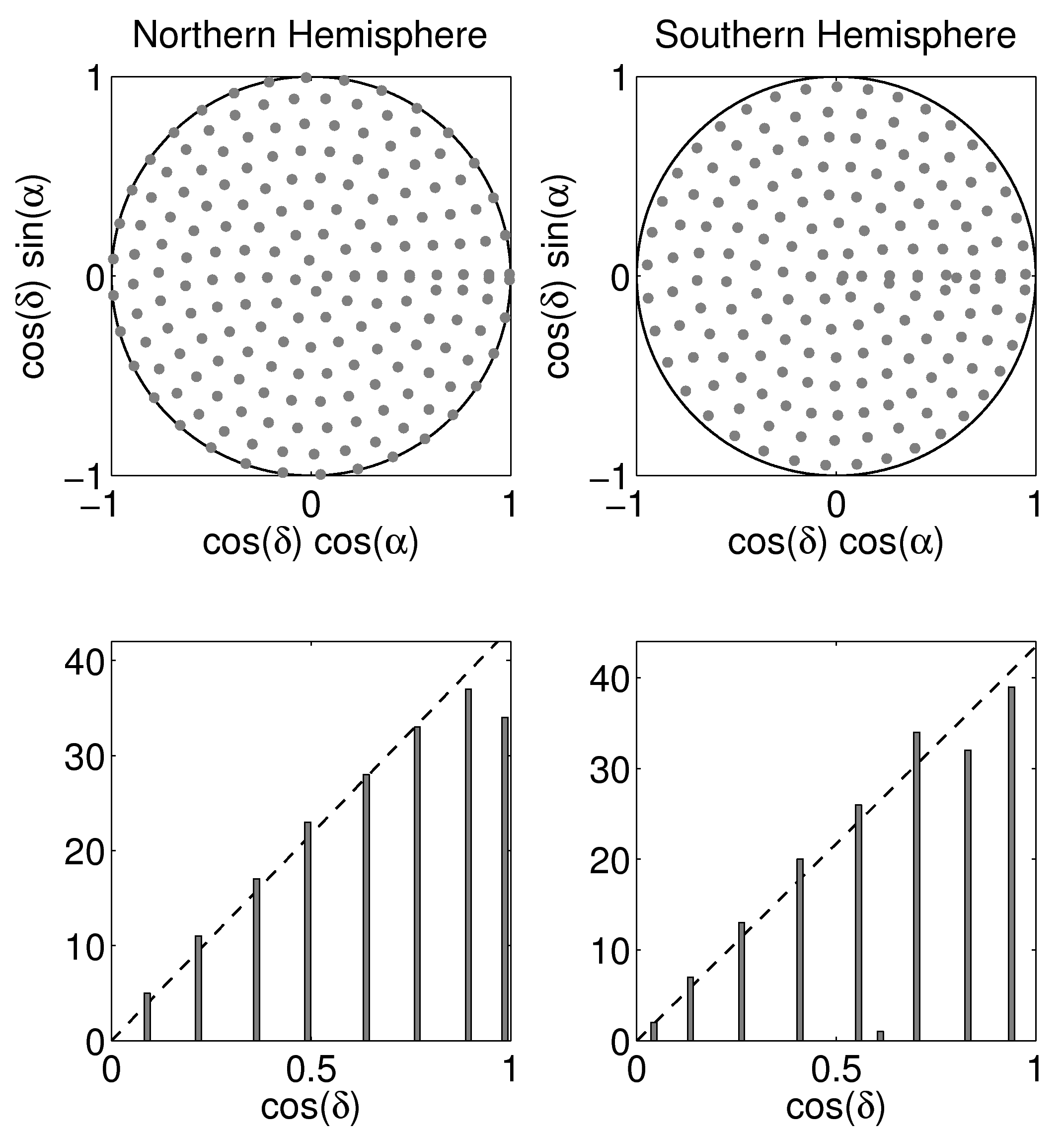}}\\
  \caption{Example sky-grid and its projection onto the equatorial plane.
  This sky-grid corresponds to the data segment $j=7$ used in 
  the  frequency range from $60\,\rm Hz$ to $70\,\rm Hz$. The top plot shows
  a Hammer-Aitoff projection of the sky-grid. The middle plots show the projection
  of the sky-grid points in the northern hemisphere (left column) and in
  the southern hemisphere (right column) onto the equatorial plane. 
  The bottom plots show histograms of $\cos (\delta)$ and the dashed
  line represents a linear fit to the distribution showing its uniformity.
    \label{f:uniform skygrid}}
\end{figure}

Determining the optimal size for the coincidence cells in the sky
coordinate directions is more difficult. Each of the 17~different data
segments uses a different sky-grid, as illustrated in
Figure~\ref{f:skygrids}. Ideally the size of the coincidence cells in
these sky directions must be just large enough to enclose one
parameter space grid point from each of the 17~different sky-grids. A
practical solution to determine the coincidence cells, which is close
to optimal, makes use of an observation concerning the parameter-space
metric that first appears in~\cite{BC00}.

To understand the properties of the parameter-space metric, first
consider the relative orders-of-magnitude of the different frequency
modulation effects.  The fractional Doppler shift due to the Earth's
annual orbital motion about the Sun has magnitude 
$|\vec{v}_{\rm orbital}|/c = 10^{-4}$ and the fractional Doppler shift due to
the detector's daily motion about the Earth rotation axis has
magnitude $|\vec{v}_{\rm rotational}|/c = 10^{-6}$.  For the
$\Tspan \approx 40\, {\rm h}$ period of a single coherent integration,
one can approximate the motion of the Earth's center of mass as motion
with constant acceleration (along a parabolic trajectory) rather than
as circular motion.  The neglected term in the fractional Doppler
shift has magnitude $|{{\ddot{\vec{v}}_{\rm orbital}}}|
\Tspan^2/2c \approx |\vec{v}_{\rm orbital}|
|\vec{\omega}|^2\Tspan^2 /2c \approx 4\times 10^{-8}$, where
$|\boldsymbol{\omega}|=2\pi/{\rm year}$ is the magnitude of the
Earth's orbital angular velocity about the sun.  This term is a factor
of 25 smaller than $|\vec{v}_{\rm rotational}|/c$ and hence can
be neglected.  With this approximation, the orbital motion of the
Earth is simply responsible for an apparent shift in the frequency $f$
and spin-down rate $\dot f$ of a source: the effects of the Earth's
center of mass motion are degenerate with a shift in frequency and
spin-down.  So the Earth's orbital motion causes a signal only to
shift to a different template in $f$ and $\dot f$; the Earth's
rotation has a period of one sidereal day and can not be modeled by a
shift in $f$ or $\dot f$.  Note that terms are neglected {\it only} in
determining where to place search grid points in parameter space
(because the neglected terms have an insignificant effect on where the
grid points are placed). The actual filtering of the data uses
``exact'' barycentering routines (total timing errors below $3\mu$s).

The search grid in parameter space is a Cartesian product of a
frequency grid, a spin-down grid, and a two-dimensional sky grid.
Since the search maximizes the detection statistic over frequency and
spin-down, the metric used to place grid points on the
sky~\cite{EHDoc} may be obtained by minimizing the four-dimensional
metric over frequency and spin-down and projecting it into the sky
directions.  As shown in the previous paragraph, this two-dimensional
projected sky metric is well-approximated by assuming that the Earth
is spinning about its axis but has its center of mass at rest.  If the
coherent integration period is an integer number of days, then by
symmetry the two-dimensional metric on the sky is invariant under
rotation about Earth's axis ($\partial_\alpha$ is a
Killing vector).  This is still an approximate symmetry for the search
described here, since the coherent integration period and $\Tspan$ are
longer than the rotation period (one day).

One can easily find coordinates in which this approximate sky metric
(the four-dimensional metric, minimized over frequency and spin-down
and projected onto the sky directions) is proportional to ${\rm
  diag}(1,1)$.  These new sky-coordinates are obtained by
perpendicular projection of a point on the two-sphere (celestial
sphere) vertically down into the unit radius disk that lies in the
equatorial plane. If $\ \hat{\vec{n}}$ denotes a unit vector pointing
from the SSB to the source the new coordinates are the components
of~$\ \hat{\vec{n}}$ in the equatorial plane: $n_x = \cos \delta\,
\cos \alpha$, $n_y = \cos \delta\, \sin \alpha$.  Points which are
equally spaced in these coordinates correspond to equal spacing in
Doppler shift, since source Doppler shift due to the Earth's rotation
is just proportional to the component of the source direction vector
in the equatorial plane.  It then follows from rotational invariance
that (with these approximations) the projected sky metric in these
coordinates is proportional to ${\rm diag}(1,1)$ ~\cite{jk4}.  The
effect may be immediately seen in Figure~\ref{f:uniform skygrid}: the
grid of ``equally-spaced'' points forms a (roughly) uniform square
grid on the unit radius disk in the equatorial plane. Computing the
Jacobian of the coordinate transformation shows that in the
original coordinates $(\alpha,\delta)$ the coordinate-space density of
grid points should be proportional to $|\cos\delta\,\sin\delta| =
|\sin(2\delta)|$.

\begin{figure}
	\includegraphics[scale=0.47,angle=0]{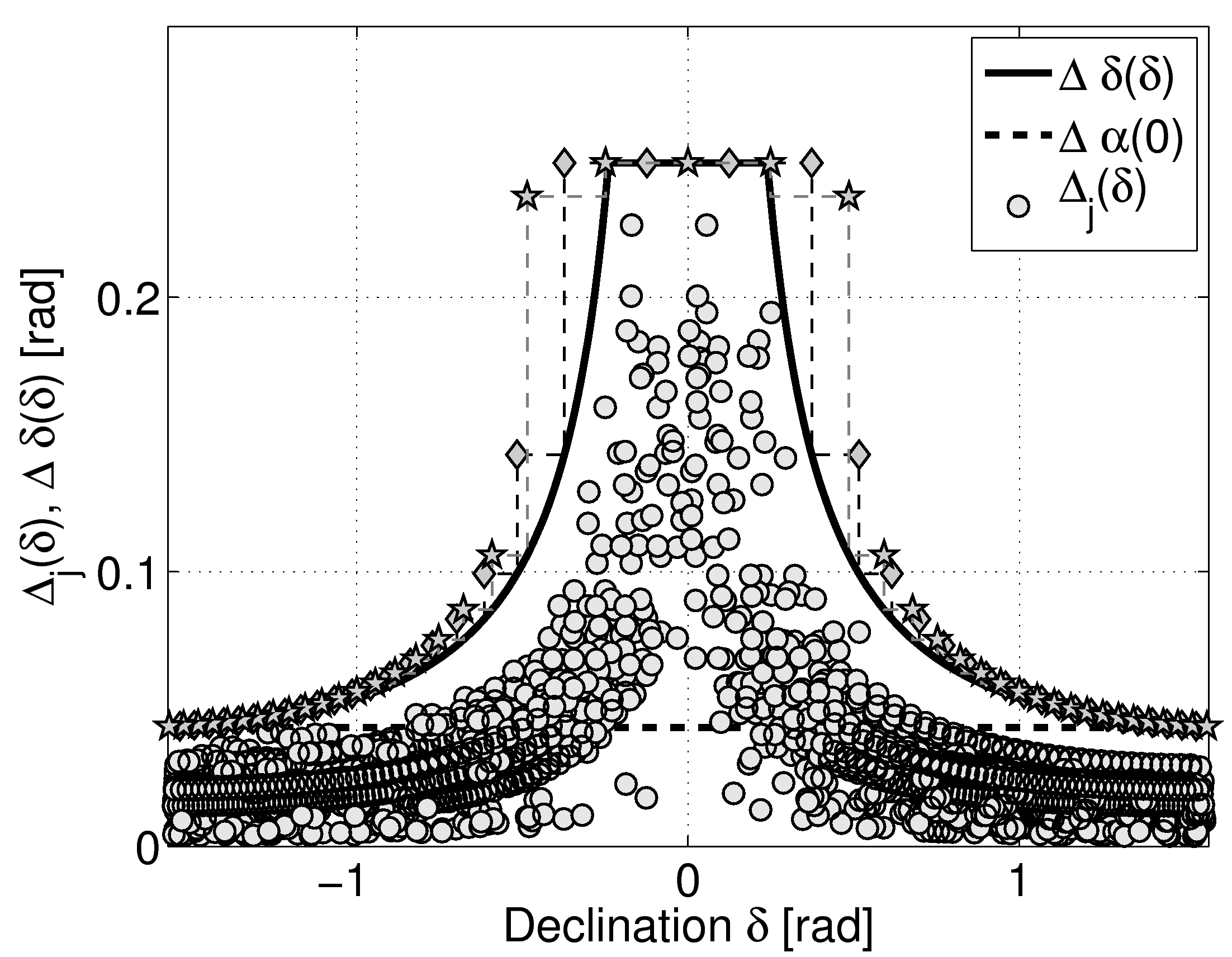}
	\caption{ The sky coincidence-window model for the frequency
          band from $510-520\,{\rm Hz}$.  The horizontal axis shows
          the declination $\delta$ in radians.  On the vertical axis,
          the circles labeled $\Delta_j(\delta)$ correspond to the the
          maximum distance in radians to neighboring $\delta$-points
          on either side. The solid curve shows the declination
          coincidence-window model $\Delta \delta(\delta)$ with the
          parameters $\Delta \delta(0) = 0.2489$, $\Delta \alpha(0) =
          0.0433$, and $\kappa = 1.5$ used in this frequency band.  It
          lies just above the largest declination separations
          shown. The stars denote the borders of the declination
          coincidence cell-grid and the diamonds represent the borders
          of the shifted declination coincidence cell-grid.}
\label{f:FMmodel}
\end{figure}

This simple behavior of the projected sky metric guides the construction of the
coincidence-windows in the sky directions.  Define polar
coordinates ($r,\alpha)$ on the unit radius disk in the equatorial
plane by $r=\cos \delta$.  The coordinate boundaries of uniformly
distributed coincidence cells containing a single parameter-grid point
are then given by $r\,d\alpha = dr = {\rm const}$.  When written in
terms of the original coordinates this becomes
\begin{equation}
  \cos(\delta) \,d\alpha = |\sin(\delta)| \,d\delta = {\rm const}.
\end{equation}
This is
not directly useful, because it is singular as $\delta \to 0$, but
suggests a coincidence window size which varies with declination
according to the model
\begin{eqnarray} 
\Delta \alpha(\delta) & = &\Delta \alpha(0) / \cos (\delta)\label{e:alphawin} \\ 
\Delta \delta (\delta)& = &\Biggl\{
\begin{array}{lc}
  \Delta \delta(0)& \mbox{ if } |\delta| < \delta_{\rm c},\\
  \Delta \alpha(0) / |\sin(|\delta| - \kappa\,  \Delta \alpha(0))| &\mbox{ if }  |\delta| \ge \delta_{\rm c}.
\end{array} \nonumber
\end{eqnarray}
To ensure continuity at $\delta = \delta_{\rm c}$, the transition
point $\delta_{\rm c}$ is defined by the condition $\Delta \alpha(0) /
|\sin(|\delta_c| - \kappa\, \Delta \alpha(0))| = \Delta \delta(0)$.
$\kappa$ is a tuning parameter of order unity, described below.  An
example of this coincidence window model is shown in
Figure~\ref{f:FMmodel}.

For each of the 145~different $10\,{\rm Hz}$ bands, the window size is
determined by the three constants $\Delta \alpha(0)$, $\Delta
\delta(0)$ and $\kappa$.  For each sky-grid $p$ these values are
directly determined from the sky-grids used in the search as
follows. For each $10\,{\rm Hz}$ frequency band the maximum distances
between adjacent declination points to either side are calculated for
each of the 17~sky-grids as a function of declination $\delta$. In this
way, 17~different overlaying curves $\Delta_j(\delta)$ (one per data
segment) are obtained. These are indicated by the circles in
Figure~\ref{f:FMmodel} for a representative $510\,{\rm Hz}-520\,{\rm
  Hz}$ frequency band as illustration.  Then the parameter $\Delta
\delta(0)$ is obtained by considering the maximum separation to either
side between all neighboring declination grid points \textit{and}
between the 17~different sky-grids, increased by a $10\%$ safety
factor as
\begin{equation} 
	\Delta \delta(0) = 1.1\, \max_{j,\delta} \, \{\Delta_j (\delta)\}\;.
	\label{e:Delta delta zero}
\end{equation}
The largest separations near the poles ($1.4 < |\delta| \le \pi/2$)
are then found and increased by a safety factor of $20\%$ to determine
the parameter $\Delta \alpha(0)$ via
\begin{equation} 
	\Delta \alpha(0) = 1.2\, \max_{j,\delta} \, \{\Delta_j ( |\delta| > 1.4 )\}\;.
	\label{e:Delta alpha zero}
\end{equation}
Finally, the parameter $\kappa$ was chosen by visually examining
diagrams similar to Figure~\ref{f:FMmodel} for all 145 of the 10~Hz
bands.  A $\kappa$ value of $1.5$ was found to be sufficient in most
cases, while some bands required somewhat higher or lower values.
For each triple of sky-coincidence parameters, tests were
then performed to check that each sky-cell contained at least one
sky-point from each data segment.  In Figure~\ref{f:FMmodel} the
complete declination coincidence window model given by
Equation~(\ref{e:alphawin}) is represented by the solid black curve.

\begin{figure}
	\includegraphics[scale=0.47,angle=0]{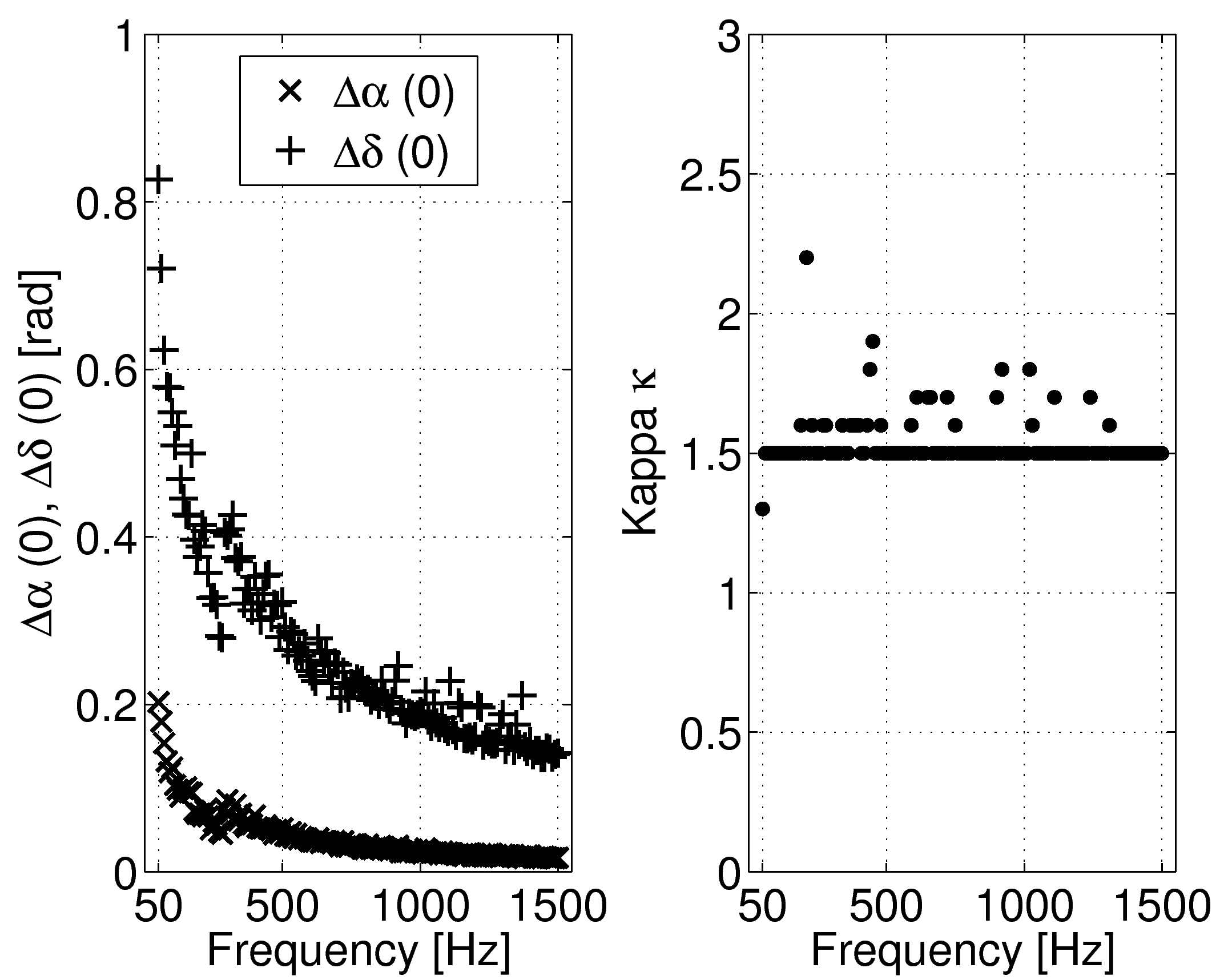}
	\caption{ The parameters $\Delta \alpha(0)$, $\Delta
          \delta(0)$ and $\kappa$ of the sky coincidence-window model
          as a function of the 10~Hz frequency band.}
\label{f:SkyCoinParams}
\end{figure}

The three parameters for all sky-grids as a function of frequency 
are shown in Figure~\ref{f:SkyCoinParams}. As stated above, 
the sky-grids are constant for $10\,{\rm Hz}$-wide steps in
frequency, and so these parameters vary with the same step-size.


\subsection{Output of the coincidence search and significance of a candidate}

The output of the post-processing pipeline is a list of the most
coincident candidates.  In each frequency band of coincidence-window
width $\Delta f$, the coincidence cell containing the largest number
of candidate events is found.  Thus for each frequency band the
pipeline finds the most coincident candidate maximized over the sky
and over the spin-down parameter-range.  The pipeline outputs the
average frequency of the coincidence cell, the average sky position
and spin-down of the candidate events, the number of candidate events
in the coincidence cell, and the significance of the candidate.

The ``significance'' of a candidate was first introduced
in~\cite{S2FstatPaper}.  A candidate, consisting of the candidate
events $1,\dots,Q$, has significance
\begin{equation}
 \mathcal{S} (1,\dots,Q) = \sum_{q=1}^{Q} (\F_q - \ln(1+\F_q))
 \label{e:significance} \; ,
\end{equation}
where $Q \le 17$ is the number of candidate events in the same coincidence
cell. To understand the meaning of the significance, consider the case
of pure Gaussian noise with no signal present.  In this case the
values of $2\F$ have a central $\chi^2$ distribution with four degrees
of freedom.  The corresponding probability density function $p_0$ of
 $2\F$ is given by
\begin{equation}
p_0(2\F) = \frac{\F}{2}\;e^{-\F} \;.
\end{equation}
The false alarm probability $P_0$ that $2\F$ exceeds a certain threshold
$2\F_0$ when there is no signal present has the form
\begin{equation}
P_0(2\F_0) = (1 + \F_0) \; e^{-\F_{0}} \;.
\label{e:P0}
\end{equation}
The joint false alarm probability of candidate events $1,\dots,Q$  
can be written as $ \prod_{q=1}^{Q} P_0(2\F_q)$. Therefore,
in this analysis candidates are ranked according to
\begin{equation}
1 - \prod_{q=1}^{Q} P_0(2\F_q) = 1 - e^{-\mathcal{S}}\;,
\end{equation}
where $\mathcal{S}=\sum_{q=1}^{Q} -\ln P_0(2\F_q)$ is exactly the
significance defined in Equation~(\ref{e:significance}).  Thus ranking
candidates by $\mathcal{S}$ is equivalent to ranking them by false
alarm probability: candidates with large positive
significance would not be expected to occur in Gaussian random noise.
As will be described later in Section~\ref{sec:Results} the significance
is used to rank equally coincident candidates within the same narrow 
frequency-band. In such cases the candidate with the largest 
significance is considered.

The post-processing pipeline has been validated by internal testing,
and also using
simulated CW signals created via so-called ``software
injections''~\cite{PletschMS}.  In addition,
Section~\ref{sec:HardwareInjections} presents realistic end-to-end
testing of the analysis pipeline using ``hardware injections'', where
simulated isolated-pulsar signals are physically added into the
interferometer control systems to produce instrumental signals that
are indistinguishable from those that would be produced by physical
sources of gravitational waves.

\section{Estimated Sensitivity\label{sec:ExpectSen}}

The sensitivity of the search is estimated using Monte-Carlo methods
for a population of simulated sources.  The goal is to find the strain
amplitude $h_0$ at which 10\%, 50\%, or 90\% of sources uniformly
populated over the sky and in their ``nuisance parameters'' (described
below) would be detected.  As previously discussed, the false alarm
probability (of getting 7 or more coincidences in a $0.5\,\Hz$
frequency-band) is of order $10^{-3}$. In this analysis,
``detectable'' means ``produces coincident events in 12 or more
distinct data segments''.  The false alarm probability for obtaining
12 or more coincidences in a $0.5\,\Hz$ band is of order $10^{-14}$,
making it extremely unlikely for candidate events from random noise to
show up consistently in 12 or more segments of data.  This is therefore an
estimate of the signal strength required for high-confidence
detection. The pipeline developed for this purpose operates in
$0.5\,\Hz$ frequency bands and consists of testing a large number of
distinct simulated sources (trials) to see if they are
detectable. A ``trial" denotes a single simulated source
which is probed for detection.

\subsection{Iteration method}

For every trial, source parameters are randomly chosen independent
of the previous trial, except for the intrinsic amplitude $h_0$.  For
the first trial $h_0$ is set to a starting value $30 \sqrt{S_h/30\;
{\rm hours}}$.  The rule for varying $h_0$ depends upon the last
$N_{\rm last}$ trials, where $N_{\rm last}^{10\%} = 100$, $N_{\rm
last}^{50\%} = 20$, and $N_{\rm last}^{90\%} = 100$.  In the past
$N_{\rm last}$ trials, if more than 10\%, 50\%, or 90\% of simulated
sources have been detected then $h_0$ is {\it decreased} by $0.25\;h_0 /
n_{\rm trial} $ for the following trial, where $n_{\rm trial}$ is an
integer in the range $ 0 \le n_{\rm trial} \le 1000$ that is
incremented with each additional trial.  On the other hand, if less
than 10\%, 50\%, 90\% of simulated sources have been detected then
$h_0$ is {\it increased} by $0.25 \; h_0/ n_{\rm trial}$ for the next trial.
This process is followed until $h_0$ has converged to a stationary
range after 1000 trials. Then the median of $h_0$ is found using
the $h_0$-values starting from that trial, where the desired
detection level has been reached the first time
during the $N_{\rm last}$ trials.
The following describes the pipeline for a single trial.

\subsection{Population of simulated sources}

For each trial, a random source frequency is chosen from a uniform
distribution in a given $0.5\,{\rm Hz}$ frequency band and a
frequency-derivative is drawn from a uniform distribution in the range
covered by the Einstein@Home search.  A sky position is chosen from a
uniform distribution on the celestial sphere, and random values are
chosen for the pulsar ``nuisance parameters''. These are the inclination
parameter $\textrm{cos} (\iota)$, initial phase $\Phi_0$, and polarization
angle $\psi$ as defined in~\cite{jks}, and are all drawn from the
appropriate uniform distributions.

\subsection{Determination of $2\F$ values for a single simulated source}

The noise floors of the different SFTs are estimated at the source's
frequency intervals using a running median with a window of $\pm
25$~frequency bins.  Figure~\ref{f:noisefloors} showed the average of
these for the data segments used from each instrument.

Then for each set of parameters the detection statistic~$2\F$ is
estimated using a semi-analytic method. From the estimated noise floor at the
simulated source's frequency and given the other source parameters, 
the expectation value of the $\F$-statistic is calculated analytically
as given in~\cite{jks}.  A random number is then drawn from a 
non-central $\chi^2$ distribution with four degrees of freedom and 
with the corresponding mean value.

These random numbers, drawn from the appropriate distribution of~$2\F$
values, would be sufficient to determine the sensitivity of the
search, if the template grid in parameter space were very closely
spaced, so that the template bank always contained at least one
waveform that was a very close match to the putative signal.  However
the grid in parameter space used in this search is quite ``coarse'',
corresponding to a mismatch of 20\% below $290\,{\rm Hz}$ and 50\%
above $300\,{\rm Hz}$, so that the $2\F$ value
that would be returned by the search might be significantly lower than
the value drawn from the distribution above.  To account for this
effect, the sensitivity prediction considers the mismatch between the
parameters of the simulated signals (determined by a random number
generator) and the template grid points of the search (fixed as
described earlier).  For each simulated source, the search grid point
that is nearest in the sense of the metric is located.  Then, using
the parameter-space metric, the mismatch between the simulated signal
and the closest search template is computed.  This gives the
fractional amount by which the $2\F$ value is reduced.
 
From this ensemble of $2\F$ values, one can determine the number of
coincidences that would be produced by each simulated source.  As
previously described, the post-processing sets an effective lower
threshold on the $\F$-statistic of the retained candidate events.  For
each simulated source, these thresholds are determined by examining
the exact workunits that would have contained the corresponding
signal.  Then the number of data segments for which the estimated
$2\F$ values are above threshold is counted.  If the $2\F$ values are
above threshold in 12 or more of the 17 data segments, the simulated
source is labeled as ``detected'', else it is labeled as
``undetected''.

\subsection{Search sensitivity, estimated errors, and comparison with
  the expected sensitivity}

\begin{figure}
   \includegraphics[scale=0.468,angle=0]{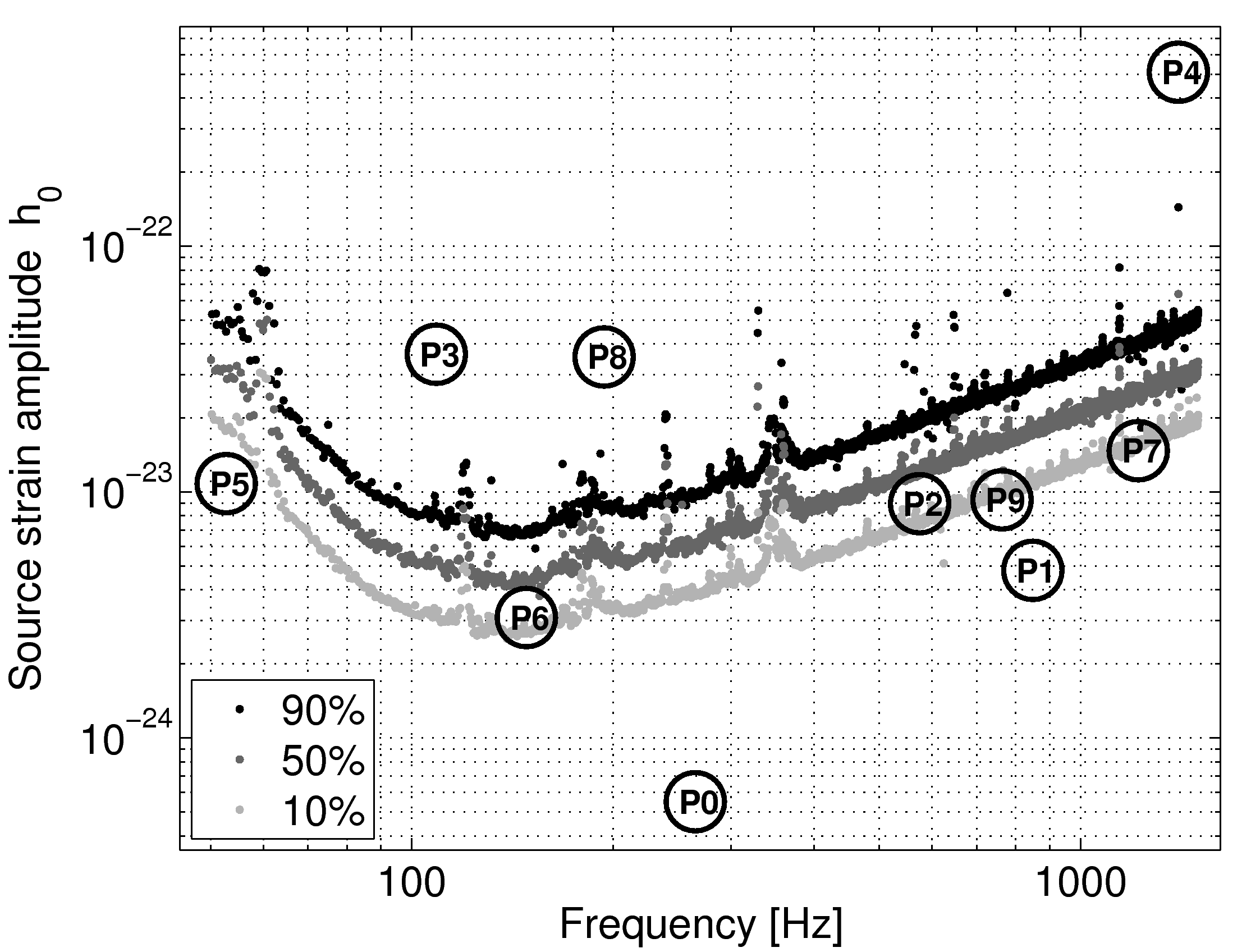}
   \caption{Estimated sensitivity of the Einstein@Home search for isolated CW
     sources in the LIGO S4 data set.  The set of three curves shows
     the source strain amplitudes $h_0$ at which 10\% (bottom), 50\%
     (middle) and 90\% (top) of simulated sources would be confidently
     detected by the Einstein@Home pipeline (appearing in coincidence
     in 12 or more of the 17 data segments).  The centers of the
     circles labeled P0 to P9 give the strain amplitudes of the S4
     hardware injections as listed in
     Table~\ref{t:HWInjectionsParams}. Based on this curve, one would
     expect that the simulated signals P3, P4 and P8 could be
     confidently detected, and that P0, P1 and P5 would not be detected.}
   \label{f:estULs}
\end{figure}

Shown in Figure~\ref{f:estULs} are the resulting search sensitivity
curves as functions of frequency. Each data point on the plot denotes
the results of 1000 independent trials.  These show the values of
$h_0$ as defined in~\cite{jks} such that 10\%, 50\%, and 90\% of
simulated sources produce 12 or more coincidences in the
post-processing pipeline.
The dominant sources of error in this sensitivity curve are uncertainties 
in the magnitudes of the LIGO detector response functions (calibration errors). 
Details of these frequency-dependent uncertainties may be 
found in reference~\cite{S4CalibrationNote}.
The uncertainties are typically of order 5\% (L1) and 8\% (H1) in the frequency band 
from 50-1500 Hz, and are always less than 10\%. Systematic errors, which arise because 
of the finite number of Monte-Carlo trials and similar effects, are less than $\pm 2$\%. 
These can be added in quadrature to the uncertainties given in \cite{S4CalibrationNote} to 
obtain frequency-dependent error bounds in the sensitivity curve.  The resulting 
error in this sensitivity plot is below 10\% at all frequencies.

The behavior of the curves shown in Figure~\ref{f:estULs} essentially reflect the
instrument noise given in Figure~\ref{f:noisefloors}. 
One may fit the curves obtained in Figure~\ref{f:estULs} to the strain noise power 
spectral density $S_h(f)$ and then
describe the three sensitivity curves in Figure~\ref{f:estULs} by
\begin{equation}
  h_0^{\D} (f) \approx R_{\D} \; \sqrt{\frac{S_h(f)}{30\; {\rm hours}}}\;,
  \label{e:sensitivity}
\end{equation}
where the pre-factors $R_{\D}$ for different detection probabilities
levels $\D=90\%$, $50\%$, and $10\%$ are well fit below $300\,\Hz$ by
$R_{90\%} = 31.8$, $R_{50\%} = 20.1$, and $R_{10\%} = 12.6$, above
$300\,\Hz$ by $R_{90\%} = 33.2$, $R_{50\%} = 21.0$, and $R_{10\%} =
12.9$.

Some other published CW searches were done at $95\%$ detection
confidence.  For comparison in the next section, the sensitivity of
{\it this} search at that confidence is $R_{95\%}=36.2$ below 300~Hz
and $R_{95\%}=37.9$ above 300~Hz.  The iteration method previously
described used $N_{\rm last}^{95\%} = 200$.

\subsection{Comparison with sensitivity of other search and upper limit methods}

The methods used here would be expected to yield very high confidence if
a strong signal were present.  It is interesting to compare
the sensitivity of this detection method with the sensitivity of
CW upper limits such as that of reference~\cite{S2FstatPaper}.
The sensitivity of the high-confidence detection method
used here is well-described by Equation~\ref{e:sensitivity}.  
The same equation describes the results of the S2 $\F$-statistic
loudest-event upper limit analysis~\cite{S2FstatPaper}, but in that
work the 95\% detection confidence curve has a pre-factor
$R_{95\%}=29.5$.  It is useful to understand the source of this apparent
difference in sensitivity (a factor of $37.9/29.5=1.28$).
There are three main contributors to this difference.

The most significant difference between the two analyses is the
spacing of the search grid templates.  In this search, the templates
are significantly farther apart (worst-case 50\% loss of
signal-to-noise ratio, or expected $2\F$) than in~\cite{S2FstatPaper},
where the worst-case mismatch was negligible.  This effect of
employing different mismatches has been studied by running the
sensitivity estimation pipeline using simulated sources only at the
template grid points, and reduces $R_{95\%}$ in
Equation~(\ref{e:sensitivity}) by a factor of 1.17.

Another difference between the two analyses is the detection criteria. 
In this work, detection requires a signal to produce 12
or more coincidences between the 17 different data segments.
This corresponds to a false alarm probability (in Gaussian noise) of order
$10^{-14}$ per $0.5\,\Hz$ frequency-band. This is different 
from~\cite{S2FstatPaper}, where simulated signals are 
compared against the loudest candidate found (largest $2\F$).  
An equivalent detection criterion for {\it this} work would be 
to compare the simulated signals against the loudest
candidates (per 0.5~Hz band). These typically had 7 or 8 coincidences,
corresponding to a Gaussian noise false alarm probability of order
$10^{-3}$ and $10^{-5}$, respectively.  To estimate the effect on the
sensitivity, the sensitivity estimation pipeline was rerun, but now
requiring the signal to exceed the $2\F$-thresholds in only 7 of the
17 data segments.  This reduced $R_{95\%}$ in
Equation~(\ref{e:sensitivity}) by an additional factor of 1.14.

The least important difference between the two analyses is the
effective threshold on the $\F$-statistic.  As explained in
Sections~\ref{ssec:select-cands} and~\ref{ssec:cands-retained}, only a
subset of candidate events with the largest values of $2\F$ are
retained in the post-processing, fixing the false alarm probability.
The smallest $2\F$-value on this list is typically around $28$ or
slightly higher.  In~\cite{S2FstatPaper} a fixed threshold of $2\F
=20$ has been used.  Then, events with a combined significance below
$\mathcal{S}=64.4$ [see Equation~(\ref{e:significance})] were also
dropped.  While it is difficult to compare these two criteria, they
seem to be fairly close.

Taken together, the differences in grid spacing and detection
thresholds are responsible for, and explain, the sensitivity
difference in the two analyses (a factor of $1.17 \times 1.14 = 1.33
\approx 1.28 $).

\section{Vetoing of instrumental line artifacts\label{sec:VetoMethod}}

When the instrument data was prepared and cleaned, narrow-band
instrumental line features of known origin were removed, as previously
described in Section~\ref{sec:DataPrep}.  However, the data also
contained stationary instrumental line features that were not
understood, or were poorly understood.  These features were {\it not}
removed from the input data for the search.  As a consequence, the
output from the post-processing pipeline contains instrumental
artifacts that in some respects mimic CW signals.  But these artifacts
tend to cluster in certain regions of parameter space, and in many
cases they can be automatically identified and vetoed.  
In previous incoherent searches for CW sources in LIGO 
data~\cite{S4IncoherentPaper} the S-veto method has been
employed, which excludes the regions of parameter space where there 
is little or no frequency modulation from the Earth's motion, 
leading to a relatively stationary detected frequency. 
This cannot directly be applied to a coherent matched-filtering 
search using the $\F$-statistic.
Thus the method used here will be similar, but arises from
a conceptually different approach that is appropriate for
an $\F$-statistic search.

\subsection{Parameter space locations of instrumental lines}

For a coherent observation of $30\,{\rm hours}$ the parameter-space 
regions where instrumental lines tend to appear are determined by the 
global-correlation hypersurfaces~\cite{PletschGC} in parameter space.
The global-correlation hypersurface on which stationary instrumental 
lines preferably produce candidate events is shown in~\cite{PletschGC} 
to be described by
\begin{equation}
  \dot f + \frac{(\vec{{\boldsymbol{ \omega}}} 
  \times \vec{v}_{\rm av}) \cdot \hat{\vec{n}} }{c} f(t_{\rm{fiducial}})  =0 \;,
  \label{e:NoiseHypersurface}
\end{equation}
where $c$ denotes the speed of light, $ \hat{\vec{n}}$ is a unit vector pointing
to the source's sky-location in the SSB frame and relates to the equatorial 
coordinates~$\alpha$ and~$\delta$ by  
$ \hat{\vec{n}} = (\cos \delta \, \cos \alpha, \cos \delta \, \sin \alpha, \sin \delta)$, 
$\vec{\boldsymbol{\omega}}$ is the angular velocity vector of
the Earth as it orbits around the Sun ($|\vec{\boldsymbol{\omega}}| \approx
2\pi/\rm year$) and $\vec{v}_{\rm av}$ is the average velocity of the
Earth ($|\vec{v}_{\rm av}| \approx 9.9\times 10^{-5} \,c$).
This equation can also be understood on simple physical grounds.  The l.h.s. of Equation (\ref{e:NoiseHypersurface})
is the rate of change of detector output frequency, for a source
whose SSB frequency and spin-down are $f$ and $\dot f$.  An instrumental line,
which has fixed detector frequency, mimics such a source when the l.h.s. vanishes.

The potential CW sources whose locations in parameter space are 
consistent with Equation~(\ref{e:NoiseHypersurface}) 
will not produce a modulation pattern that would distinguish 
them from an instrumental line.  As the resolution in
parameter space is finite, the post-processing analysis 
eliminates (vetoes) candidates that satisfy the condition 
\begin{equation}
  \left| \,\dot f + \frac{(\vec{{\boldsymbol{ \omega}}} 
  \times \vec{v}_{\rm av}) \cdot \hat{\vec{n}} }{c} 
  f(t_{\rm{fiducial}}) \, \right| < \epsilon \;,
  \label{e:VetoCond}
\end{equation}
where the parameter $\epsilon > 0$ accounts for a certain tolerance
needed due to the parameter-space gridding. This tolerance-parameter
can be understood as
\begin{equation}
  \epsilon = \frac{\Delta f}{\Delta T} N_{\rm cell} \;,
\end{equation}
where $\Delta f$ denotes width in frequency (corresponding
to the coincidence-cell width in the post-processing) up to
which candidate events can be resolved during the 
characteristic length of time $\Delta T$, and $N_{\rm cell}$
represents the size of the vetoed or rejected region, measured in coincidence cells .
In this analysis $\Delta T = 2\,122\,735\, {\rm s}$ 
($\approx 24\,{\rm days}$) is the total time interval
spanned by the data
\begin{equation}
	\Delta T =  |\max_{j} \, t_j^{\textrm{end}}  
	- \min_{j} \, t_{j} | = t_{11}^{\textrm{end}} - t_{4} \; .
\end{equation}
For potential sources that satisfy~(\ref{e:VetoCond}), the
modulation due to the Earth's motion does not make the signal 
appear in more than $N_{\rm cell}$ coincidence cells during $\Delta T$.

\subsection{Fraction of parameter space excluded by the veto}

\begin{figure}
\includegraphics[scale=0.46,angle=0]{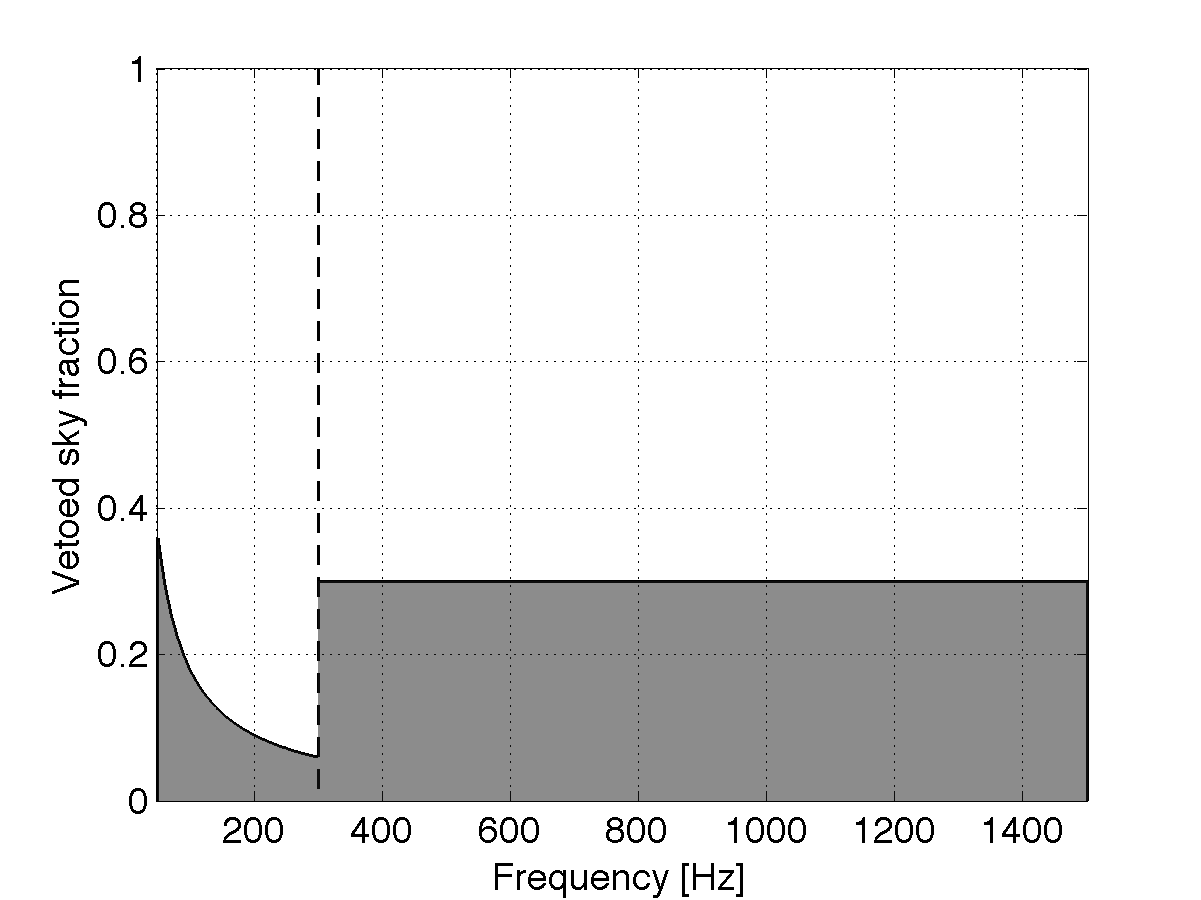}
\caption{ The average fraction of sky excluded by the veto method as a
  function of frequency, uniformly averaged over the searched
  spin-down range.\label{f:skyfraction} }
\end{figure}

One can visualize and calculate the volume of the region in
four-dimensional parameter space that is excluded by this veto.  For
a given source sky position, Equation~(\ref{e:VetoCond}) 
is linear in $f$ and $\dot f$. Thus, for fixed sky position
$\hat{\vec{n}}$, the veto condition~(\ref{e:VetoCond}) defines two
parallel lines in the $(f, \dot f)$-plane. Candidate events that lie
in the region between the lines are discarded (vetoed).  Candidates
that lie outside this region are retained (not vetoed).  The
locations of these two lines in the $(f,\dot f)$ plane depend upon
the sky position.  The fractional volume excluded by the veto depends
upon whether or not (as the source position varies over the sky) the
excluded region between the lines lies inside or outside of the
boundaries of the search, or intersects it.  In this work, for the
search region $-f/\tau < \dot f < 0.1\,f/\tau$ described in the
Abstract, the excluded region lies entirely within the parameter space
above 300~Hz, but crosses the boundaries below 300~Hz.  This is
because a wider range of spin-down ages is searched below 300~Hz.

The fractional volume of the region in parameter space excluded by the
veto may be easily calculated. The details of the calculation are
in Appendix~\ref{sec:fractionveto}.  The resulting fraction of sky 
excluded by the veto (uniformly averaged over spin-down) as a function 
of frequency is shown in Figure~\ref{f:skyfraction}.
In this search, the fraction of the sky excluded for frequencies $f \in [300,
1500)\,{\rm Hz}$ has been fixed at the constant fraction $30\%$. 
In this search, the fraction of the sky excluded for frequencies $f \in
[50, 300)\,{\rm Hz}$ has been chosen to depend upon the values of 
$f$~and~$\dot f$, where the uniform average of the excluded sky fraction 
over the spin-down range considered in this analysis is 
$36\%$ at $50\,{\rm Hz}$ and $6\%$ just below $300\,{\rm Hz}$.
Finally, Figure~\ref{f:VetoedCands} shows a conclusion diagram
illustrating which of the candidates have been vetoed in this search.

\section{Hardware-Injected Signals \label{sec:HardwareInjections}}

\begin{figure*}
   \subfigure{\includegraphics[scale=0.21,angle=0]{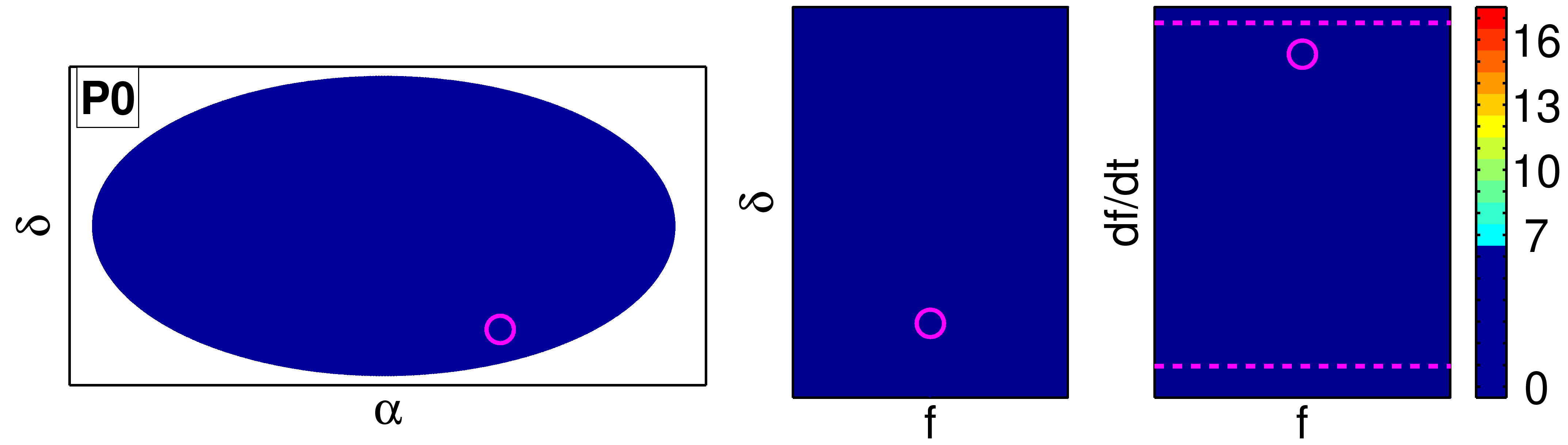}}
     \qquad
   \subfigure{\includegraphics[scale=0.21,angle=0]{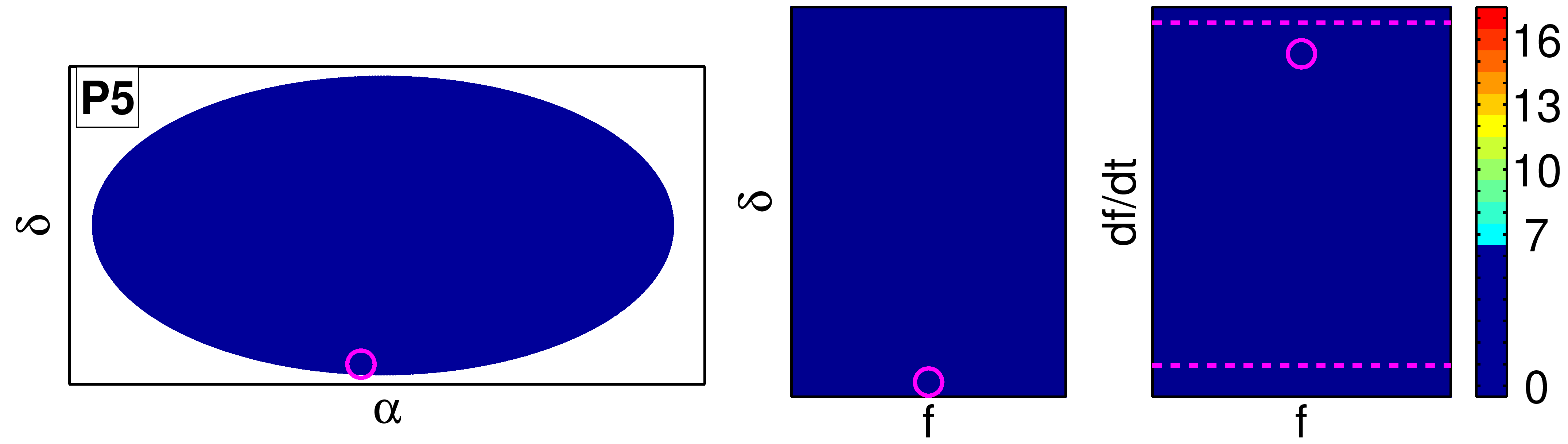}}\\
   
   \subfigure{\includegraphics[scale=0.21,angle=0]{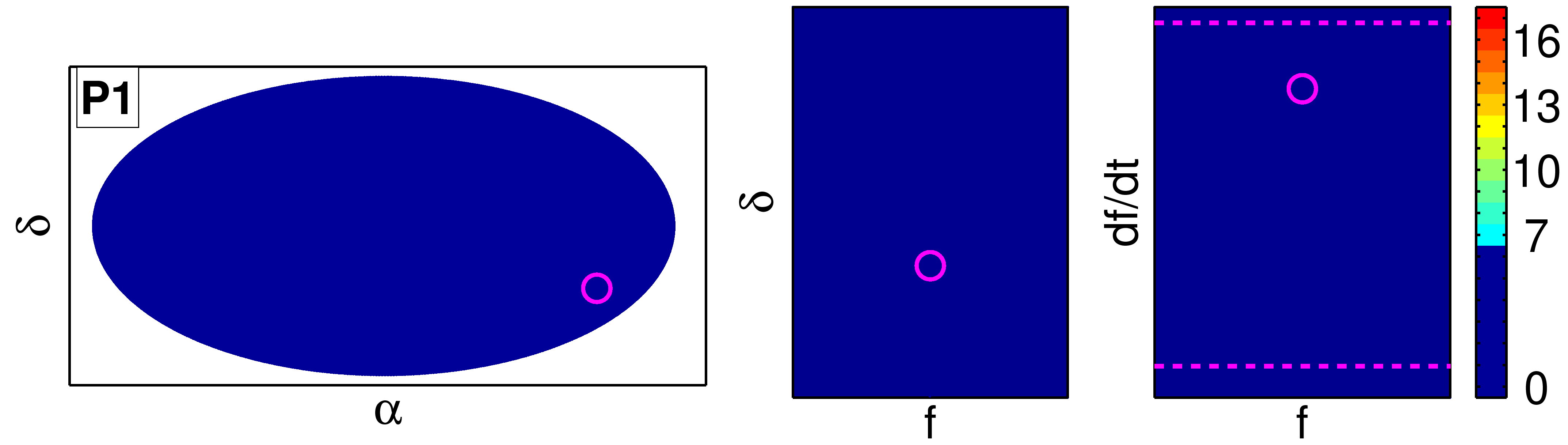}}
     \qquad
   \subfigure{\includegraphics[scale=0.21,angle=0]{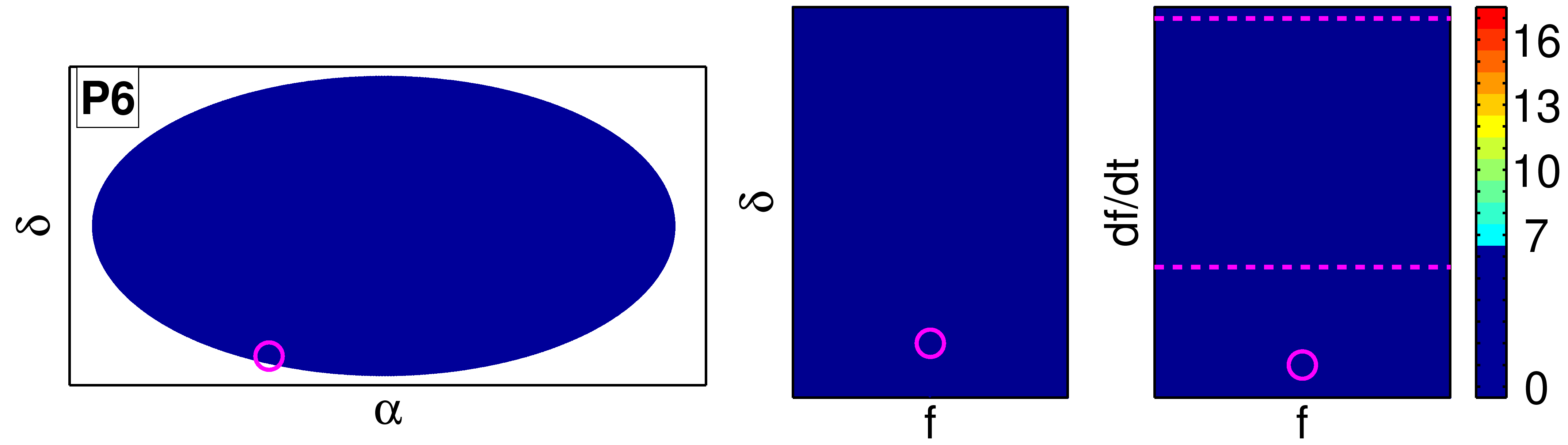}}\\
   
   \subfigure{\includegraphics[scale=0.21,angle=0]{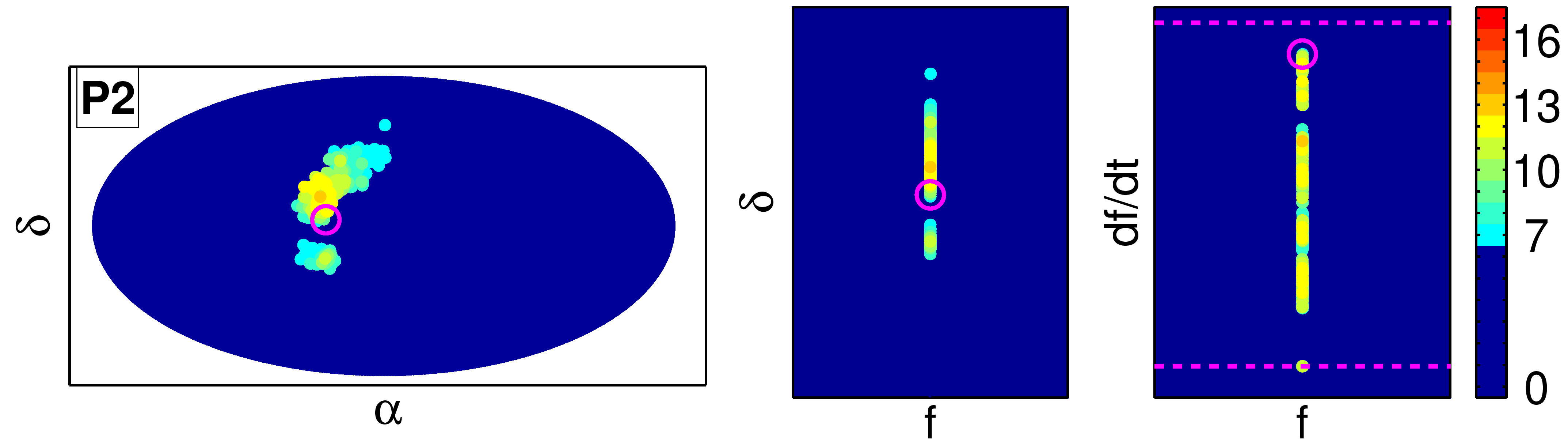}}
     \qquad
   \subfigure{\includegraphics[scale=0.21,angle=0]{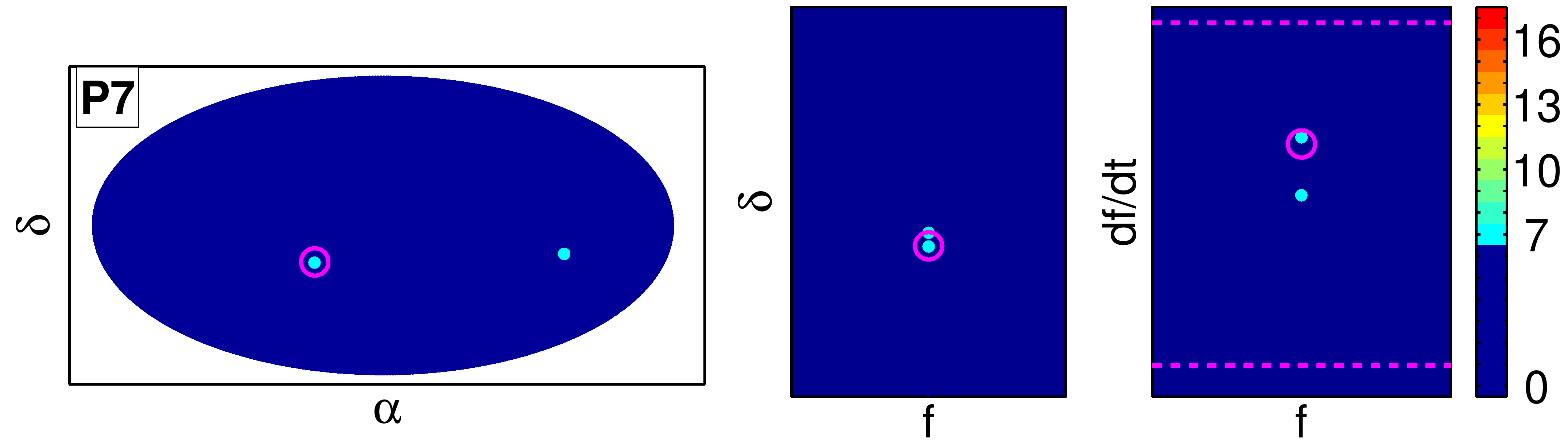}}\\
   
   \subfigure{\includegraphics[scale=0.21,angle=0]{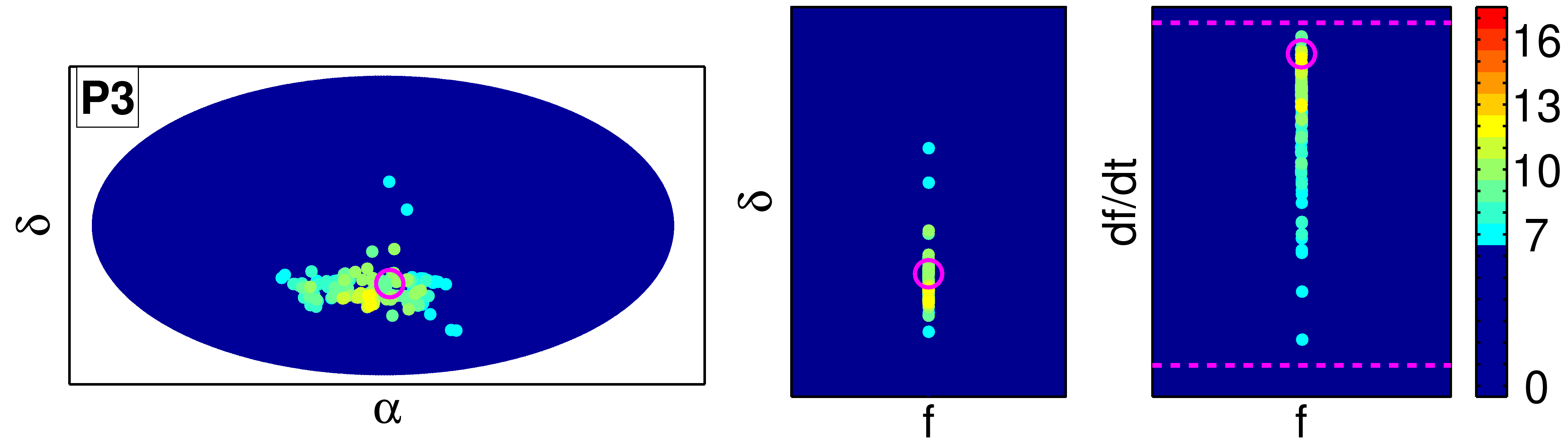}}
     \qquad
   \subfigure{\includegraphics[scale=0.21,angle=0]{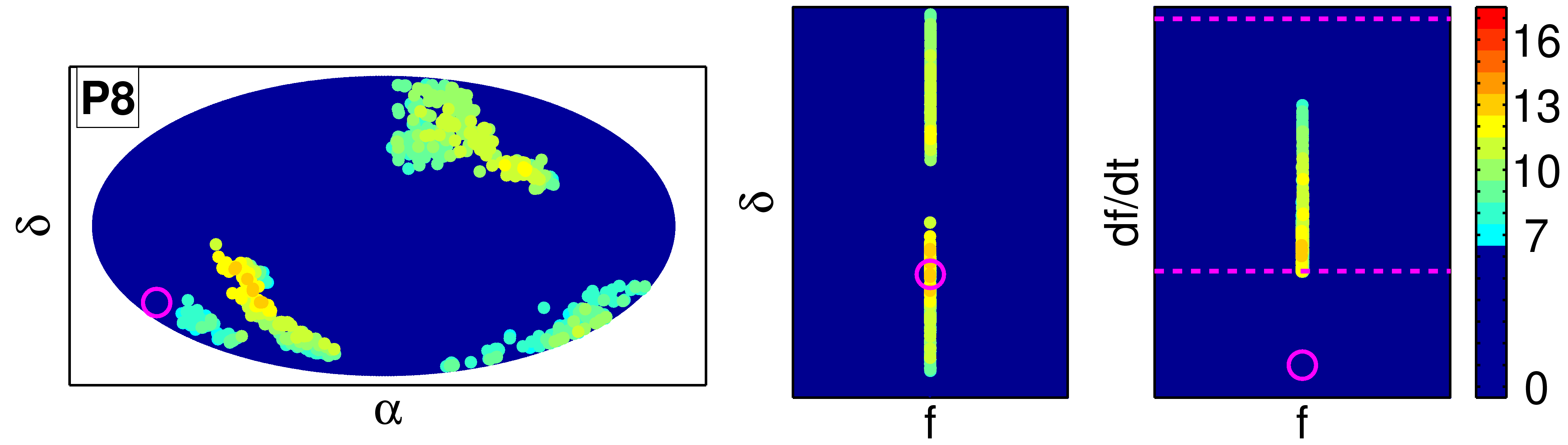}}\\
   
   \subfigure{\includegraphics[scale=0.21,angle=0]{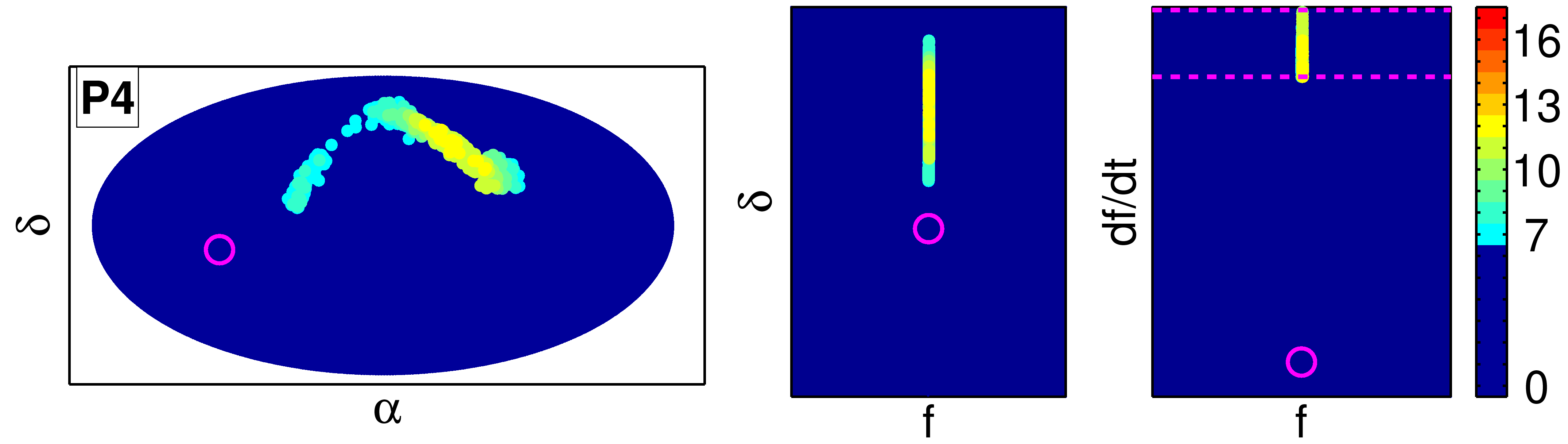}}
     \qquad
   \subfigure{\includegraphics[scale=0.21,angle=0]{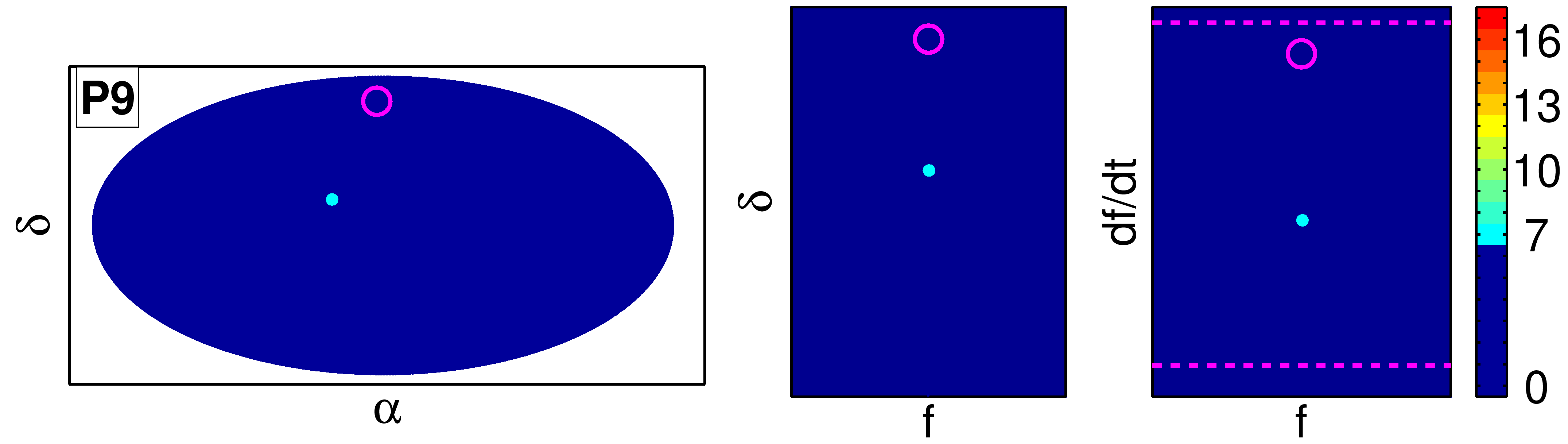}}
   
   \caption{Einstein@Home results showing the 10 hardware-injected
     pulsar signals labeled P0 to P9.  Here, a narrow band of width
     $2\times 10^{-4}\;f$ to either side of each injection's frequency
     $f$ is considered. The color-bar in each plot indicates the
     number of coincidences.  As shown in the color-scale, only
     candidates having 7 or more coincidences appear.  For each
     hardware injection a group of three different sub-plots are given
     representing different projections of the parameter space.  The
     left sub plot is a Hammer-Aitoff projection of the entire sky.
     The middle sub-plot shows declination $\delta$ versus frequency
     $f$.  The right sub-plot shows spin-down $\dot f$ versus
     frequency $f$, where the region between the two horizontal
     magenta dashed lines refers to searched range of spin-downs.  The
     center of a magenta circle represents the location of the
     injection.  P4 and P8 appear at the wrong sky position because
     their intrinsic spin-downs lie outside the searched range.  
     Table~\ref{t:HWresults} shows a comparison with the expectations
     for these simulated signals.
     \label{f:HWresults}}
\end{figure*}

\begin{table*}
\begin{tabular}{ccrcrccrrcrcc} \hline
 Name  & &$f(t_{\rm ref})$ [Hz] & & $\dot f$ [$\mathrm{Hz}~\mathrm{s}^{-1}$] & &$\alpha$ [rad] & $\delta$ [rad] & $\psi$ [rad] & $\Phi_0$ [rad] & $\cos \iota$ [rad] & & $h_0$ \\
\hline \hline
 Pulsar0 & & $265.57693318$ & &$-4.15\times 10^{-12}$ & &$1.248817$ & $-0.981180$ & $0.770087$ & $2.66$ & $0.794905$ & &$4.93 \times 10^{-25}$ \\
 Pulsar1 & & $849.07086108$ & &$-3.00\times 10^{-10}$ & &$0.652646$ & $-0.514042$ & $0.356036$ & $1.28$ & $0.463799$ & &$4.24 \times 10^{-24}$ \\
 Pulsar2 & & $575.16356732$ & &$-1.37\times 10^{-13}$ & &$3.756929$ & $0.060109$ & $-0.221788$ & $4.03$ & $-0.928575$ & &$8.04 \times 10^{-24}$ \\
 Pulsar3 & & $108.85715940$ & &$-1.46\times 10^{-17}$ & &$3.113189$ & $-0.583579$ & $0.444280$ & $5.53$ & $-0.080666$ & &$3.26 \times 10^{-23}$ \\
 Pulsar4 & & $1402.11049084$ & &$-2.54\times 10^{-08}$ & &$4.886707$ & $-0.217584$ & $-0.647939$ & $4.83$ & $0.277321$ & &$4.56 \times 10^{-22}$ \\
 Pulsar5 & & $52.80832436$ & &$-4.03\times 10^{-18}$ & &$5.281831$ & $-1.463269$ & $-0.363953$ & $2.23$ & $0.462937$ & &$9.70 \times 10^{-24}$ \\
 Pulsar6 & & $148.44006451$ & &$-6.73\times 10^{-09}$ & &$6.261385$ & $-1.141840$ & $0.470985$ & $0.97$ & $-0.153727$ & &$2.77 \times 10^{-24}$ \\
 Pulsar7 & & $1220.93315655$ & &$-1.12\times 10^{-09}$ & &$3.899513$ & $-0.356931$ & $0.512323$ & $5.25$ & $0.756814$ & &$1.32 \times 10^{-23}$ \\
 Pulsar8 & & $193.94977254$ & &$-8.65\times 10^{-09}$ & &$6.132905$ & $-0.583263$ & $0.170471$ & $5.89$ & $0.073904$ & &$3.18 \times 10^{-23}$ \\
 Pulsar9 & & $763.8473216499$ & &$-1.45\times 10^{-17}$ & &$3.471208$ & $1.321033$ & $-0.008560$ & $1.01$ & $-0.619187$ & &$8.13 \times 10^{-24}$ \\
\hline
\end{tabular}
\caption{
  Parameters for hardware-injected CW signals during the S4 run,
  labeled Pulsar0 to Pulsar9.  The parameters are defined at the GPS
  reference time $t_{\rm ref}= 793130413\,{\rm s}$ in the
  Solar System Barycenter.  These are the frequency $f(t_{\rm ref})$, the spin-down
  $\dot f$, the sky position right ascension $\alpha$ and declination
  $\delta$, the polarization angle $\psi$, the initial phase $\Phi_0$,
  the inclination parameter $\cos \iota$, and the dimensionless strain
  amplitude $h_0$. Because the calibration was only accurately determined
  after S4 was finished, the H1 strain amplitudes should be multiplied
  by the correction factor $1.12$.  The L1 amplitudes should be
  multiplied by $1.15$ for Pulsar1, $1.18$ for Pulsar9, and $1.11$ for
  the others.
}
\label{t:HWInjectionsParams}
\end{table*}

A good way to test and validate search algorithms and code is to add
simulated signals into the detector's data stream.  This can either be
done while the experiment is in progress (real-time injections) or
after the data has been collected (software injections).  If it is
done while the experiment is in progress, the simulated signals can
either be added into the hardware (into feedback and error-point
control signals) or after data acquisition.

At the time that the S4 run was carried out, ten simulated CW signals
were injected at the hardware level: using magnetic coil actuators,
the interferometer mirrors were physically made to move as if a
gravitational wave was present.  

\subsection{Parameters of hardware injections}

Table~\ref{t:HWInjectionsParams} shows the parameters of the hardware
injections that were carried out at the LIGO detectors during the S4
run, mimicking gravitational-wave signals from ten different isolated
pulsars with different frequencies, sky locations, and frequency
derivatives.  The ten artificial pulsars are denoted Pulsar0 to
Pulsar9.  At the time of the injections, lack of complete knowledge of
the instrument's response function (calibration) meant that the actual
hardware injections did not actually have the intended strain
amplitudes as given in the Table.  The effective strain amplitudes may
be computed from correction factors provided in
reference~\cite{S4IncoherentPaper}.  These factors are $1.12$ for all
simulated pulsars in the H1 detector.  In the L1 detector, the
correction factor is $1.11$ for all simulated pulsars, except for
Pulsar1 ($1.15$) and Pulsar9 ($1.18$).

\begin{table}
\begin{center}
\lineup
\begin{tabular}{ccccccc}
\hline
 $j$ && Detector & & Overlapping  && Fractional  \\
 & &  & & Duration [s]  && Overlap \\
\hline\hline
1 & &H1&& 107201 && 99.3 \% \\
2 & &H1&&107554 && 99.6 \% \\
3 & &H1&&107272 && 99.3 \% \\
4 & &H1&&0 && 0 \\
5 & &H1&&$\099799$ && 92.4 \% \\
6 & &H1&&0 && 0 \\
7 & &H1&&101991 && 94.4 \% \\
8 & &H1&& $\021268$ && 19.7 \% \\
9 & &H1&&100773 && 93.3 \% \\
10 & &H1&&0 && 0 \\
11 & &L1&&$\023164$ && 21.5 \% \\
12 & &L1&&106760 && 98.9 \% \\
13 & &L1&&107294 && 99.4 \% \\
14 & &L1&&102711 && 95.1 \% \\
15 & &L1&&0 && 0 \\
16 & &L1&&0 && 0 \\
17 & &L1&&98696 && 91.4 \% \\
\hline
\end{tabular}
\caption{ \label{t:HWoverlap} The time overlap between
  the Einstein@Home data segments and the hardware injections.  The
  hardware injections were only turned on about 2/3 of the time.  The
  columns are data segment index $j$, detector, the duration of the
  overlap, and the fractional overlap (obtained by dividing the third
  column by $30 \,\rm hours = 108\,000 \,\rm s$).}
\end{center}
\end{table}

\subsection{Duty cycle of hardware injections}

During S4 the hardware injections were not active all of the time.
Table~\ref{t:HWoverlap} shows the fractional overlap between the times
when the hardware injections were active and the times of the S4
Einstein@home data segments.  As can be seen from the table, the
hardware injections were only turned on during twelve of the data
segments analyzed in this paper, and for two of those twelve data data
segments, the injections were only turned on for about 20\% of the
data taking time.  In the remaining ten data segments, the hardware
injections were turned on for almost the entire segment. This needs to
be taken into account when analyzing the Einstein@Home search results
for these injections. Because of this, the maximum possible number of
coincidences expected from these simulated signals is $12$, even
though $17$ data segments are searched.

\begin{table}
\begin{center}
\lineup
\begin{tabular}{ccccc}
\hline
Name &   Predicted   &  Measured & Predicted & Measured \\
            &  $\mathcal{C}$  & $\mathcal{C}$   & $\mathcal{S}$     & $\mathcal{S}$  \\
\hline\hline
Pulsar2  & $12$ & $13$ &  $\0\0263.1$ & $\0\0249.3$ \\
Pulsar3 & $12$ & $12$ &  $\03160.9$ & $\02397.5$ \\
Pulsar4& $12$ & $12$ &  $35108.2$ & $\01749.6$ \\
Pulsar7  & $\06$ & $\07$ & $\0\0\093.2$ & $\0\0100.0$ \\
Pulsar8 & $12$ & $13$ &  $\03692.6$ & $\02263.6$ \\
Pulsar9 & $\07$ & $\07$ &  $\0\0131.2$ & $\0\0\098.9$\\
\hline
\end{tabular}
\caption{ \label{t:HWresults} The estimated (predicted) and obtained
  (measured) results for the hardware-injected pulsar signals. For
  each simulated signal the predicted number of coincidences
  $\mathcal{C}$ and a predicted value for the significance
  $\mathcal{S}$ is given, as well as the measured number of
  coincidences and measured value for the significance from the
  Einstein@Home search. The measured values are obtained by
  maximizing over a narrow band of $2\times 10^{-4} \,f$ on either
  side of the injection frequency, the whole sky and the entire
  spin-down range.  As explained in the text, Pulsar4 and Pulsar8 are
  not expected to have the correct significance. Pulsar0, Pulsar1,
  Pulsar5 and Pulsar6 are not listed.  They are so weak that they
  produce less than 7 coincidences, consistent with random
  noise containing no signal.}
\end{center}
\end{table}

\subsection{Results from the hardware injections}

For each hardware-injected pulsar signal Table~\ref{t:HWresults}
compares a prediction for the outcome of the Einstein@Home search to
the actual results found through the Einstein@Home analysis pipeline.
The predicted values given in Table~\ref{t:HWresults} are obtained by
feeding the sensitivity-estimation pipeline, which was described in
Section~\ref{sec:ExpectSen}, with the parameters of the simulated
pulsars and only considering data segments where the hardware
injections were active.

As shown in Table~\ref{t:HWresults} and consistent with
Figure~\ref{f:estULs}, the hardware-injected signals Pulsar0, Pulsar1,
Pulsar5 and Pulsar6 are too weak to be confidently detected by the
search. In contrast, Pulsar2, Pulsar3, Pulsar4 and Pulsar8 are clearly
detected. 
The parameters of Pulsar7 and Pulsar9 are such that in both
cases the search pipeline found $7$ coincidences, but this is
consistent with the level of coincidences that would result from
Gaussian noise with no signal present, and so these are not
confidently detected.

Figure~\ref{f:HWresults} presents the results of the search for
all hardware injections. Small subspaces of the search
parameter space around the hardware injections are shown, as well as the
locations of the artificial signal parameters. The subspaces considered
in Figure~\ref{f:HWresults} and also for the (measured) results presented 
in Table~\ref{t:HWresults} are constrained to a band of 
$2\times 10^{-4}\;f$ to either side of the injected frequency.
This choice of frequency-bandwidth is motivated by the
global parameter-space correlations~\cite{{PrixItoh},{PletschGC}}
and represents the approximate maximum extension in frequency-direction
of a global-correlation hypersurface generated by a single signal.

Moreover, the global correlations also explain the significant sky position
offset between the a priori location of the simulated source and the
location where the search located the source 
with respect to the detected signals Pulsar4 and Pulsar8.
This arises because for Pulsar4 and Pulsar8, the spin-down range 
that is searched (region between dashed lines in the far right column) 
is too small to include the actual spin-down value used in creating the simulated
signals.  Due to the global parameter-space correlations the offset between 
the actual and detected spin-down value gives rise to the offset in the 
sky position. The observed structure of large-coincident events in the sky
is consistent with a (sky-projected) global-correlation
hypersurface first found analytically in~\cite{PletschGC}.
This is also why Pulsar4 shows a considerable discrepancy 
between the significance that
would have been expected if the search-grid had also covered 
the a priori parameters, and the significance that was actually 
observed in the search, as shown in Table~\ref{t:HWresults}.

\section{Results \label{sec:Results}}

\begin{table*}
\begin{tabular}{crrrrrrcrcccc} \hline
&$f_{\rm cand}$ [Hz] & $f_{\rm start}$ [Hz] & $\Delta f_{\rm cand}$ [Hz] &$\delta_{\rm cand}$ [rad] & 
 $\alpha_{\rm cand}$ [rad] & $\dot f_{\rm cand}$ [$\mathrm{Hz}~\mathrm{s}^{-1}$] & $\mathcal{C}_{\rm cand}$ & $\mathcal{C}^{\rm H1}_{\rm cand}$ & $\mathcal{C}^{\rm L1}_{\rm cand}$ & \; $\mathcal{S}_{\rm cand}$ \;& Information\\

\hline \hline
& $193.9276$ & $193.9263$ & $0.040112$ & $-0.583514$ & $4.723595$ & $-5.6001\times 10^{-09}$ & $13$ &  $7 $  &  $ 6$  & $2263.6$ & Pulsar 8 & \\ 
& $575.1681$ & $575.1562$ & $0.030612$ & $0.285505$ & $3.834511$ & $-5.0913\times 10^{-10}$ & $13$ &  $7 $  &  $ 6$  & $249.3$ & Pulsar 2 & \\ 
& $1128.1147$ & $1128.0336$ & $0.220321$ & $-1.395918$ & $0.744273$ & $-3.4249\times 10^{-09}$ & $13$ &  $10 $  &  $3 $  & $219.3$ & H1 MC 2/3 & \\ 
& $108.8549$ & $108.8522$ & $0.008158$ & $-0.705729$ & $3.361465$ & $-4.4362\times 10^{-11}$ & $12$ &  $9 $  &  $3 $  & $2397.5$ & Pulsar 3 & \\ 
& $329.6107$ & $329.5507$ & $0.066447$ & $1.027320$ & $1.336051$ & $-5.7799\times 10^{-10}$ & $12$ &  $10 $  &  $ 2$  & $3127.1$ &  Demod &\\ 
& $545.9973$ & $545.9929$ & $0.10958$ & $-0.293877$ & $4.849960$ & $-1.5782\times 10^{-09}$ & $12$ &  $10 $  &  $ 2$  & $893.3$ & H1 MC 2/3 & \\ 
& $566.0868$ & $566.0490$ & $0.105853$ & $-1.367663$ & $0.665233$ & $-1.626\times 10^{-09}$ & $12$ &  $ 10$  &  $2 $  & $2340.8$ & H1 MC 2/3 & \\ 
& $568.0886$ & $567.9893$ & $0.165769$ & $-1.323532$ & $0.726729$ & $-1.7149\times 10^{-09}$ & $12$ &  $ 10$  &  $ 2$  & $4137.7$ & H1 MC 2/3 &  \\ 
& $648.8288$ & $648.6930$ & $0.206223$ & $-1.232868$ & $1.005733$ & $-1.0298\times 10^{-09}$ & $12$ &  $ 10$  &  $ 2$  & $1870.8$ & H1 MC 1 & \\ 
& $1143.9976$ & $1143.9182$ & $0.232221$ & $-1.491264$ & $1.314456$ & $-7.7434\times 10^{-10}$ & $12$ &  $10 $  &  $2 $  & $1028.8$ & H1 Cal & \\ 
& $1144.5198$ & $1144.4533$ & $0.228407$ & $-1.535248$ & $4.497733$ & $-2.5257\times 10^{-11}$ & $12$ &  $10 $  &  $2 $  & $989.8$ &H1 Cal & \\ 
& $1289.6769$ & $1289.5081$ & $0.242915$ & $1.461093$ & $0.266878$ & $-2.0949\times 10^{-09}$ & $12$ &  $ 10$  &  $2 $  & $493.7$ & H1 MC 1 & \\ 
& $1402.2838$ & $1402.2677$ & $0.063117$ & $1.025583$ & $2.502838$ & $-3.8482\times 10^{-09}$ & $12$ &  $6 $  &  $ 6$  & $1749.6$ & Pulsar 4 & \\ 
& $329.7593$ & $329.7396$ & $0.066078$ & $-1.536179$ & $4.887048$ & $-5.5375\times 10^{-10}$ & $11$ &  $ 10$  &  $1 $  & $3038.3$ & Demod & \\ 
& $335.7735$ & $335.7100$ & $0.065415$ & $0.469606$ & $0.955884$ & $-1.0646\times 10^{-09}$ & $11$ &  $10 $  &  $1 $  & $298.5$ &EM Interference & \\ 
& $545.9232$ & $545.8662$ & $0.063608$ & $-1.060735$ & $1.078303$ & $-8.032\times 10^{-10}$ & $11$ &  $ 10$  &  $1 $  & $196.8$ &  H1 MC 2/3 & \\ 
& $564.1219$ & $564.0096$ & $0.113783$ & $0.386877$ & $1.111355$ & $-1.6868\times 10^{-09}$ & $11$ &  $10 $  &  $0$  & $1069.3$ & H1 MC 2/3 & \\ 
& $646.3758$ & $646.3206$ & $0.127884$ & $-1.281366$ & $0.897933$ & $-1.8931\times 10^{-09}$ & $11$ &  $10 $  &  $ 1$  & $3202.7$ & H1 MC 1 & \\ 
& $1092.1387$ & $1091.9671$ & $0.217482$ & $-0.523866$ & $1.302500$ & $-6.4347\times 10^{-11}$ & $11$ &  $10 $  &  $ 1$  & $196.7$ &H1 MC 2/3 & \\ 
& $1136.2217$ & $1136.1460$ & $0.168345$ & $-1.216945$ & $0.935876$ & $-3.4811\times 10^{-09}$ & $11$ &  $ 10$  &  $ 1$  & $165.6$ & H1 MC 2/3& \\ 
& $1142.8210$ & $1142.7200$ & $0.23173$ & $-1.310037$ & $1.114563$ & $-3.5022\times 10^{-09}$ & $11$ &  $ 10$  &  $1 $  & $250.7$ & H1 Cal & \\ 
& $1145.8318$ & $1145.6515$ & $0.231067$ & $1.330065$ & $0.976422$ & $-2.0297\times 10^{-10}$ & $11$ &  $10 $  &  $1 $  & $256.4$ & H1 Cal & \\ 
& $1376.7370$ & $1376.4697$ & $0.271536$ & $0.201677$ & $1.282354$ & $-2.3875\times 10^{-09}$ & $11$ &  $9 $  &  $2 $  & $165.0$ & TM violin & \\ 
& $1388.6402$ & $1388.4070$ & $0.279967$ & $1.176082$ & $0.850794$ & $-2.8907\times 10^{-09}$ & $11$ &  $ 10$  &  $1 $  & $200.0$ & TM violin & \\ 
& $56.9966$ & $56.9966$ & $$ & $-0.935903$ & $0.150238$ & $-1.5029\times 10^{-09}$ & $10$ &  $ 8$  &  $ 2$  & $136.7$ & EM Interference& \\ 
& $329.4918$ & $329.4784$ & $0.021358$ & $-1.307440$ & $4.692056$ & $-5.2405\times 10^{-10}$ & $10$ &  $ 10$  &  $0$  & $1137.8$ & Demod & \\ 
& $392.8322$ & $392.8322$ & $$ & $-1.210088$ & $1.268596$ & $-1.069\times 10^{-09}$ & $10$ &  $ 9$  &  $ 1$  & $150.9$ & H1Cal& \\ 
& $393.4060$ & $393.4057$ & $0.000342$ & $0.632053$ & $1.270922$ & $-1.1043\times 10^{-09}$ & $10$ &  $ 9$  &  $ 1$  & $154.7$ & H1Cal & \\ 
& $646.7224$ & $646.7224$ & $0.002174$ & $-1.446520$ & $0.825633$ & $-1.8813\times 10^{-09}$ & $10$ &  $ 3 $  &  $ 7 $  & $2774.4$ & L1 MC 1 & \\ 
& $648.4132$ & $648.4132$ & $0.024291$ & $1.319729$ & $1.033730$ & $-1.8479\times 10^{-09}$ & $10$ &  $ 3$  &  $ 7$  & $5067.5$ & L1 MC 1 & \\ 
& $658.6353$ & $658.6353$ & $0.000055$ & $-0.470832$ & $4.762475$ & $-1.6992\times 10^{-09}$ & $10$ &  $3 $  &  $7 $  & $261.4$ & EX +15v & \\ 
& $777.9202$ & $777.8377$ & $0.117087$ & $1.511859$ & $4.010213$ & $-5.6101\times 10^{-10}$ & $10$ &  $ 3$  &  $ 7$  & $1951.7$ & EM Interference & \\ 
& $1296.4962$ & $1296.4962$ & $$ & $-0.993190$ & $4.557370$ & $-1.0022\times 10^{-09}$ & $10$ &  $3 $  &  $ 7$  & $247.1$ & L1 MC 1 & \\ 

\hline
\end{tabular}
\caption{
  The post-processing candidates that have 10 or more
  coincidences. The frequency $f_{\rm cand}$ corresponds to the most
  coincident candidate in the band.  The lowest frequency of a candidate
  in the band is labeled by $f_{\rm start}$.  The difference from the
  highest frequency is given by $\Delta f_{\rm cand}$.  The parameters
  $\delta_{\rm cand}$, $\alpha_{\rm cand}$, $\dot f_{\rm cand}$,
  $\mathcal{C}_{\rm cand} = \mathcal{C}^{\rm H1}_{\rm cand} +
  \mathcal{C}^{\rm L1}_{\rm cand}$ and  $\mathcal{S}_{\rm cand}$ are for the most
  significant most coincident candidate within the frequency band, where
  $ \mathcal{C}^{\rm H1}_{\rm cand}$ and  $\mathcal{C}^{\rm L1}_{\rm cand}$
  denote the number of coincidences contributing to $\mathcal{C}_{\rm cand} $
  from detector H1 and L1, respectively.
  The  column ``Information'' lists information about the source.  The following are
  understood sources of narrow-band line noise in the instrument:
  ``Demod'' are the electronics boards that demodulate the
  signal at the antisymmetric port of the interferometer, ``H1 (or L1) MC 1'' is
  a violin mode resonance of the first mode cleaner mirror, ``H1 MC
  2/3'' are violin mode resonances of the second and third mode cleaner
  mirrors,``TM violin'' are harmonics of the test mass violin modes,
   ``EX +15v'' is a fifteen volt power supply at the end
  station of the X arm, ``EM Interference'' is electromagnetic interference, 
  ``H1 Cal'' are side-bands of  calibration lines at $393.1\,\Hz$ and $1144.3\,\Hz$.
   \label{t:Outliers10}}
\end{table*}

\begin{table*}
\begin{tabular}{crrrccrcccccc} \hline
&$f_{\rm cand}$ [Hz] & $f_{\rm start}$ [Hz] & $\Delta f_{\rm cand}$ [Hz] &$\delta_{\rm cand}$ [rad] & 
 $\alpha_{\rm cand}$ [rad] & $\dot f_{\rm cand}$ [$\mathrm{Hz}~\mathrm{s}^{-1}$] & $\mathcal{C}_{\rm cand}$ & $\mathcal{C}^{\rm H1}_{\rm cand}$ & $\mathcal{C}^{\rm L1}_{\rm cand}$ & \; $\mathcal{S}_{\rm cand}$ \;& Information\\
\hline \hline
& $193.9276$ & $193.9261$ & $0.040646$ & $-0.583514$ & $4.723595$ & $-5.6001\times 10^{-09}$ & $13$ &  $7$  &  $6$  & $2263.6$ & Pulsar 8 & \\ 
& $575.1681$ & $575.1562$ & $0.039394$ & $0.285505$ & $3.834511$ & $-5.0913\times 10^{-10}$ & $13$ &  $7$  &  $6$  & $249.3$ & Pulsar 2 & \\ 
& $108.8549$ & $108.8518$ & $0.008506$ & $-0.705729$ & $3.361465$ & $-4.4362\times 10^{-11}$ & $12$ &  $9$  &  $3$  & $2397.5$ & Pulsar 3 & \\ 
& $1402.2838$ & $1402.2488$ & $0.08678$ & $1.025583$ & $2.502838$ & $-3.8482\times 10^{-09}$ & $12$ &  $6$  &  $6$  & $1749.6$ & Pulsar 4 & \\ 
& $545.9987$ & $545.9568$ & $0.141563$ & $-0.398855$ & $5.013332$ & $-4.6693\times 10^{-10}$ & $11$ &  $10 $  &  $1 $  & $794.1$ & H1 MC 2/3 & \\ 
& $56.9966$ & $56.9963$ & $0.000933$ & $-0.935903$ & $0.150238$ & $-1.5029\times 10^{-09}$ & $10$ &  $8$  &  $2$  & $136.7$ & EM Interference & \\ 
& $329.4849$ & $329.4833$ & $0.005843$ & $-0.344739$ & $5.171401$ & $-5.6694\times 10^{-10}$ & $10$ &  $ 10$  &  $0 $  & $1024.0$ &  EM Interference & \\ 
& $329.6040$ & $329.6040$ & $$ & $-0.439100$ & $1.006331$ & $-5.2546\times 10^{-10}$ & $10$ &  $ 9$  &  $ 1$  & $2625.6$ & EM Interference & \\ 
& $329.7434$ & $329.7413$ & $0.032463$ & $-0.338712$ & $5.025108$ & $-5.6923\times 10^{-10}$ & $10$ &  $9 $  &  $ 1$  & $2490.4$ & EM Interference & \\ 
& $567.9984$ & $567.9984$ & $0.051768$ & $-0.353846$ & $5.116972$ & $-1.5532\times 10^{-09}$ & $10$ &  $9 $  &  $ 1$  & $409.3$ & EM Interference & \\ 
& $69.6964$ & $69.6964$ & $$ & $-1.223613$ & $4.232687$ & $-5.4823\times 10^{-10}$ & $9$ &  $9$  &  $0 $  & $130.3$ &EM Interference & \\ 
& $317.4207$ & $317.4207$ & $$ & $1.389330$ & $2.663214$ & $-8.0338\times 10^{-10}$ & $9$ &  $3$  &  $6$  & $157.8$ &  EM Interference & \\ 
& $329.5615$ & $329.5615$ & $$ & $-1.027976$ & $3.822726$ & $-6.3014\times 10^{-10}$ & $9$ &  $7 $  &  $2 $  & $2176.0$ & Demod & \\ 
& $335.7541$ & $335.7141$ & $0.056927$ & $1.395059$ & $3.271989$ & $-6.362\times 10^{-10}$ & $9$ &  $9 $  &  $ 0$  & $259.3$ & EM Interference& \\ 
& $795.4783$ & $795.4783$ & $$ & $0.245291$ & $3.211417$ & $-1.4374\times 10^{-09}$ & $9$ &  $7$  &  $2$  & $110.7$ & EM Interference & \\ 
& $1092.1564$ & $1092.1564$ & $$ & $-0.252089$ & $1.099873$ & $-2.1099\times 10^{-11}$ & $9$ &  $8 $  &  $ 1$  & $147.1$ & H1 MC 2/3 & \\ 
& $1117.3032$ & $1117.3032$ & $$ & $-0.207300$ & $4.051169$ & $-3.3192\times 10^{-09}$ & $9$ &  $4$  &  $5$  & $116.1$ & Unknown& \\ 
& $1145.6678$ & $1145.6678$ & $$ & $-0.247554$ & $5.067301$ & $-3.1679\times 10^{-09}$ & $9$ &  $ 6$  &  $ 3$  & $168.7$ & H1 Cal & \\

\hline
\end{tabular}
\caption{
  Post-processing candidates that have 9 or more coincidences and
  that are not excluded by the veto. The frequency $f_{\rm cand}$ corresponds to the most
  coincident candidate in the band.  The lowest frequency of a candidate
  in the band is labeled by $f_{\rm start}$.  The difference from the
  highest frequency is given by $\Delta f_{\rm cand}$.  The parameters
  $\delta_{\rm cand}$, $\alpha_{\rm cand}$, $\dot f_{\rm cand}$,
  $\mathcal{C}_{\rm cand} = \mathcal{C}^{\rm H1}_{\rm cand} +
  \mathcal{C}^{\rm L1}_{\rm cand}$ and  $\mathcal{S}_{\rm cand}$ are for the most
  significant most coincident candidate within the frequency band, where
  $ \mathcal{C}^{\rm H1}_{\rm cand}$ and  $\mathcal{C}^{\rm L1}_{\rm cand}$
  denote the number of coincidences contributing to $\mathcal{C}_{\rm cand} $
  from detectors H1 and L1, respectively.
  The  column ``Information'' lists information about the source.  The following are
  understood sources of narrow-band line noise in the instrument:
  ``Demod'' are the electronics boards which demodulate the
  signal at the antisymmetric port of the interferometer, ``H1 MC
  2/3'' are violin mode resonances of the second and third mode cleaner
  mirrors,``EM Interference'' is electromagnetic interference, 
  ``H1 Cal'' are side-bands of a $1144.3\,\rm Hz$  calibration line. 
  For the single candidate labeled ``Unknown" in the last column no instrumental
  source could be confidently identified, however the $9$ coincidences are far below the 
  confident-detection threshold.
  \label{t:Outliers9}}
\end{table*}

 \begin{figure}
	\includegraphics[scale=0.456,angle=0]{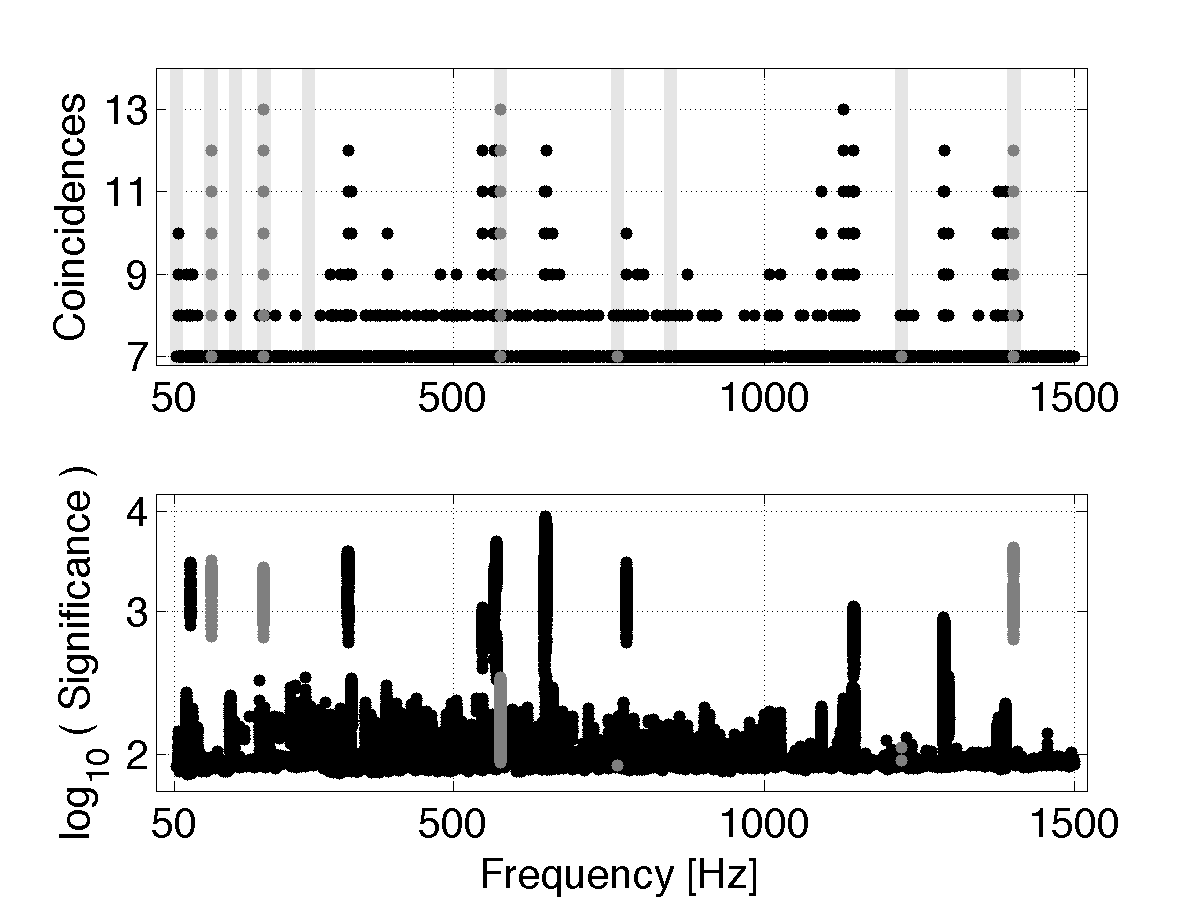}
        \caption{
          Numbers of coincidences greater than $7$ (top) and
          the significance (bottom) of all candidates found in the
          Einstein@Home post-processing, shown as functions of
          frequency. The light-gray shaded rectangular regions
          highlight the S4 hardware injections, listed in
          Table~\ref{t:HWInjectionsParams}. The data points colored in
          dark-gray show the candidates resulting from the
          hardware-injected CW signals.
          \label{f:specplot}
        }
 \end{figure}
      
 \begin{figure}
	\subfigure{\includegraphics[scale=0.425,angle=0]{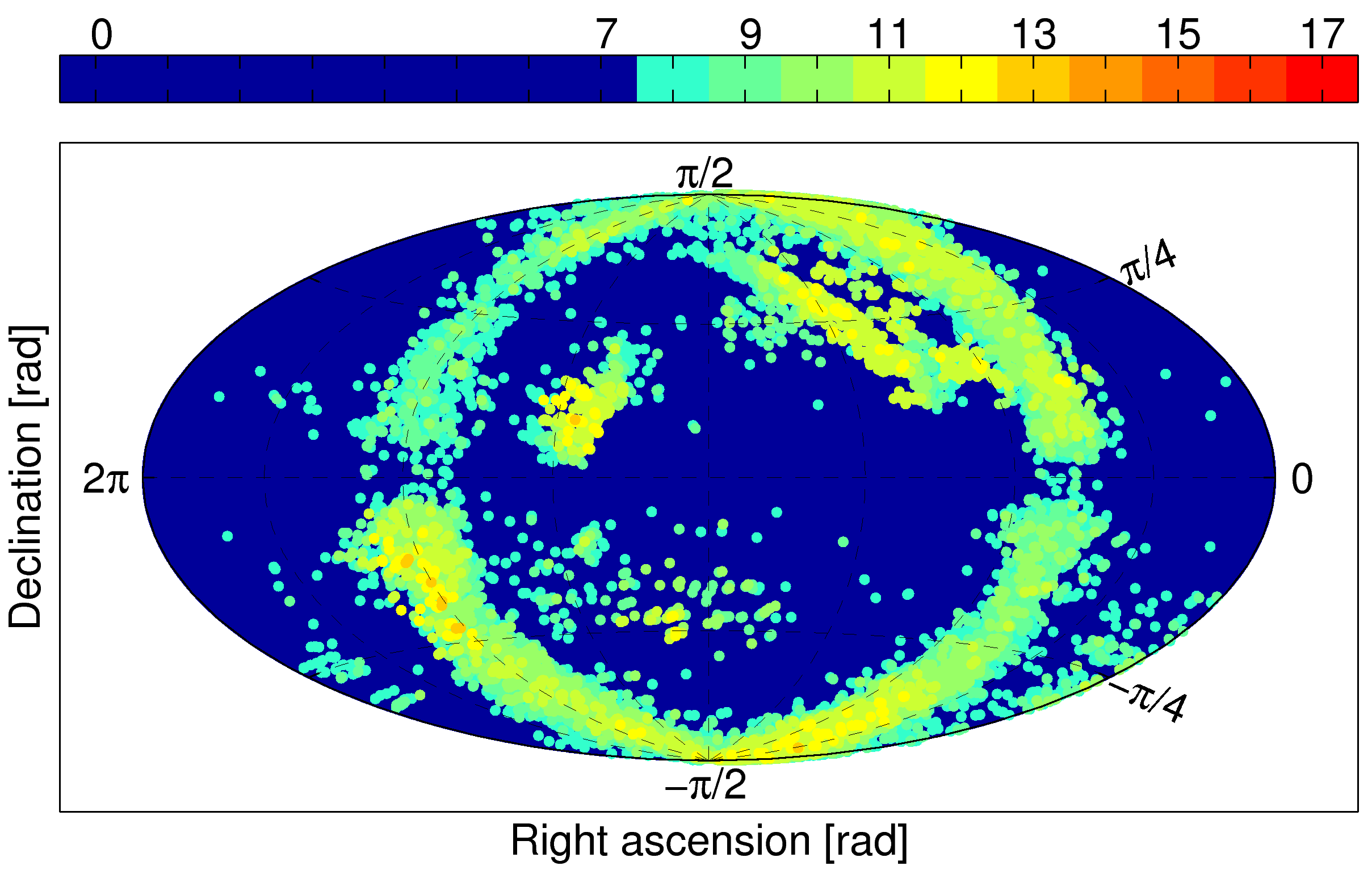}}\\
	\subfigure{\includegraphics[scale=0.425,angle=0]{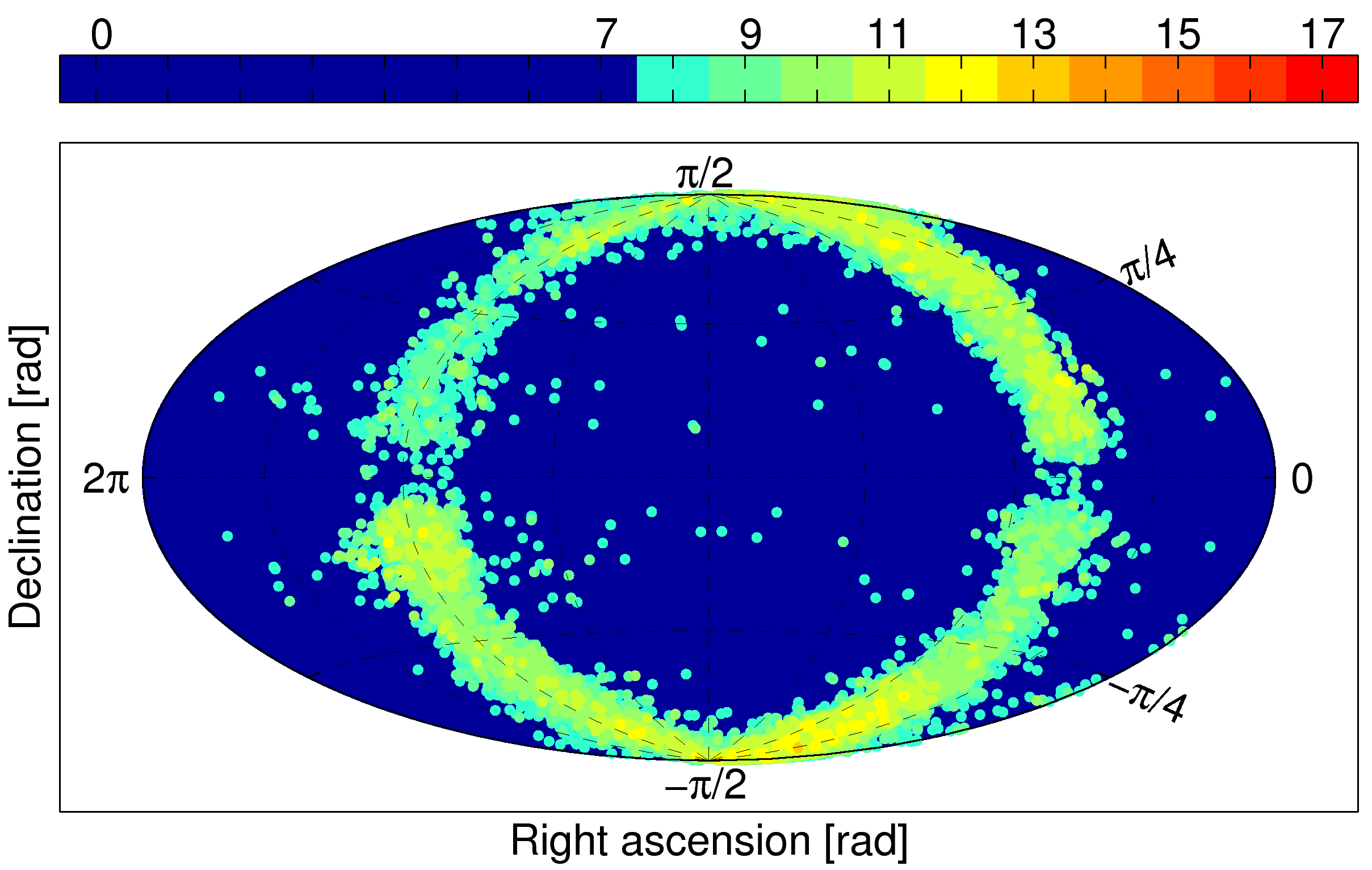}}
        \caption{ All candidates obtained from the post-processing
          that have more than 7 coincidences, shown in
          Hammer-Aitoff projections of the sky.  The color-bar
          indicates the number of coincidences of a particular
          candidate (cell). The upper plot includes the S4
          hardware-injected pulsars.  In the lower plot, bands of
          $2\times 10^{-4} \;f$ width to either side of the hardware
          injections' frequencies $f$ have been removed. 
          \label{f:AllResults} }
\end{figure}

\begin{figure}
  \subfigure{\includegraphics[scale=0.425,angle=0]{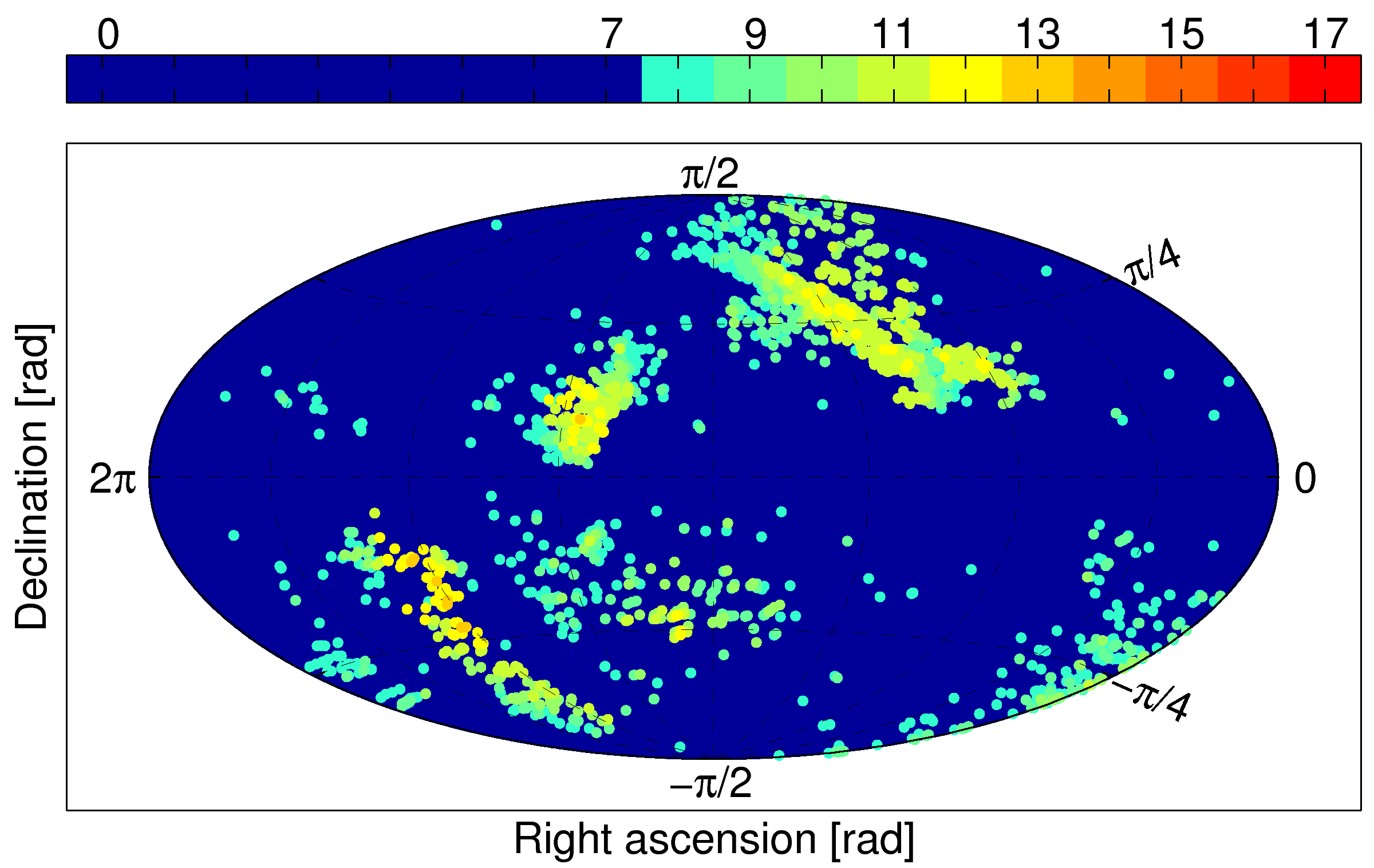}}\\
  \subfigure{\includegraphics[scale=0.425,angle=0]{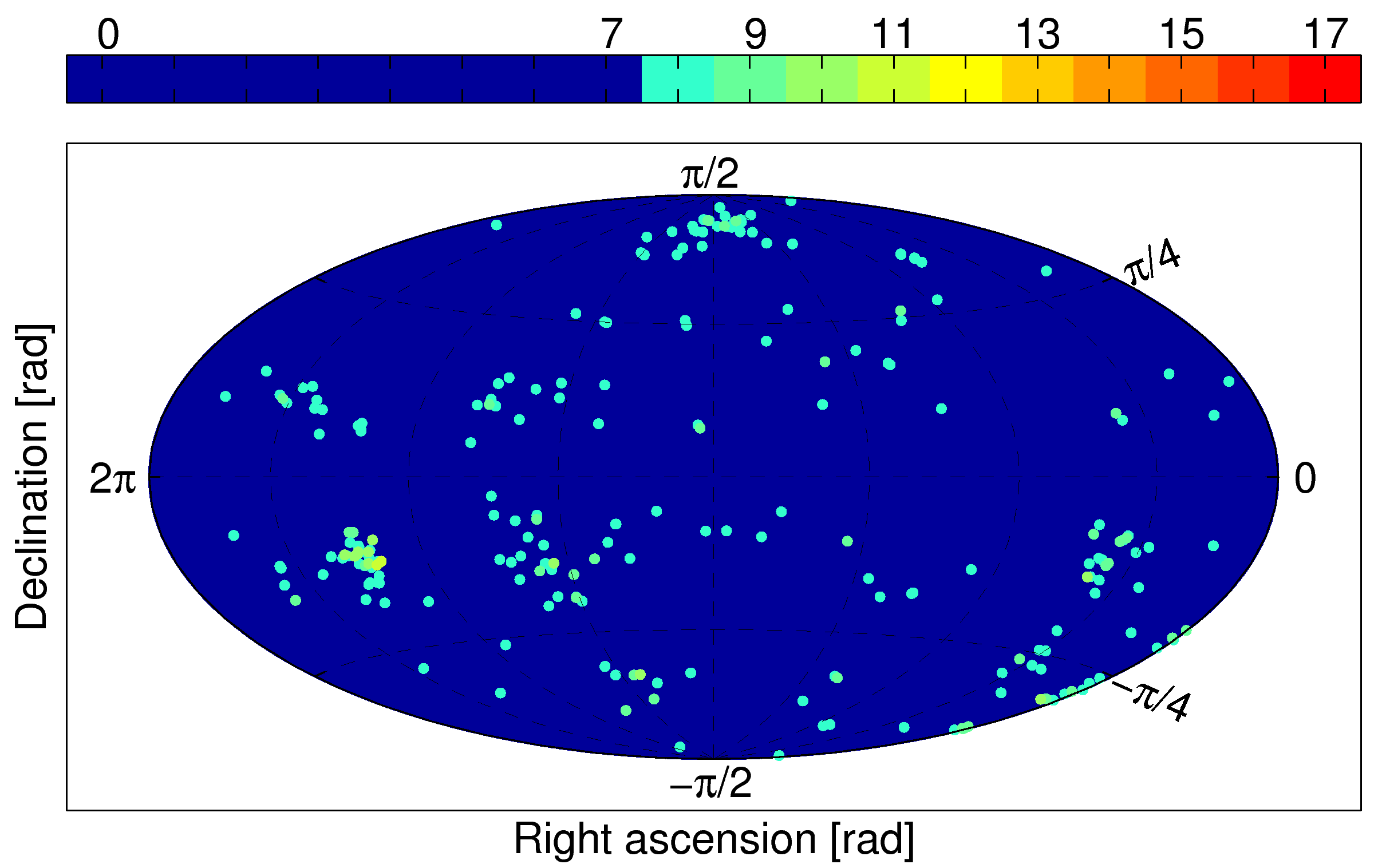}}
  \caption{ Candidates not eliminated by the veto. This shows
    Hammer-Aitoff sky projections of all candidates obtained from
    post-processing that had more than 7 coincidences and that
    passed the veto.  The upper plot includes the S4 hardware
    injections.  The lower plot removes bands of $2\times 10^{-4}\;f$
    width to either side of the hardware injections' frequencies $f$.
    In comparison to Figure~\ref{f:AllResults}, after excluding the
    hardware injections, the veto rejects $99.5\%$ of all
    candidates. 
    \label{f:AllResultsAfterVeto} }
\end{figure}

\begin{figure}
  \includegraphics[scale=0.45,angle=0]{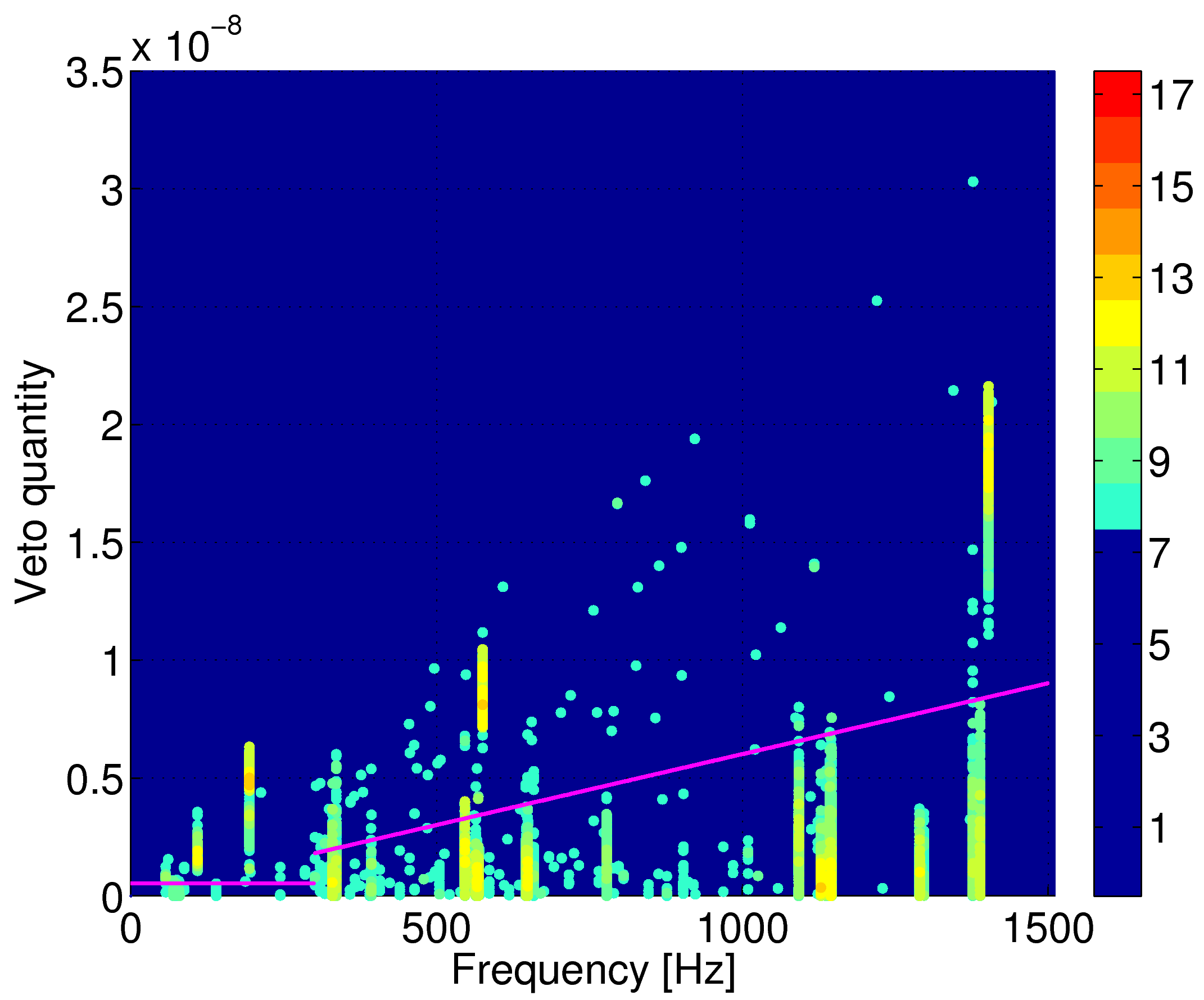}
  \caption{ Conclusion diagram of candidates discriminated by the veto 
    method.  All candidate cells obtained from post-processing that have 
    more than 7 coincidences are shown, where the color-bar indicates 
    the number of coincidences of a particular cell. The vertical axis represents
    the veto quantity on the left-hand side of~(\ref{e:VetoCond}), 
    as a function of frequency. Candidates located below the 
    magenta line are eliminated by the veto. The four accumulations of highly
    coincident cells above the magenta line are the hardware injected
    pulsars, which are not eliminated by the
    veto. \label{f:VetoedCands}}
\end{figure}

This section presents the results of the Einstein@Home S4 CW search.
Figures~\ref{f:specplot}~and~\ref{f:AllResults} give a summary of all
post-processing results, from $50$ to $1500$~Hz.  In
Figure~\ref{f:specplot} the coincidences and significance of all
candidates that have $7$ or more coincidences are shown as functions
of frequency.  Figure~\ref{f:AllResults} presents the same information
as given in Figure~\ref{f:specplot}, but projected on the sky, and showing all cells that have
more than 7 candidate events.

In Figure~\ref{f:AllResults}  the number of coincidences is maximized over
the entire sky and full spin-down range. The color indicates the
numbers of coincidences, where the same color-scale has been used in
each plot. The maximum possible number of coincidences ranges from a 
minimum of 0 to a maximum of 17 (the number of data segments analyzed). 
The meaning of 0~coincidences is that there is no candidate event found, 
1~coincidence means a single candidate events is found 
(which is always coincident with itself).

Four illustrative examples of different types of typical post-processing 
results in particular $10\,\rm Hz$ bands are shown in Appendix~\ref{sec:example-results} 
by Figures~\ref{f:results340},~\ref{f:results570},~\ref{f:results110},
and~\ref{f:results640}.  

Table~\ref{t:Outliers10} shows all candidates (cells) which have $10$
or more coincidences. In cases where a set of candidates is clustered
together at slightly different frequencies, Table~\ref{t:Outliers10}
lists the bandwidth in frequency covered by these candidates and shows
the parameters of the most coincident candidate.  If candidates within
these narrow frequency-bands have the same number of coincidences,
then the candidate with the largest significance is shown.

Table~\ref{t:Outliers9} shows the same information {\it after} the
veto method described in Section~\ref{sec:VetoMethod} has been
applied, for candidates with $9$ or more coincidences.  There are no
candidates that exceed the predefined detection threshold of
appearing in 12 or more data segments.  (Note that this would be a
threshold for initiating a more extensive investigation of the
candidate event, not a threshold for announcing a discovery!)

Figure~\ref{f:AllResultsAfterVeto} shows all candidates from the
post-processing results that have not been discriminated by the veto
introduced in Section~\ref{sec:VetoMethod}.
Figure~\ref{f:VetoedCands} illustrates the fraction of candidates
that has been excluded by the veto. Removing fractional bands of $2\times
10^{-4}\;f$ around the frequencies $f$ of the S4 hardware injections, 
the veto discriminates $99.5\%$ of all
candidates that have more than $7$ coincidences.

\section{Conclusion\label{sec:Conclusion}}

These are the first published results from the Einstein@Home project,
which was launched in February 2005.  While no credible CW sources
were found in this search of LIGO S4 data, the results clearly
establish that this type of distributed computing project can carry
out a credible and sensitive search for such signals.

In retrospect, it probably would have been a good idea to employ
identical grids on the four-dimensional parameter space for all 17
data segments.  This would have required more CPU time on the part of
participants, but would have greatly simplified and sped up the
development of the post-processing pipeline and would also have greatly
simplified the interpretation of the results.

A similar search (also with a 30-hour time baseline) has already been
completed using 660 hours of data from the beginning of the S5 science
run.  The post-processing of that data set is currently underway,
using methods identical to those employed here.

Future Einstein@Home searches overcome some of the sensitivity
limitations discussed at the end of Section~\ref{sec:ExpectSen} by
doing the incoherent step (called ``post-processing'' in this paper)
on the host machines.  This allows the use of the optimal threshold of
$2\F \sim 5$, so those searches are expected to be the most sensitive blind CW
searches that will be possible using LIGO data.  Results from those
searches should become available within the next one to two years, and
are expected to offer more than one order of magnitude improvement in
strain sensitivity compared with the work presented here.

In the longer term, further increases in sensitivity will probably
result mostly from improvements in the detectors rather than from
improvements in the data analysis methods.  In 2009 LIGO is expected
to begin its S6 run with an ``enhanced'' detector configuration that
should improve on S5 sensitivity by at least a factor of two.  By
2014, an advanced LIGO detector configuration should give at least
another factor of five improvement.  By combining these data sets with
those from LIGO's international partner projects Virgo and GEO, there
is real hope that the first direct CW detection can be made using
methods like the ones described here.

\section{Acknowledgments\label{sec:Ack}}
\input acknowledgements.tex


This document has been assigned LIGO Laboratory document number
LIGO-P080021-01-Z.

\begin{appendix}
\section{Fraction of parameter space excluded by the veto method 
\label{sec:fractionveto}}

The fractional volume of the region in parameter space excluded by the
veto method presented in~\cite{PletschGC} and used in Section~\ref{sec:VetoMethod} 
may be easily calculated. Since the time $\Delta T$
is small compared to one year, one may use the following approximation
\begin{equation}
(\vec{{\boldsymbol{ \omega}}} \times \vec{v}_{\rm av}) \cdot
\hat{\vec{n}} \approx | \vec{\boldsymbol{\omega}}| |\vec{v}_{\rm av}|
\cos \theta \;,
\end{equation}
 where $\theta \in [0, \pi]$ is the angle between the
SSB-to-Earth vector and the source sky position $\hat{\vec{n}}$.  The
veto condition~(\ref{e:VetoCond}) may then be rewritten as
\begin{equation}
  | \dot f + \gamma f \, \cos \theta| < \epsilon \;,
\end{equation}
where $\gamma$ is defined as $\gamma = | \vec{\boldsymbol{\omega}}| |\vec{v}_{\rm av}|/c$.
For fixed values of $f$ and $\dot f$ the situation is depicted in
Table~\ref{t:excluded}.  Depending upon the values of 
$(\pm \epsilon -\dot f)/\gamma f$, a part of the sky might be excluded 
by the veto.  As shown in the Table, there are six possible cases, 
determined by the values of 
\begin{eqnarray}
  \cos \theta_- &=& \frac{\epsilon - \dot f}{\gamma f}  \quad \rm and \\
  \cos \theta_+ &=& \frac{-\epsilon - \dot f}{\gamma f} \;. 
\end{eqnarray}
For example in the case (labeled case 4 in Table~\ref{t:excluded}) where 
both $\cos \theta_-$ and $\cos \theta_+$ lie in the
range $[-1,1]$ then the excluded region of the sky is an annulus
defined by $0 \le \theta_- < \theta < \theta_+ \le \pi$, and the excluded
solid angle is 
\begin{eqnarray}
\Omega_{\rm excluded} &=& \int_0^{2 \pi} d\phi
\int_{\theta_-}^{\theta_+} d\theta \sin \theta \\ 
&=& 2\pi(\cos \theta_- - \cos \theta_+) \;.
\end{eqnarray}
 The fraction of sky excluded in this case is then
 \begin{eqnarray}
 \frac{\Omega_{\rm excluded}}{4\pi} &=& (\cos \theta_- - \cos \theta_+)/2 \\
 &=& \frac{\epsilon}{\gamma f} \\
 &=& \epsilon\, {c \over | \vec{\boldsymbol{\omega}}|\,
  |\vec{v}_{\rm av}|} \, {1 \over f}\;.
\end{eqnarray}
In the other cases listed in Table~\ref{t:excluded} the excluded 
region of the sky might be a cap about $\theta=0$ or about 
$\theta=\pi$ or the null set, or the entire sky.

\begin{table*}
\begin{center}
\begin{tabular}{cccc}
\hline
      & Range of                                            & Range of                                          & Excluded sky fraction \\
Case  &  $\quad \quad \quad$ $\cos \theta_+ = {-\epsilon - \dot f \over \gamma f} $   $\quad \quad \quad$ &  $\quad \quad \quad$  $\cos \theta_- = {\epsilon - \dot f \over \gamma f} $ $\quad \quad \quad$  & $\Omega_{\rm excluded}/4\pi $ \\
\hline\hline
1     & $(-\infty,-1)$                                      & $(-\infty,-1)$                                    & 0 \\
2     & $(-\infty,-1)$                                      & $[-1,1]$                                          & $(\cos\theta_-+1)/2 = ( 1+ {\epsilon - \dot f \over \gamma f})/2$ \\
3     & $(-\infty,-1)$                                      & $(1,\infty)$                                      & 1 \\
4     & $[-1,1]$                                            & $[-1,1]$                                          & $ (\cos \theta_- - \cos \theta_+)/2 = \epsilon/\gamma f$  \\
5     & $[-1,1]$                                            & $(1,\infty)$                                      & $(1-\cos\theta_+)/2 = ( 1+ {\epsilon + \dot f \over \gamma f})/2  $ \\
6     & $(1,\infty)$                                        & $(1,\infty)$                                      & 0 \\
\hline
\end{tabular}
\caption{\label{t:excluded} For given values of frequency $f$ and
  spin-down $\dot f$, this shows the fractional volume of the sky
  excluded by the veto~(\ref{e:VetoCond}). There are six possible
  cases, depending upon the values of $\cos \theta_- $ and $\cos
  \theta_+$. (There are six rather than nine cases because
  $\cos\theta_+$ is never greater than $\cos \theta_-$.) For the
  ranges of $f$ and $\dot f$ considered in this work, case 4 applies
  above 300~Hz.  Between 50~Hz and 300~Hz, because of the wider range
  of $\dot f$ considered, the three cases 4, 5 and 6 are found.
  Values of $\cos \theta$ outside the range $[-1,1]$ correspond to
  imaginary (unphysical) values of $\theta$.  In such cases the upper
  and/or lower limits of integration are replaced by $\theta=\pi$ or
  $\theta=0$ respectively, as can be seen from the final column of
  this table.}
\end{center}
\end{table*}

In this search, the fraction of the sky excluded for frequencies $f \in [300,
1500)\,{\rm Hz}$ has been fixed at the constant fraction $\Omega_{\rm
  excluded} / 4 \pi=30\%$.  This is equivalent to choosing $\epsilon$
to be a linear function of frequency
\begin{equation}
  \epsilon =  0.3  \frac{|\vec{\boldsymbol{\omega}}||\vec{v}_{\rm av}|}{c} \, f \;.
\end{equation} 

In this search, the fraction of the sky excluded for frequencies $f \in
[50, 300)\,{\rm Hz}$ has been chosen to depend upon the value of $\dot f$.  
The instruments allow (e.g. compare with Figure~\ref {f:VetoedCands}) the use 
of a frequency-independent value $\epsilon=5.4 \times 10^{-10}\,\Hz/\rm s$ 
which corresponds to $N_{\rm cell} = 1.5$. Within the region of parameter space 
which is searched 
($-f/\tau < \dot f < 0.1\,f/\tau$ for $\tau=1000$~years) 
cases 4, 5, or 6 from Table~\ref{t:excluded} 
occur depending of the spin-down value $\dot f$. If
\begin{equation}
  \dot f >  \epsilon  - {|\vec{\boldsymbol{\omega}}| |\vec{v}_{\rm av}| f \over c}  \;,
\end{equation}
then case 4 of Table~\ref{t:excluded} applies, and 
the fraction of sky excluded is given by
\begin{equation}
  {{\Omega}_{\rm excluded} \over 4 \pi} 
  = \epsilon\, {c \over | \vec{\boldsymbol{\omega}}|\,
  |\vec{v}_{\rm av}|} \, {1 \over f}\;.
  \label{e:ExcludedFraction}
\end{equation}
This fraction ranges from $52\%$ at $50\,{\rm Hz}$ to $8.7\%$ at
$300\,{\rm Hz}$.
If $\dot f$ is in the interval
\begin{equation}
  \dot f \in \biggl[ - \epsilon - {|\vec{\boldsymbol{\omega}}||\vec{v}_{\rm av}| f \over c},
 \epsilon - {|\vec{\boldsymbol{\omega}}||\vec{v}_{\rm av}| f \over c} \biggr],
\end{equation}
then case 5 of Table~\ref{t:excluded} applies, and 
the fraction of sky excluded is given by
\begin{equation}
  {{\Omega}_{\rm excluded} \over 4 \pi} = {1 \over 2}
\biggl(
{1 + {\epsilon +   \dot f \over f  |\vec{\boldsymbol{\omega}}| |\vec{v}_{\rm av}|/c }} \biggr).
\end{equation}
Finally, if 
\begin{equation}
\dot f <  - \epsilon - {|\vec{\boldsymbol{\omega}}||\vec{v}_{\rm av}| f \over c}  \;,
\end{equation}
then case 6 applies and 
none of the sky is excluded by the veto: ${\Omega}_{\rm  excluded}~=~0$.  
Below $300\,{\rm Hz}$, one can compute a uniform
average of the excluded sky fraction over the spin-down range
considered in this analysis.  As shown in Figure~\ref{f:skyfraction} 
this gives an excluded sky fraction of
$36\%$ at $50\,{\rm Hz}$ and $6\%$ just below $300\,{\rm Hz}$.\\\\

\section{Illustrative examples of typical post-processing results
\label{sec:example-results}}

\begin{figure}[ht]
  \subfigure{\includegraphics[scale=0.44,angle=0]{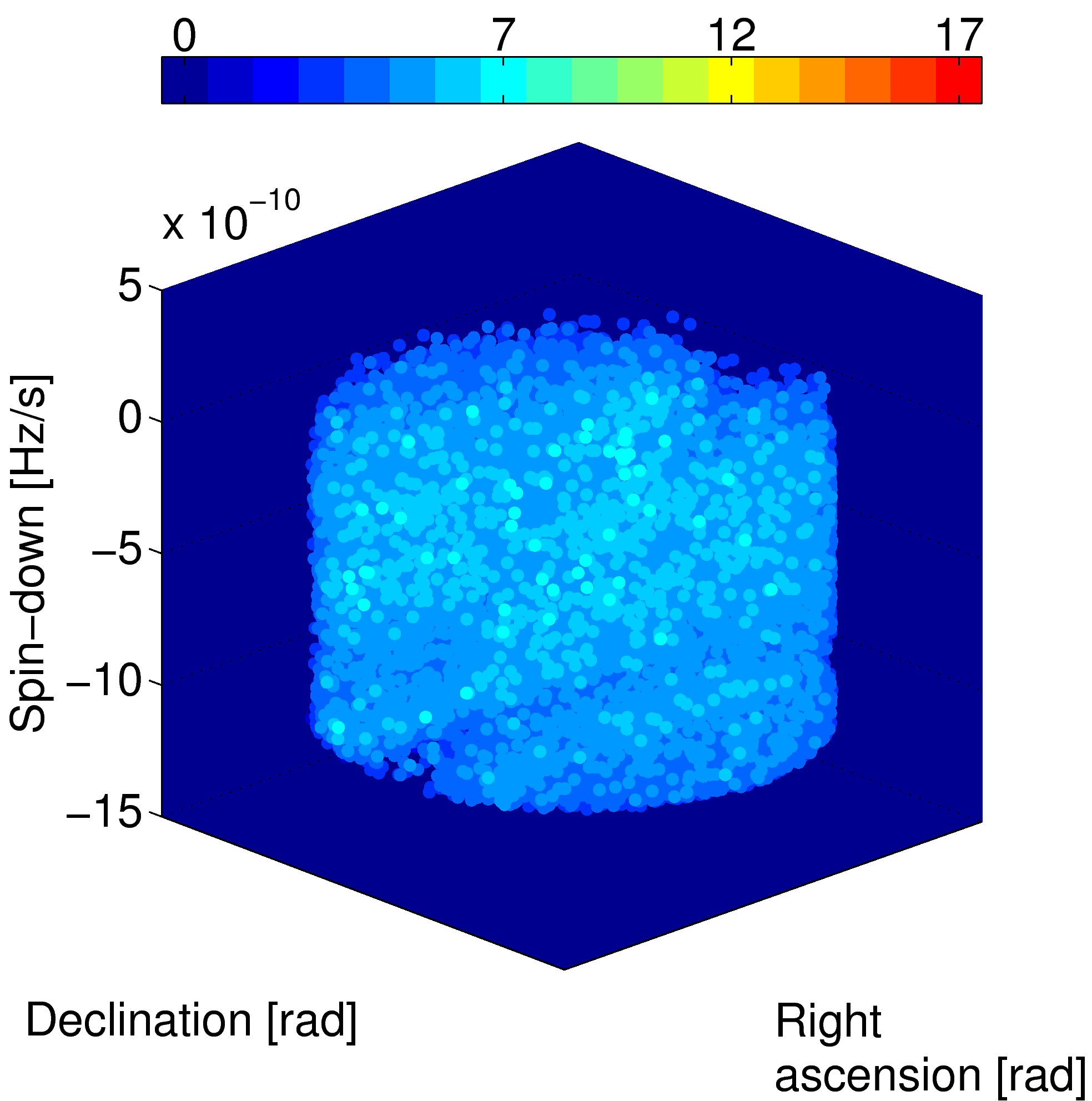}}\\
  \subfigure{\includegraphics[scale=0.35,angle=0]{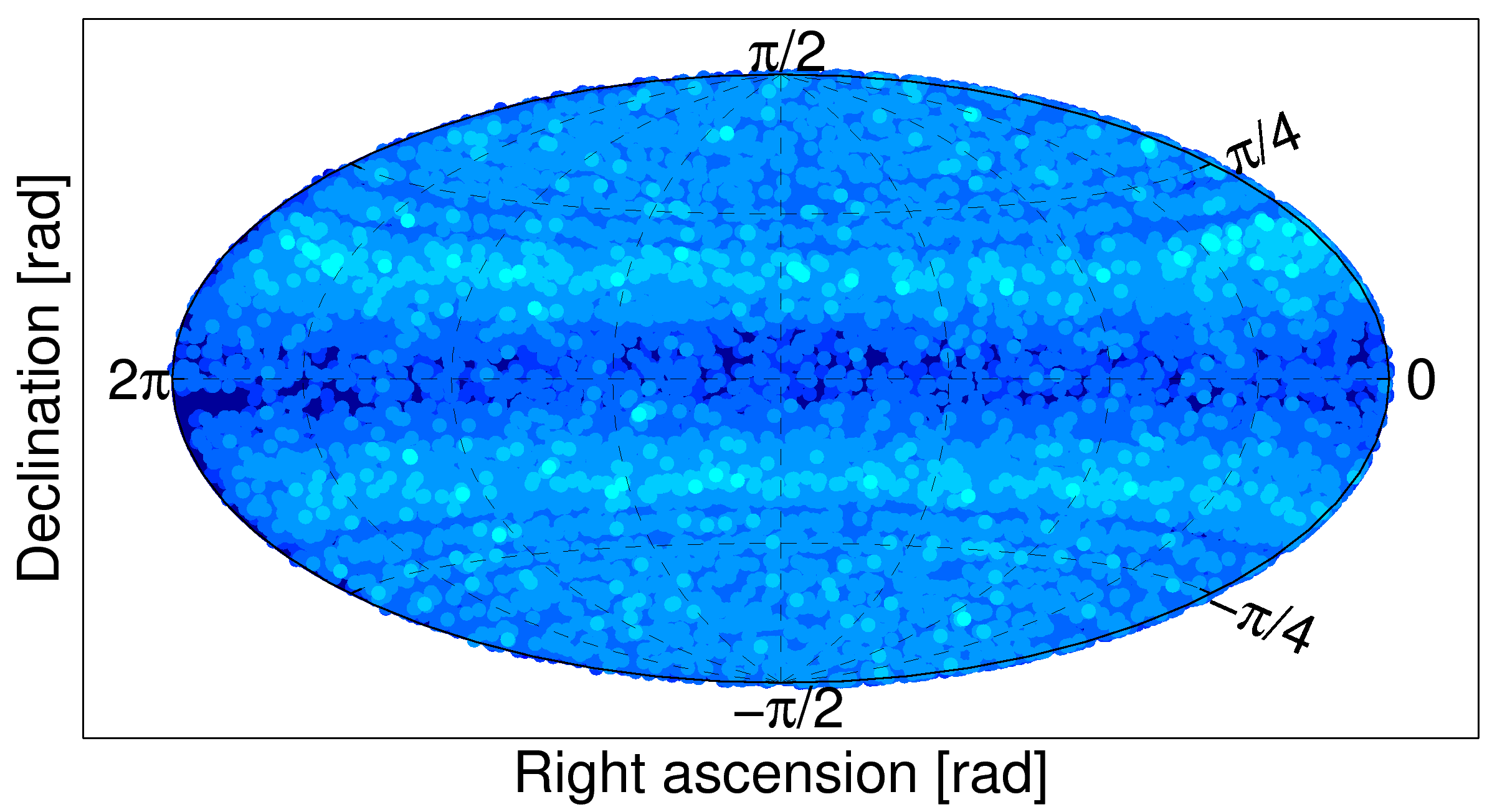}}\\
  \subfigure{\includegraphics[scale=0.44,angle=0]{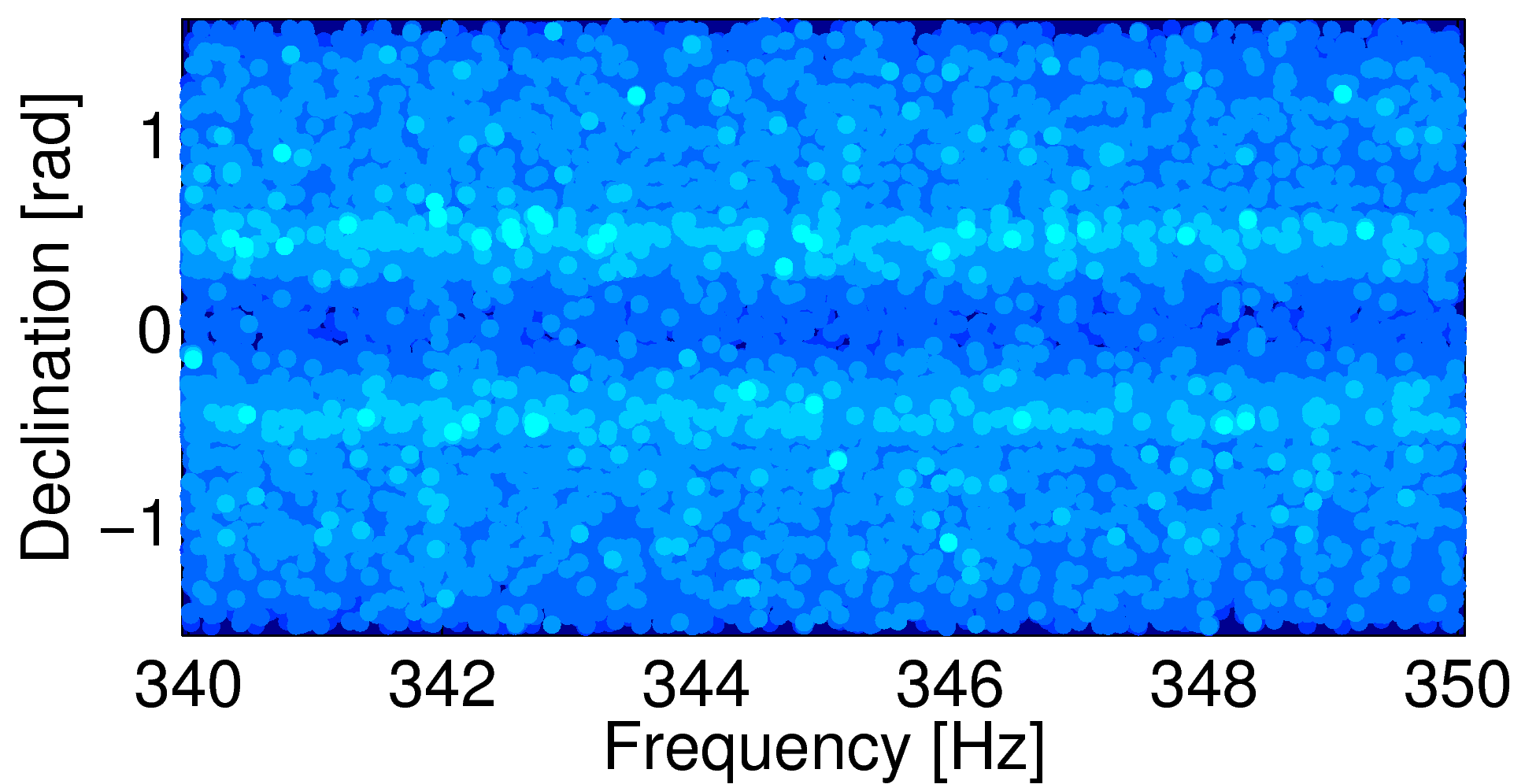}}\\
  \subfigure{\includegraphics[scale=0.44,angle=0]{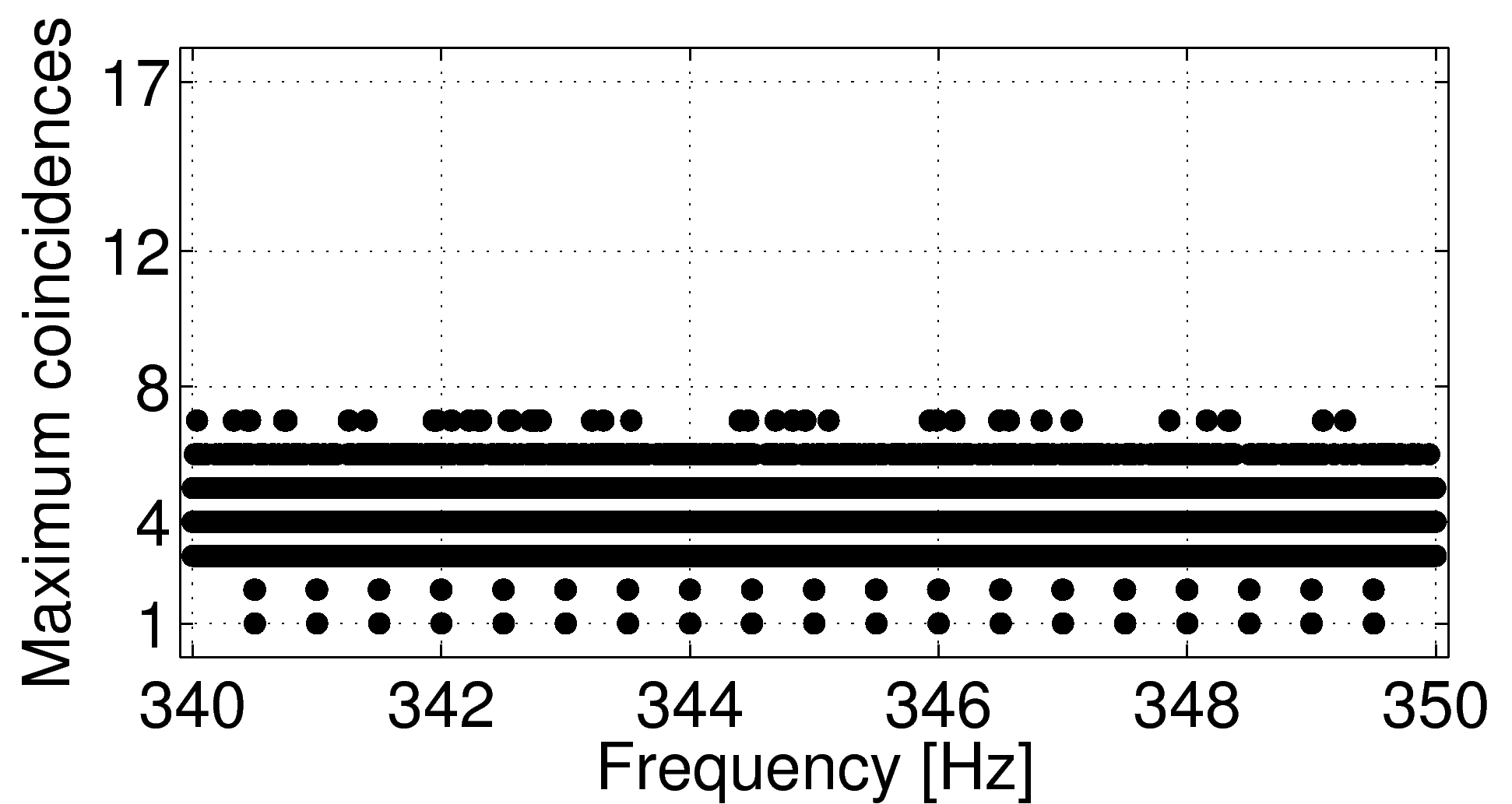}}\\
  \caption{Einstein@Home S4 Post-Processing results for the frequency
    band 340.0-350.0~Hz, which is pure Gaussian noise for L1 and
    mostly Gaussian noise for H1.  This is because in this band the
    line-cleaning process has replaced all the L1 data and most of the
    H1 data with computer-generated random numbers (see
    Table~\ref{t:lines}).  From top to bottom the different plots show
    the numbers of coincidences in a 3D map of sky and spin-down, in a
    2D Hammer-Aitoff projection of the sky, in a 2D plot of
    declination over frequency, and in a histogram as a function of
    frequency. \label{f:results340}}
\end{figure}

\begin{figure}[ht]
	\subfigure{\includegraphics[scale=0.44,angle=0]{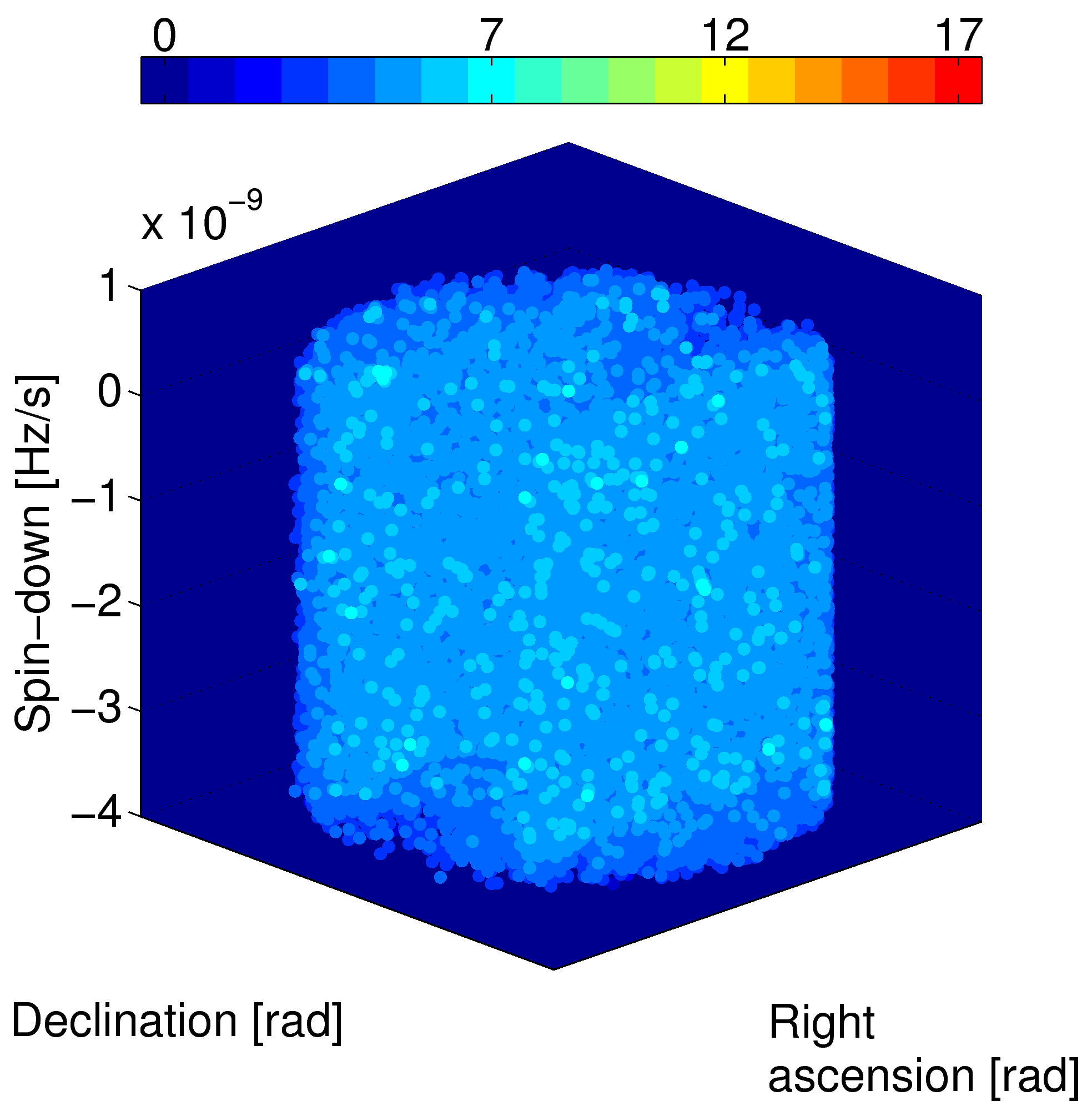}}\\
	\subfigure{\includegraphics[scale=0.35,angle=0]{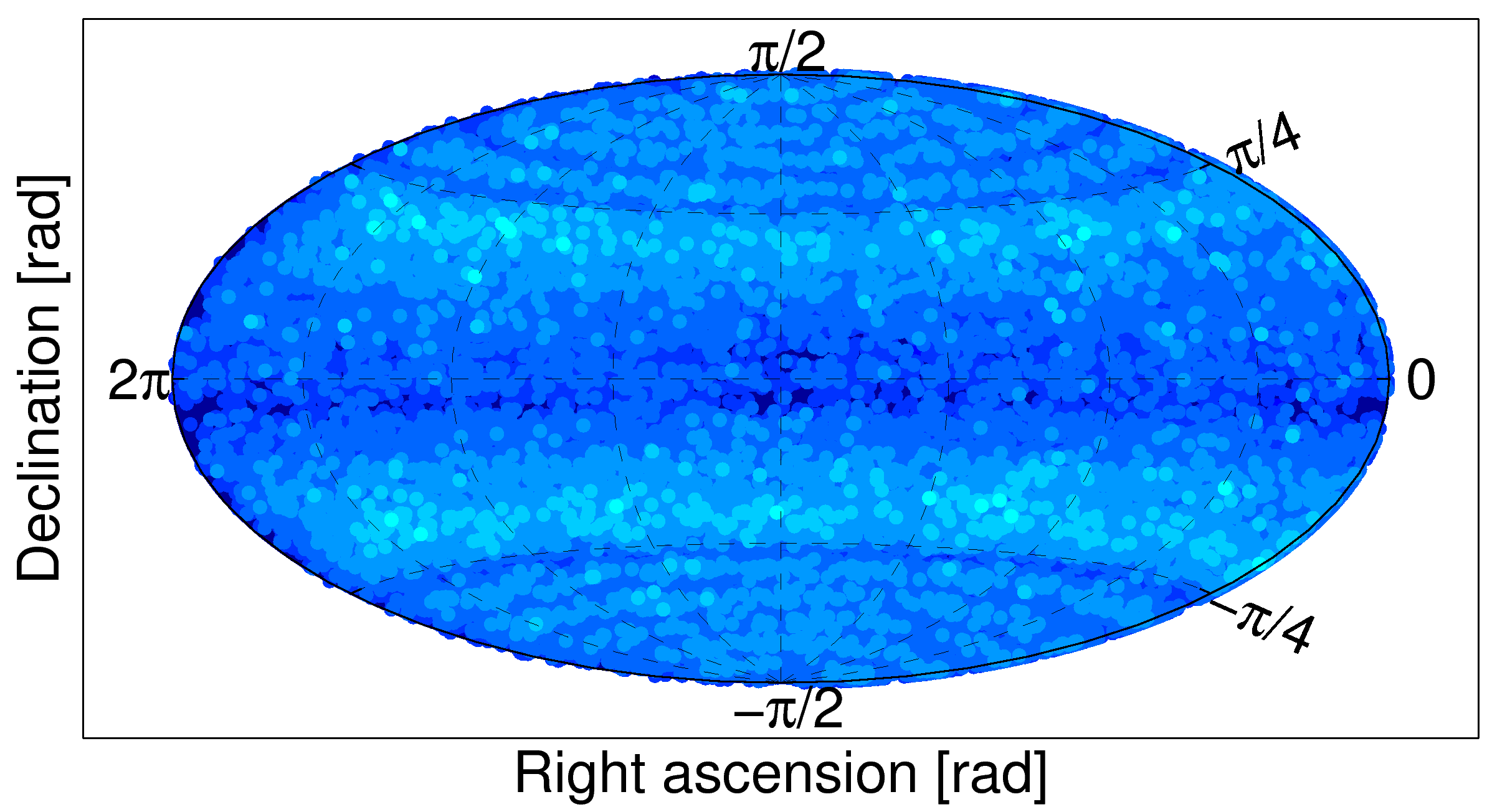}}\\
	\subfigure{\includegraphics[scale=0.44,angle=0]{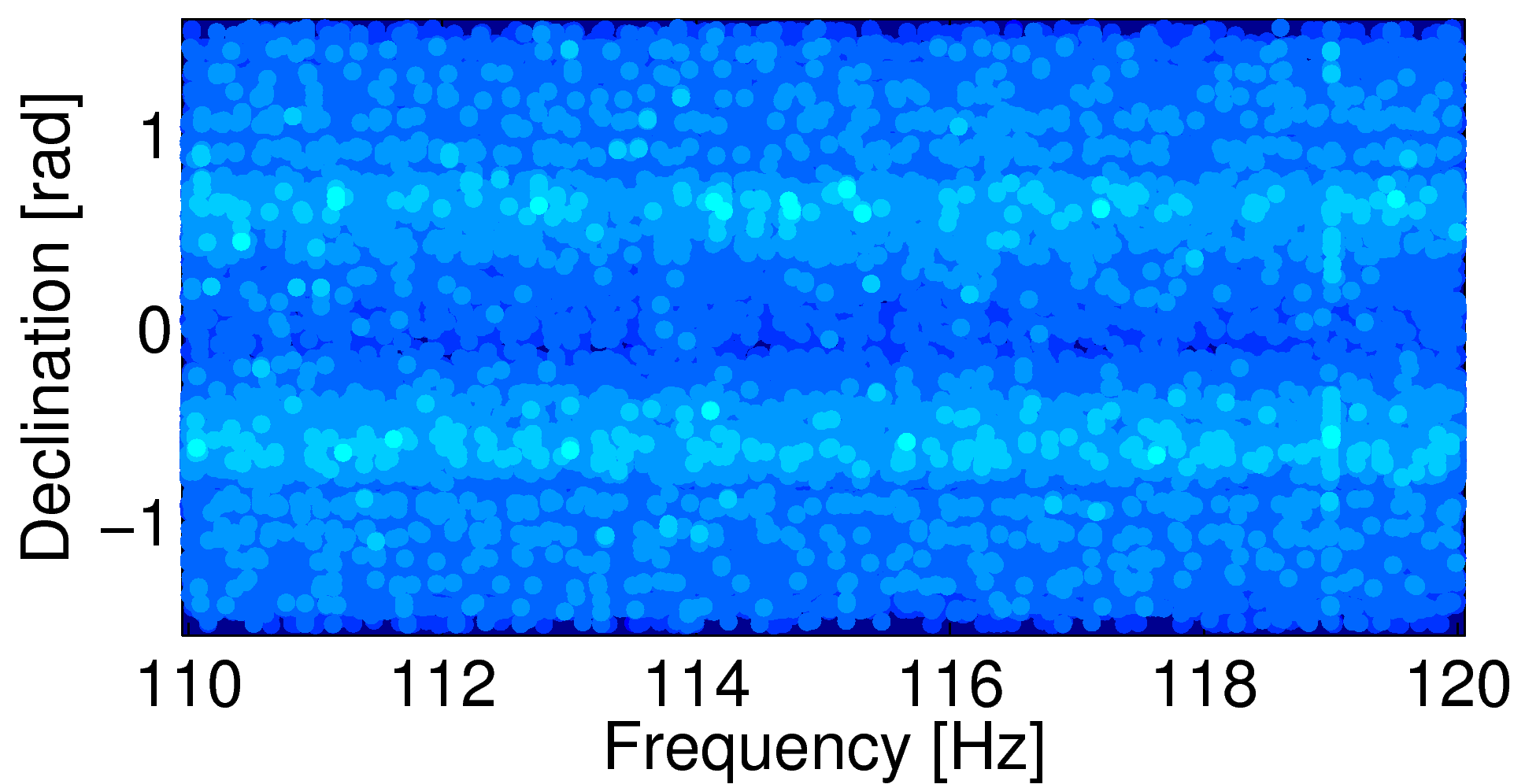}}\\
	\subfigure{\includegraphics[scale=0.44,angle=0]{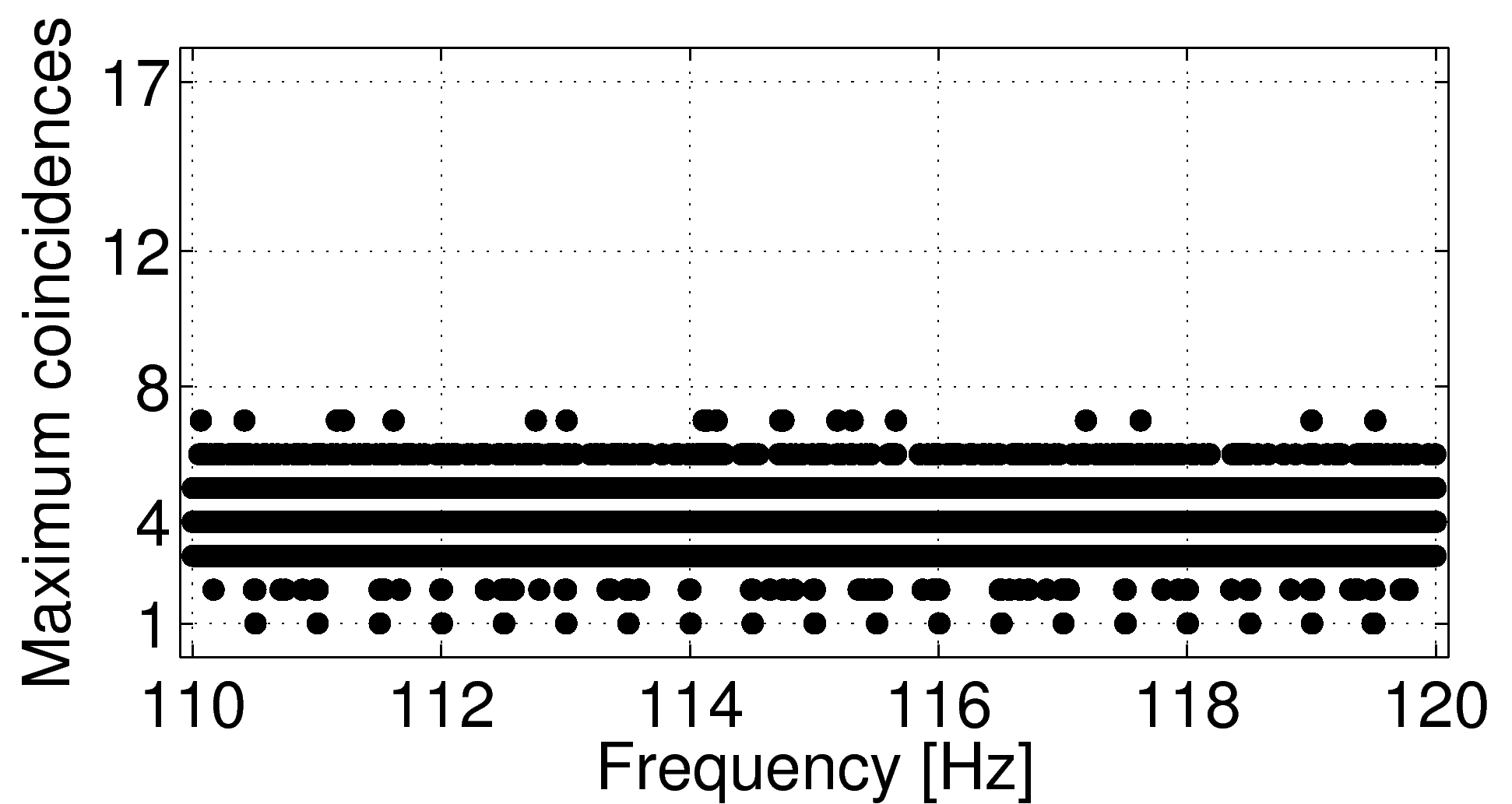}}\\
	\caption{Einstein@Home S4 Post-Processing results for a
          ``quiet'' frequency band of real instrumental data from
          110.0-120.0~Hz.  From top to bottom the different plots show
          the numbers of coincidences in a 3D map of sky and
          spin-down, in a 2D Hammer-Aitoff projection of the sky, in a
          2D plot of declination over frequency, and in a histogram as
          a function of frequency.\label{f:results110}}
\end{figure}

\begin{figure}[ht]
  \subfigure{\includegraphics[scale=0.44,angle=0]{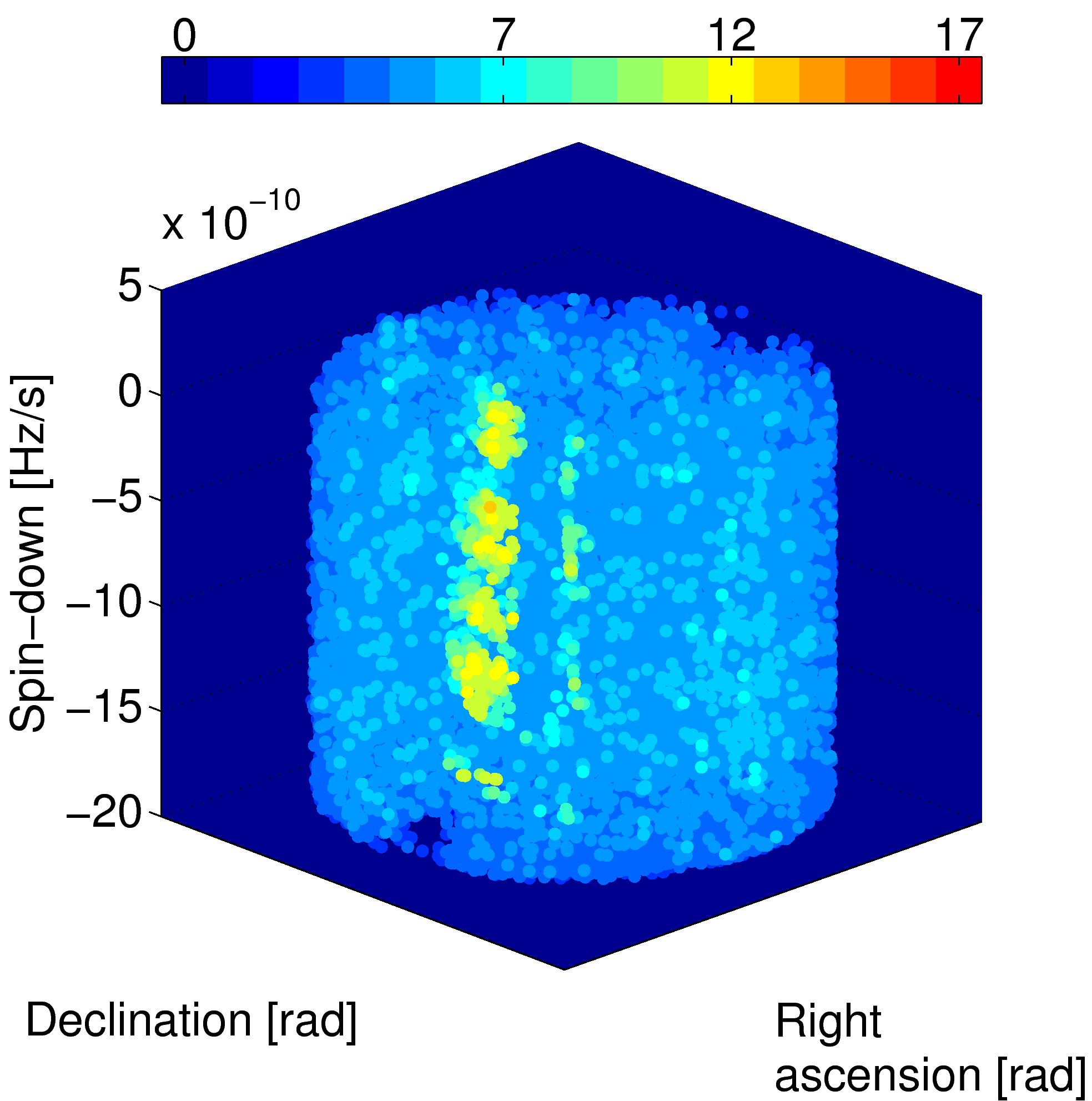}}\\
  \subfigure{\includegraphics[scale=0.35,angle=0]{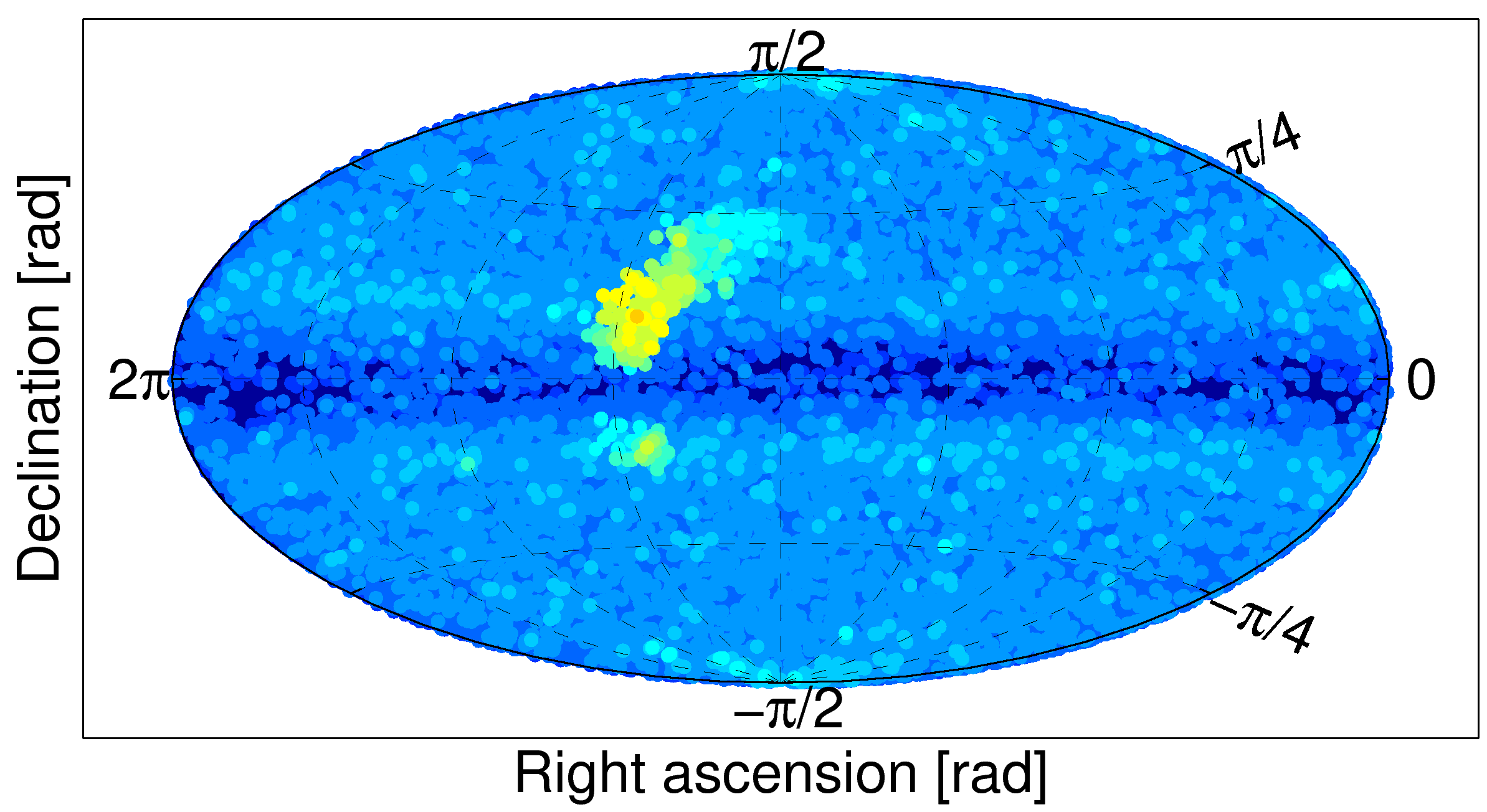}}\\
  \subfigure{\includegraphics[scale=0.44,angle=0]{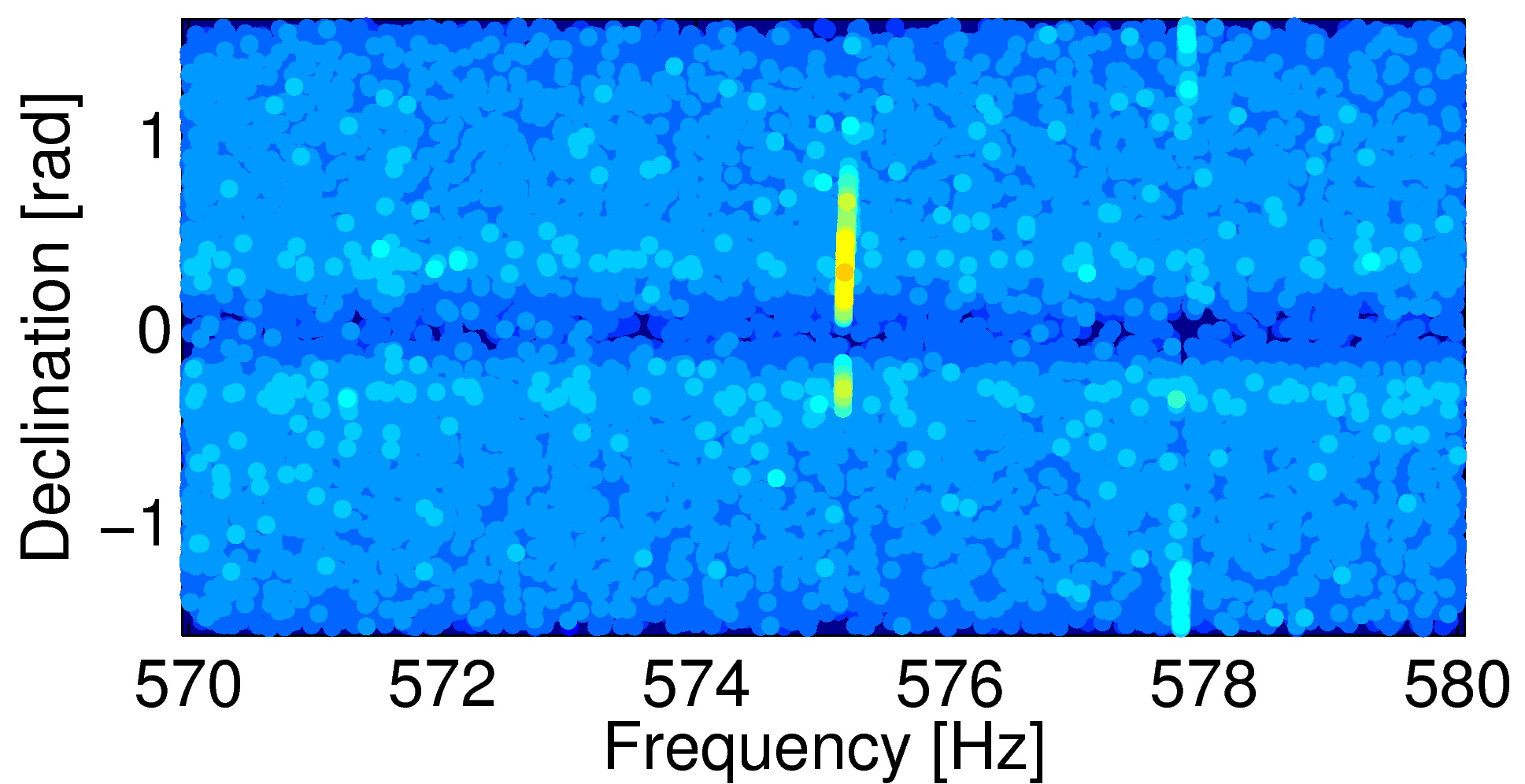}}\\
  \subfigure{\includegraphics[scale=0.44,angle=0]{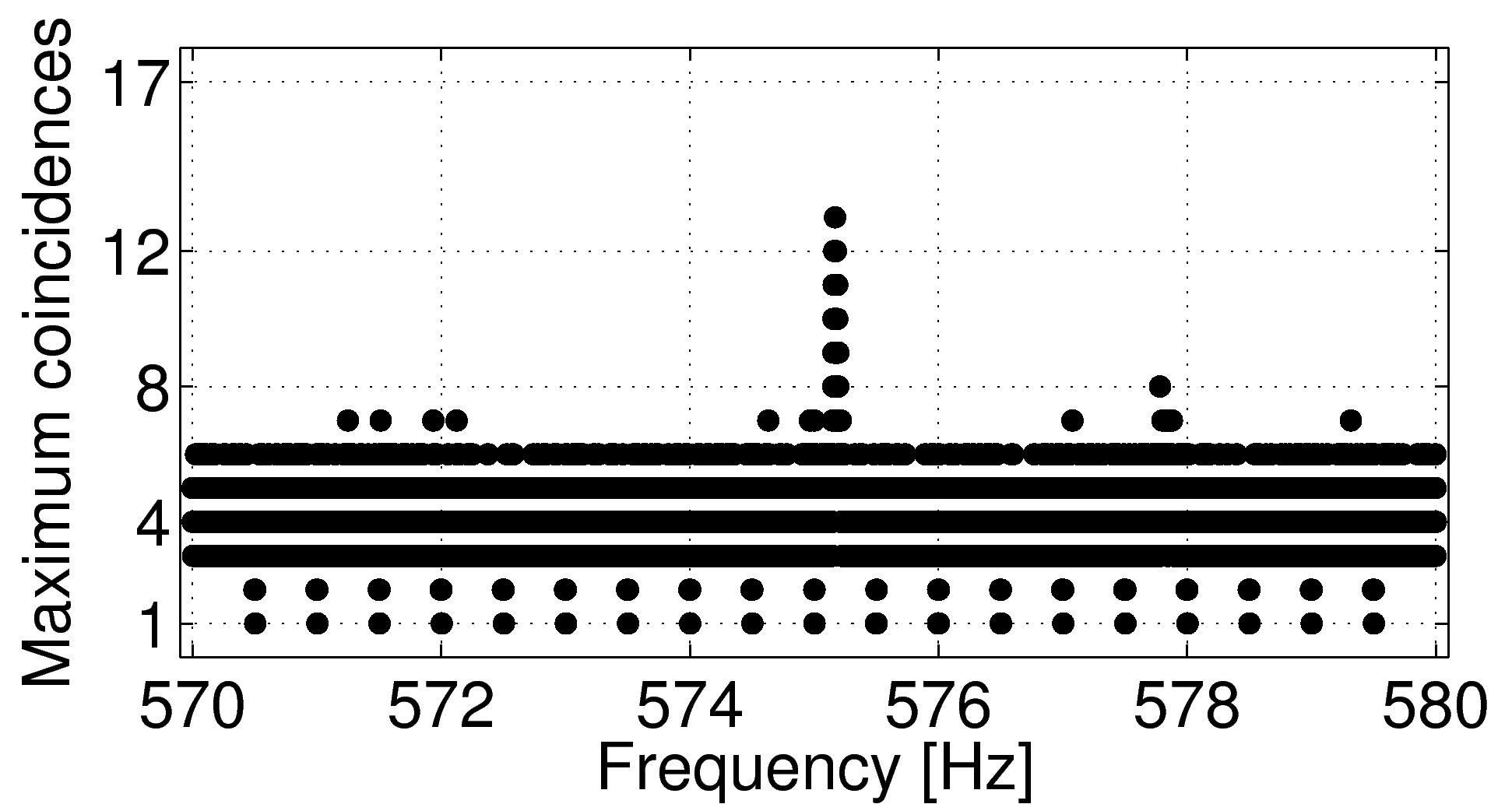}}\\
  \caption{Einstein@Home S4 Post-Processing results for the frequency
    band 570.0-580.0~Hz including a hardware injected CW signal
    (Pulsar2).  From top to bottom the different plots show the numbers of
    coincidences in a 3D map of sky and spin-down, in a 2D Hammer-Aitoff
    projection of the sky, in a 2D plot of declination over frequency, and
    in a histogram as a function of frequency.
    \label{f:results570}}
\end{figure}

\begin{figure}[ht]
	\subfigure{\includegraphics[scale=0.44,angle=0]{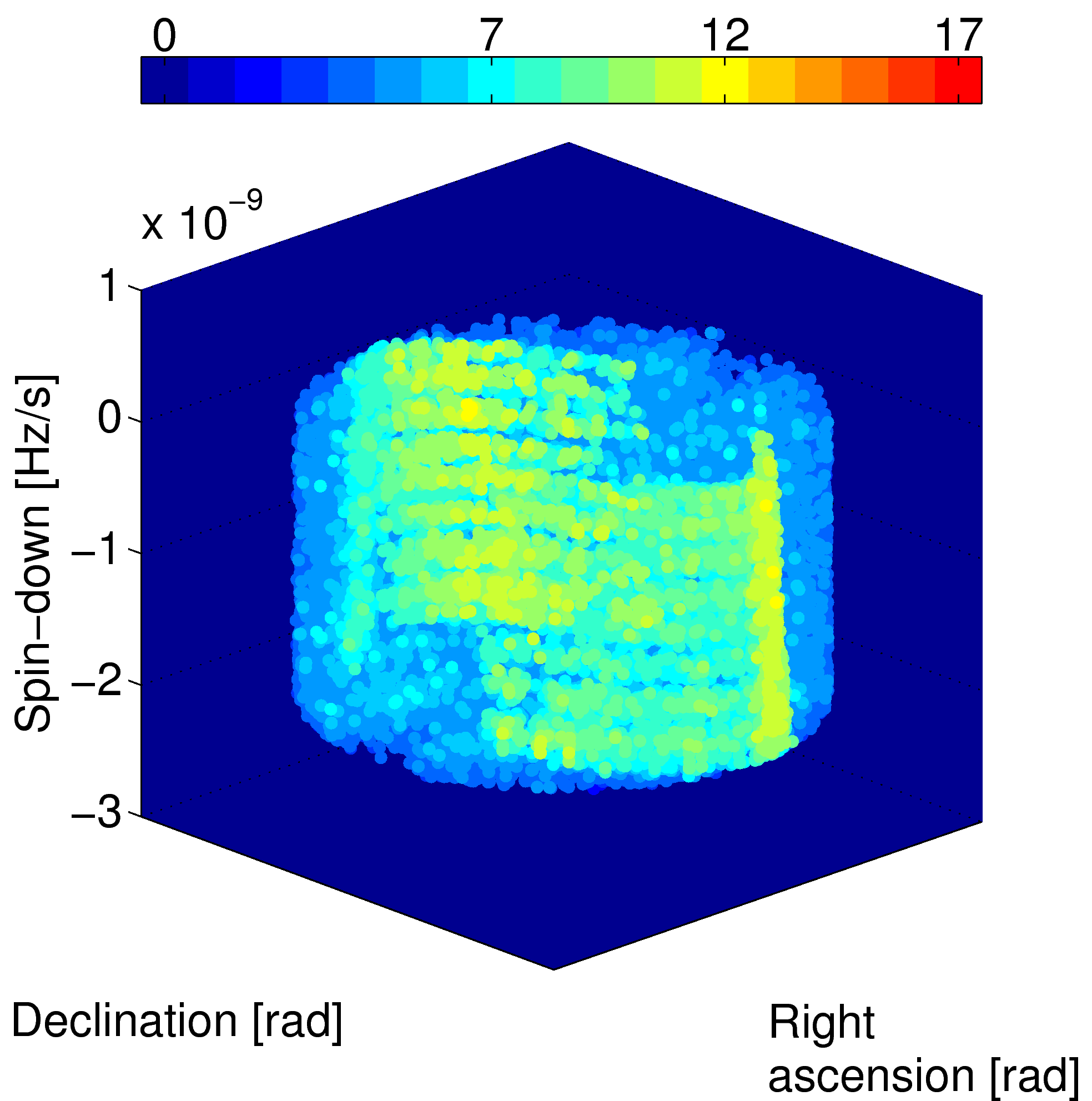}}
	\subfigure{\includegraphics[scale=0.35,angle=0]{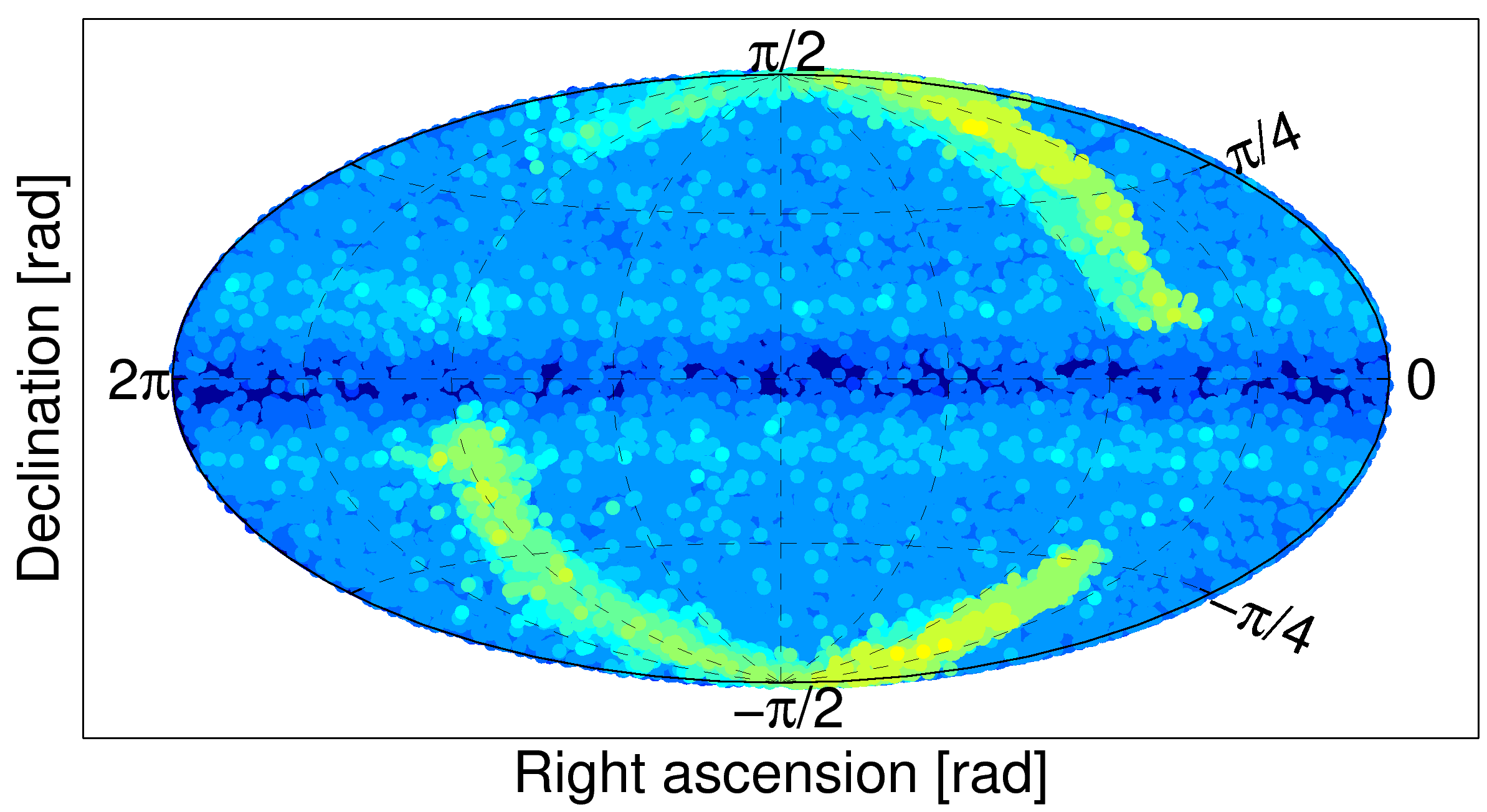}}\\
	\subfigure{\includegraphics[scale=0.44,angle=0]{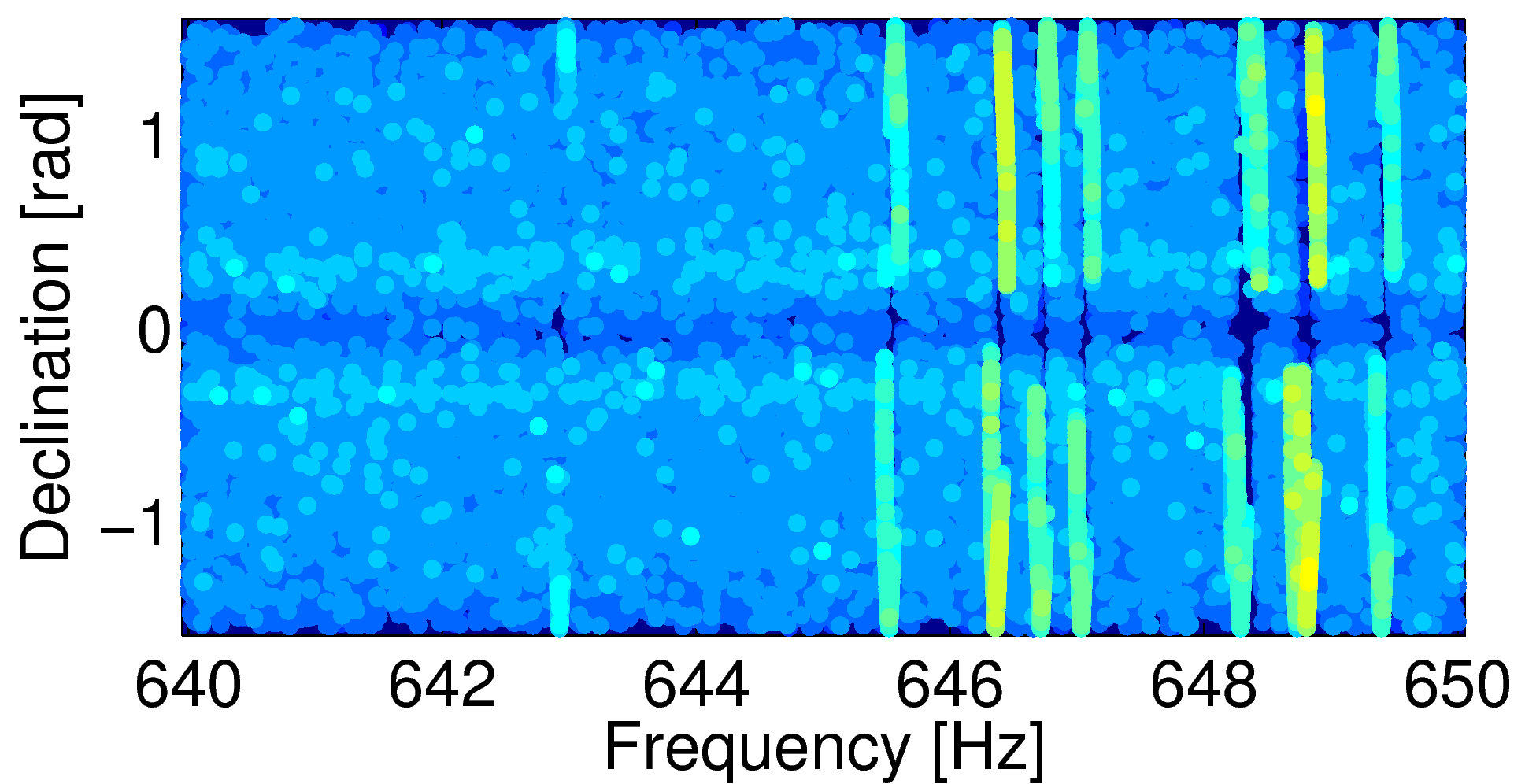}}\\
	\subfigure{\includegraphics[scale=0.44,angle=0]{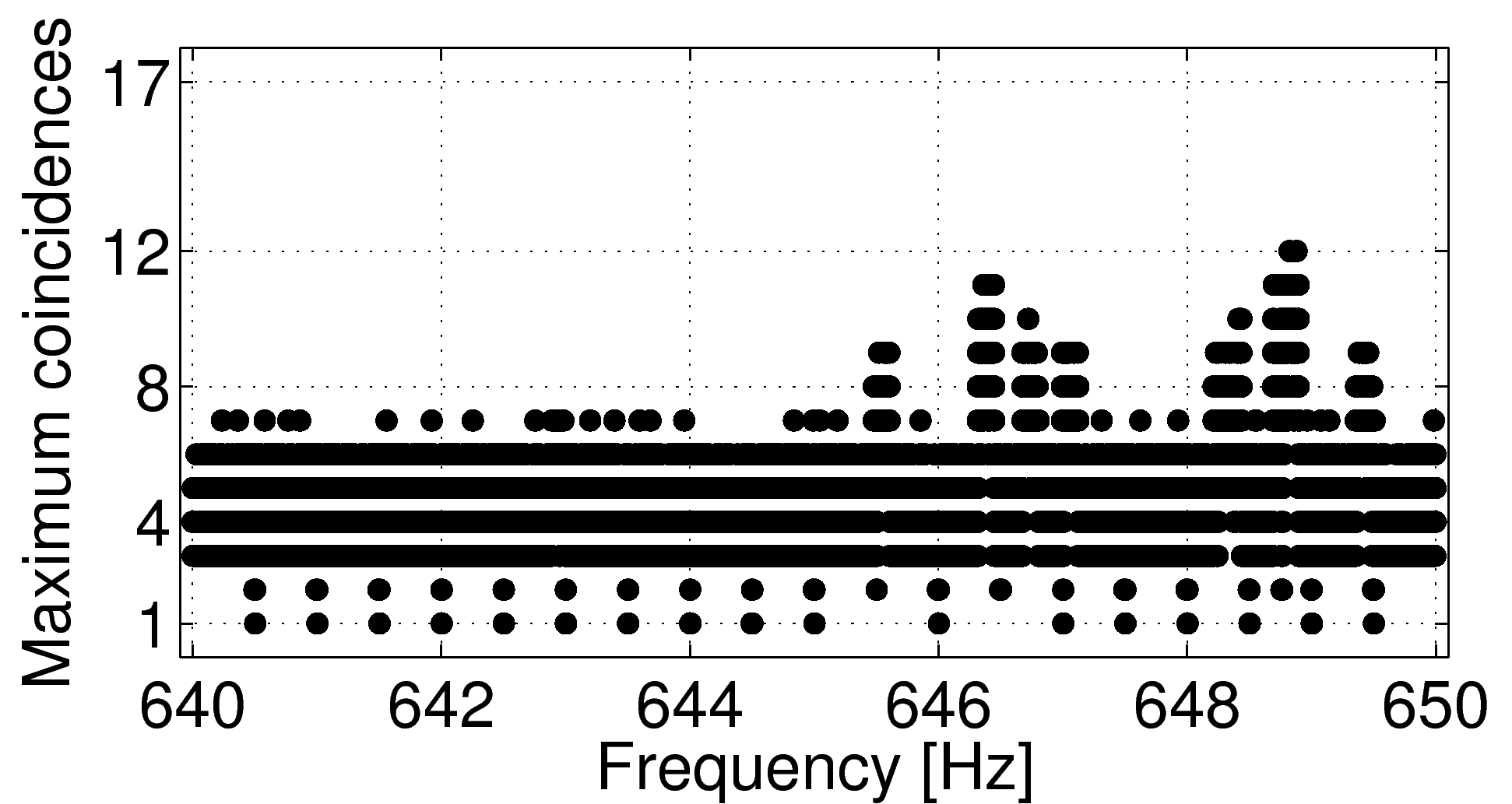}}\\
	\caption{Einstein@Home S4 Post-Processing results for a
          ``noisy'' frequency band of data polluted by instrumental
          noise artifacts from 640.0-650.0~Hz.  These spectral
          features are resonance modes of the mode cleaner optics
          suspensions.  From top to bottom the different plots show
          the numbers of coincidences in a 3D map of sky and
          spin-down, in a 2D Hammer-Aitoff projection of the sky, in a
          2D plot of declination over frequency, and in a histogram as
          a function of frequency.\label{f:results640}}
 \end{figure}

In the following, illustrative examples of different types of typical 
post-processing results in four individual $10\,\rm Hz$ bands are shown in
Figures~\ref{f:results340},~\ref{f:results570},~\ref{f:results110},
and~\ref{f:results640}.  Figure~\ref{f:results340} shows a $10\,\rm Hz$
frequency band containing pure Gaussian
noise. Figure~\ref{f:results570} shows the frequency band of the
hardware-injected signal Pulsar2. Figure~\ref{f:results110} shows a
``quiet'' $10\,\rm Hz$ band of real instrument data without any ``noisy''
lines. In contrast to this, Figure~\ref{f:results640} shows a noisy
band which is polluted by instrumental noise artifacts.

As described in Section~\ref{sec:Results}, each of these plots shows the number 
of coincidences maximized over
the entire sky and full spin-down range. The color indicates the
numbers of coincidences, where the same color-scale has been used in
each figure.

\end{appendix}



\end{document}

%% file: defs.tex
\def\bea{\begin{eqnarray}}
\def\eea{\end{eqnarray}}
\def\beq{\begin{equation}}
\def\eeq{\end{equation}}
\def\vec#1{\mathbf{#1}}
\def\be{\begin{equation}}
\def\ee{\end{equation}}

\newcommand{\ba}{\begin{eqnarray}}
\newcommand{\ea}{\end{eqnarray}}

\newcommand{\m}{\langle}

\newcommand{\bml}{\begin{mathletters}}
\newcommand{\eml}{\end{mathletters}}



%% file: authorlist.tex
%
%
%

\newcommand*{\AG}{Albert-Einstein-Institut, Max-Planck-Institut f\"ur Gravitationsphysik, D-14476 Golm, Germany}
\affiliation{\AG}
\newcommand*{\AH}{Albert-Einstein-Institut, Max-Planck-Institut f\"ur Gravitationsphysik, D-30167 Hannover, Germany}
\affiliation{\AH}
\newcommand*{\AU}{Andrews University, Berrien Springs, MI 49104 USA}
\affiliation{\AU}
\newcommand*{\AN}{Australian National University, Canberra, 0200, Australia}
\affiliation{\AN}
\newcommand*{\CH}{California Institute of Technology, Pasadena, CA  91125, USA}
\affiliation{\CH}
\newcommand*{\CA}{Caltech-CaRT, Pasadena, CA  91125, USA}
\affiliation{\CA}
\newcommand*{\CU}{Cardiff University, Cardiff, CF24 3AA, United Kingdom}
\affiliation{\CU}
\newcommand*{\CL}{Carleton College, Northfield, MN  55057, USA}
\affiliation{\CL}
\newcommand*{\CS}{Charles Sturt University, Wagga Wagga, NSW 2678, Australia}
\affiliation{\CS}
\newcommand*{\CO}{Columbia University, New York, NY  10027, USA}
\affiliation{\CO}
\newcommand*{\ER}{Embry-Riddle Aeronautical University, Prescott, AZ   86301 USA}
\affiliation{\ER}
\newcommand*{\HC}{Hobart and William Smith Colleges, Geneva, NY  14456, USA}
\affiliation{\HC}
\newcommand*{\IA}{Institute of Applied Physics, Nizhny Novgorod, 603950, Russia}
\affiliation{\IA}
\newcommand*{\IU}{Inter-University Centre for Astronomy  and Astrophysics, Pune - 411007, India}
\affiliation{\IU}
\newcommand*{\HU}{Leibniz Universit{\"a}t Hannover, D-30167 Hannover, Germany}
\affiliation{\HU}
\newcommand*{\CT}{LIGO - California Institute of Technology, Pasadena, CA  91125, USA}
\affiliation{\CT}
\newcommand*{\LM}{LIGO - Massachusetts Institute of Technology, Cambridge, MA 02139, USA}
\affiliation{\LM}
\newcommand*{\LO}{LIGO Hanford Observatory, Richland, WA  99352, USA}
\affiliation{\LO}
\newcommand*{\LV}{LIGO Livingston Observatory, Livingston, LA  70754, USA}
\affiliation{\LV}
\newcommand*{\LU}{Louisiana State University, Baton Rouge, LA  70803, USA}
\affiliation{\LU}
\newcommand*{\LE}{Louisiana Tech University, Ruston, LA  71272, USA}
\affiliation{\LE}
\newcommand*{\LL}{Loyola University, New Orleans, LA 70118, USA}
\affiliation{\LL}
\newcommand*{\MS}{Moscow State University, Moscow, 119992, Russia}
\affiliation{\MS}
\newcommand*{\ND}{NASA/Goddard Space Flight Center, Greenbelt, MD  20771, USA}
\affiliation{\ND}
\newcommand*{\NA}{National Astronomical Observatory of Japan, Tokyo  181-8588, Japan}
\affiliation{\NA}
\newcommand*{\NO}{Northwestern University, Evanston, IL  60208, USA}
\affiliation{\NO}
\newcommand*{\RA}{Rutherford Appleton Laboratory, Chilton, Didcot, Oxon OX11 0QX United Kingdom}
\affiliation{\RA}
\newcommand*{\SJ}{San Jose State University, San Jose, CA 95192, USA}
\affiliation{\SJ}
\newcommand*{\SM}{Sonoma State University, Rohnert Park, CA 94928, USA}
\affiliation{\SM}
\newcommand*{\SE}{Southeastern Louisiana University, Hammond, LA  70402, USA}
\affiliation{\SE}
\newcommand*{\SO}{Southern University and A\&M College, Baton Rouge, LA  70813, USA}
\affiliation{\SO}
\newcommand*{\SA}{Stanford University, Stanford, CA  94305, USA}
\affiliation{\SA}
\newcommand*{\SR}{Syracuse University, Syracuse, NY  13244, USA}
\affiliation{\SR}
\newcommand*{\PU}{The Pennsylvania State University, University Park, PA  16802, USA}
\affiliation{\PU}
\newcommand*{\TA}{The University of Texas at Austin, Austin, TX 78712, USA}
\affiliation{\TA}
\newcommand*{\TC}{The University of Texas at Brownsville and Texas Southmost College, Brownsville, TX  78520, USA}
\affiliation{\TC}
\newcommand*{\TR}{Trinity University, San Antonio, TX  78212, USA}
\affiliation{\TR}
\newcommand*{\BB}{Universitat de les Illes Balears, E-07122 Palma de Mallorca, Spain}
\affiliation{\BB}
\newcommand*{\UA}{University of Adelaide, Adelaide, SA 5005, Australia}
\affiliation{\UA}
\newcommand*{\BR}{University of Birmingham, Birmingham, B15 2TT, United Kingdom}
\affiliation{\BR}
\newcommand*{\BE}{University of California at Berkeley, Berkeley, CA 94720 USA}
\affiliation{\BE}
\newcommand*{\FA}{University of Florida, Gainesville, FL  32611, USA}
\affiliation{\FA}
\newcommand*{\GU}{University of Glasgow, Glasgow, G12 8QQ, United Kingdom}
\affiliation{\GU}
\newcommand*{\MD}{University of Maryland, College Park, MD 20742 USA}
\affiliation{\MD}
\newcommand*{\MA}{University of Massachusetts, Amherst, MA 01003 USA}
\affiliation{\MA}
\newcommand*{\MU}{University of Michigan, Ann Arbor, MI  48109, USA}
\affiliation{\MU}
\newcommand*{\MN}{University of Minnesota, Minneapolis, MN 55455, USA}
\affiliation{\MN}
\newcommand*{\OU}{University of Oregon, Eugene, OR  97403, USA}
\affiliation{\OU}
\newcommand*{\RO}{University of Rochester, Rochester, NY  14627, USA}
\affiliation{\RO}
\newcommand*{\SL}{University of Salerno, 84084 Fisciano (Salerno), Italy}
\affiliation{\SL}
\newcommand*{\SN}{University of Sannio at Benevento, I-82100 Benevento, Italy}
\affiliation{\SN}
\newcommand*{\SH}{University of Southampton, Southampton, SO17 1BJ, United Kingdom}
\affiliation{\SH}
\newcommand*{\SC}{University of Strathclyde, Glasgow, G1 1XQ, United Kingdom}
\affiliation{\SC}
\newcommand*{\WA}{University of Western Australia, Crawley, WA 6009, Australia}
\affiliation{\WA}
\newcommand*{\UW}{University of Wisconsin-Milwaukee, Milwaukee, WI  53201, USA}
\affiliation{\UW}
\newcommand*{\WU}{Washington State University, Pullman, WA 99164, USA}
\affiliation{\WU}

\author{B.~Abbott}    \affiliation{\CT}
\author{R.~Abbott}    \affiliation{\CT}
\author{R.~Adhikari}    \affiliation{\CT}
\author{P.~Ajith}    \affiliation{\AH}
\author{B.~Allen}    \affiliation{\AH}  \affiliation{\UW}
\author{G.~Allen}    \affiliation{\SA}
\author{R.~Amin}    \affiliation{\LU}
\author{D.~P.~Anderson} \affiliation{\BE}
\author{S.~B.~Anderson}    \affiliation{\CT}
\author{W.~G.~Anderson}    \affiliation{\UW}
% ...

 \collaboration{The LIGO Scientific Collaboration, http://www.ligo.org}
 \noaffiliation
%
%
%

%% file: linetable.tex
\begin{table}
\begin{minipage}[b]{0.48\columnwidth}%
\centering
\begin{tabular}[t]{rrr}
\hline
& H1 & \\
$f_{\textrm{Line}}$[Hz] & $N$ & $\Delta f_{\textrm{Line}}$[Hz] \\
\hline\hline
1.0 & 1450 & 0.0006 \\
60.0 & 1 & 3.0 \\
60.0 & 24 & 1.0 \\
87.4 & 1 & 0.1 \\
109.9 & 1 & 0.2 \\
122.1 & 1 & 0.3 \\
193.94 & 1 & 0.02 \\
242.1 & 1 & 0.3 \\
279.24 & 1 & 0.2 \\
279.9 & 1 & 0.3 \\
329.55 & 1 & 0.15 \\
329.77 & 1 & 0.07 \\
335.8 & 1 & 1.0 \\
346.0 & 1 & 4.0 \\
392.365 & 1 & 0.01 \\
392.835 & 1 & 0.01 \\
393.1 & 1 & 0.0 \\
393.365 & 1 & 0.01 \\
393.835 & 1 & 0.01 \\
546.05 & 1 & 0.05 \\
564.05 & 1 & 0.05 \\
566.1 & 1 & 0.1 \\
568.1 & 1 & 0.1 \\
646.38 & 1 & 0.02 \\
648.835 & 1 & 0.035 \\
688.5 & 1 & 2.0 \\
694.75 & 1 & 1.25 \\
973.3 & 1 & 0.0 \\
1030.55 & 1 & 0.1 \\
1032.18 & 1 & 0.04 \\
1032.58 & 1 & 0.1 \\
1033.855 & 1 & 0.05 \\
1034.6 & 1 & 0.4 \\
1042.5 & 1 & 1.5 \\
1142.83 & 1 & 0.11 \\
1143.57 & 1 & 0.2 \\
1144.033 & 1 & 0.2 \\
1144.3 & 1 & 0.0 \\
1144.567 & 1 & 0.2 \\
1145.033 & 1 & 0.2 \\
1145.766 & 1 & 0.11 \\
1374.44 & 1 & 0.1 \\
1377.14 & 1 & 0.1 \\
1389.06 & 1 & 0.06 \\
1389.82 & 1 & 0.07 \\
1391.5 & 1 & 0.5 \\
\hline
\end{tabular}
\end{minipage}%
\hfill
\begin{minipage}[t]{0.48\columnwidth}%
\centering
\begin{tabular}[t]{rrr}
\hline
& L1  & \\
$f_{\textrm{Line}}$[Hz] & $N$ & $\Delta f_{\textrm{Line}}$[Hz] \\
\hline\hline
1.0 & 1450 & 0.0006 \\
36.8725 & 39 & 0.8 \\
54.7 & 1 & 0.0 \\
60.0 & 25 & 1.0 \\
108.12 & 1 & 0.28 \\
115.5 & 1 & 0.1 \\
131.9 & 1 & 0.3 \\
137.65 & 1 & 0.15 \\
138.2 & 1 & 0.1 \\
141.75 & 1 & 0.05 \\
143.0 & 1 & 0.2 \\
154.0 & 1 & 0.2 \\
162.35 & 1 & 0.15 \\
168.1 & 1 & 0.3 \\
189.0 & 1 & 0.05 \\
190.683 & 1 & 0.05 \\
191.9 & 1 & 0.3 \\
197.65 & 1 & 0.15 \\
280.9 & 1 & 0.7 \\
329.3 & 1 & 0.2 \\
345.0 & 1 & 10.0 \\
396.7 & 1 & 0.0 \\
648.35 & 1 & 0.15 \\
686.5 & 1 & 1.0 \\
688.83 & 1 & 0.5 \\
693.7 & 1 & 0.7 \\
777.9 & 1 & 0.05 \\
927.7 & 1 & 0.0 \\
1000.00 & 1 & 0.05 \\
1029.5 & 1 & 0.25 \\
1031 & 1 & 0.5 \\
1033.6 & 1 & 0.2 \\
1041 & 1 & 1.0 \\
1151.5 & 1 & 0.0 \\
1372.925 & 1 & 0.075 \\
1374.7 & 1 & 0.1 \\
1375.2 & 1 & 0.1 \\
1380 & 1 & 2.0 \\
1387.4 & 1 & 0.05 \\
1388.5 & 1 & 0.5 \\
\hline
\end{tabular}
\vfill
\end{minipage}
\vfill
\caption{ 
  \label{t:lines} 
  Instrumental lines removed from the input data.
  The three columns show the frequency of the fundamental harmonic
  $f_{\textrm{Line}}$, the number of harmonics $N$, and
  the bandwidth $\Delta f_{\textrm{Line}}$ removed on either side
  of the central frequency (total bandwidth removed per harmonic is 
  $2\Delta f_{\textrm{Line}}$). 
  In total $77.92\,\Hz$ of H1 data and
  $144.29\,\Hz$ of L1 data have been excluded ab initio.
  If $\Delta f_{\textrm{Line}}=0$ then the line-cleaning algorithm replaces 
  a single Fourier bin with the average of bins on either side.  
  The spacing between Fourier bins is $1/1800\,\Hz$.
 }
\end{table}

%% file: acknowledgements.tex
The authors thank the tens of thousands of volunteers who have supported
the Einstein@Home project by donating their computer time and expertise for this
analysis. Without
their contributions, this work would not have been possible.

The authors gratefully acknowledge the support of the United States
National Science Foundation for the construction and operation of
the LIGO Laboratory and the Particle Physics and Astronomy Research
Council of the United Kingdom, the Max-Planck-Society and the State
of Niedersachsen/Germany for support of the construction and
operation of the GEO600 detector. The authors also gratefully
acknowledge the support of the research by these agencies and by the
Australian Research Council, the Natural Sciences and Engineering
Research Council of Canada, the Council of Scientific and Industrial
Research of India, the Department of Science and Technology of
India, the Spanish Ministerio de Educacion y Ciencia, The National
Aeronautics and Space Administration, the John Simon Guggenheim
Foundation, the Alexander von Humboldt Foundation, the Leverhulme
Trust, the David and Lucile Packard Foundation, the Research
Corporation, and the Alfred P. Sloan Foundation.